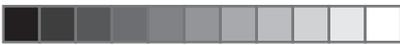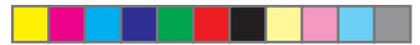

# Ciencia para la paz y en beneficio de la humanidad: El concepto del Juramento Hipocrático para Científicos[1]

## Guillermo A. Lemarchand[2]

A la memoria de Mischa Cotlar (1913-2007)

**Resumen:** *En este trabajo se muestra la importancia que ha tenido la investigación científica, el desarrollo tecnológico y la innovación en el incremento de la letalidad del armamento disponible durante el último siglo. Se describe un conjunto de iniciativas promovidas desde la comunidad científica para detener la carrera armamentista nuclear que puso en peligro la continuidad de la vida en el planeta. En este punto, se hace un relevamiento exhaustivo de los textos y propuestas de Juramentos Hipocráticos para Científicos presentados en distintas épocas. Se observa que el interés por vincular los aspectos éticos con la ciencia y la tecnología crece exponencialmente a partir de la Segunda Guerra Mundial. Se describe cómo las distintas propuestas de juramentos y compromisos éticos para científicos, ingenieros y tecnólogos se difunden siguiendo un crecimiento logístico, de la misma manera que una tecnología desincorporada en un determinado nicho. El análisis de los datos muestra que la tasa máxima de propuestas coincide con el momento histórico de mayor número de ojivas nucleares emplazas (70.586) y el mayor gasto militar global de la historia (USD 1.485.000.000.000). Posteriormente, se analiza el origen del Juramento Hipocrático para Científicos que se emplea desde hace más de dos décadas en las ceremonias de graduación de la Facultad de Ciencias Exactas y Naturales de la Universidad de Buenos Aires y se lo vincula con las circunstancias históricas de su nacimiento.*

**Abstract:** *This article shows the importance that has had the scientific research, the technological development and the innovation processes in increasing the lethality of the available weapons during the last century. A set of initiatives promoted by the scientific community to stop the nuclear arms race that threatened the continuation of life on the planet is described. At this point, a thorough survey of the texts and proposals of Hippocratic Oaths for Scientists presented at different epochs is made. It is observed that the interest in linking ethical aspects with science and technology issues shows an exponential growth behavior since the Second World War. It is shown how the several proposals of oaths and ethical commitments for scientists, engineers and technologists are disseminated following a logistic growth behavior, in the same manner as a disembodied technology in a particular niche. The data analysis shows that there is a coincidence between the maximum rate of proposals and the historical moment at which the world had deployed the largest number of nuclear warheads (70,586) as well as the largest world* *military expenditures in history (USD 1,485,000,000,000). Subsequently, the origin of the Hippocratic Oath for Scientists used for more than two decades in graduation ceremonies at the Faculty of Exact and Natural Sciences of the University of Buenos Aires is analyzed and linked with the historical circumstances of its birth.*

## Introducción

Entre el 11 y 15 de abril de 1988, un grupo de estudiantes de física pertenecientes a la Comisión de Astrofísica del Centro de Estudiantes[3], junto con la Secretaría de Extensión Universitaria de la Facultad de Ciencias Exactas y Naturales (FCEN) de la Universidad de Buenos Aires (UBA), organizaron el *Simposio Internacional sobre los Científicos, la Paz y el Desarme*. Dicha reunión estuvo auspiciada por la UNESCO, la Secretaría de Ciencia y Tecnología de la Nación (SECYT), el Rectorado de la UBA y una enorme variedad de organizaciones no gubernamentales internacionales del ámbito científico (Lemarchand y Pedace 1988).

La reunión contó con 60 expositores provenientes de una docena de países. Entre los disertantes se encontraban Jeremy Stone, Presidente de la Federación de Científicos Americanos (FAS); Jean Marie Legay, Presidente de la Federación Mundial de Trabajadores Científicos (WFSW); Gustavo Malek, Director de la Oficina Regional de Ciencia y Tecnología de la UNESCO para América Latina y el Caribe (ROSTLAC), autoridades nacionales y extranjeras, representantes de la academia y de organizaciones vinculadas a la sociedad civil de la ciencia como: AAIE, AFA,

Grupo Pugwash, IPPNW, ICSC World Laboratory, SIPRI, SBF, USPID[4].

Las presentaciones y contribuciones más sobresalientes estuvieron a cargo de científicos y humanistas de la talla de Daniel Bes, Mario Bunge, Julio César Carasales, Vittorio Canuto, Félix Cernuschi, Arthur Clarke, Mischa Cotlar, Ernst Hamburger, Amílcar Herrera, Manfred Heindler, Gregorio Klimovsky, Francesco Lenci, Carlos A. Mallmann, Luiz Pinguelli Rosa, Abdus Salam, Giorgi Stenchikov, Norma Sánchez, Fernando Souza Barros, y Héctor Torres, por nombrar solo algunos.

Las actas, con una selección de trabajos, fueron publicadas por una editorial internacional a los pocos meses[5]. Entre panelistas, estudiantes y público asistente, el simposio llegó a concentrar a un grupo de más de 500 participantes diarios. La reunión se extendió a lo largo de una semana, alternándose entre sesiones plenarias y hasta tres sesiones simultáneas. El Simposio despertó un interés público y político destacado, el cual se vió reflejado por una importante cobertura en medios locales e internacionales.

---

[3]   *El grupo de estudiantes que organizó con mucha profesionalidad el evento, tenía edades comprendidas entre 19 y 23 años. Entre ellos se encontraban Ricardo Bravo, Marcelo Castro, Flavio Colavecchia, Paula da Cunha, Jordana Dorfman, Ramón Elía, Guillermo Giménez de Castro, Elvira González Folgar, Leonardo Graciotti, Diego Lamas, Marcelo López Fuentes, Gabriela Marani, Graciela Morales, Patricia Olivella, Agnes Paterson, Gerardo Pozetti, Claudia Ramírez, Daniel Rodríguez Sierra, Eduardo Sergio Santini, Andrés Schuschny, Horacio Slavich, Alejandro Valda Ochoa y Alberto Vásquez. La coordinación estuvo a cargo de Guillermo A. Lemarchand (Comisión de Astrofísica, CECEN) y de A. Roque Pedace (Secretario de Extensión Universitaria de la FCEN-UBA).*

[4]   *AAIE es la Asociación Argentina de Investigaciones Éticas. AFA es la Asociación Física Argentina. El grupo de científicos que conforman las Conferencias Pugwash sobre Ciencia y Asuntos Mundiales, fue fundado por Bertrand Russell y Albert Einstein en 1955. En el año 1995 esta organización recibió el Premio Nobel de la Paz. IPPNW es el movimiento Internacional de Médicos para la Prevención de la Guerra Nuclear, que también recibió el Premio Nobel de la Paz en el año 1985. ICSC World Laboratory es el Laboratorio Mundial del Centro Internacional de Cultura Científica de Lausana, fundado por Antonino Zichichi y financiado con fondos que se ahorraron de la carrera armamentista. SIPRI es el Instituto Internacional sobre Investigaciones de la Paz de Estocolmo. SBF es la Sociedad Brasileña de Física. USPID es la Unión de Científicos Italianos para el Desarme.*

[5]   *G. A. Lemarchand & A.R. Pedace, eds., Scientists, Peace and Disarmament, Singapur y Londres, World Scientific, 1988.*

   

Los trabajos de preparación y organización del Simposio comenzaron durante el "Año Internacional de la Paz" declarado por las Naciones Unidas en 1986. Un año que tiene el triste récord de acaparar, por un lado el mayor gasto militar de la historia de la humanidad (USD 1.482.000.000.000 equivalentes de 2009) y por otro el de haber tenido instaladas el mayor número de ojivas nucleares (70.586). Por iniciativa de Hendrik Bramhoff de la Universidad de Hamburgo, durante ese mismo año, científicos de dos docenas de países comenzaron a celebrar la llamada "Semana Internacional de los Científicos y la Paz" (10-16 de noviembre). Este evento se repitió con regularidad en los años subsiguientes. Por esta razón, en 1988, unos meses después de celebrarse el "Simposio", una resolución de la Asamblea General de las Naciones Unidas (ver Apéndice 3) institucionaliza la *Semana Internacional de la Ciencia y la Paz*[6], para todos sus Estados Miembros.

A la distancia, se observa con mayor claridad que el momento histórico en que se desarrolló el simposio era especial. La Argentina hacía unos pocos años que había recuperado la democracia, un proceso que se estaba dando simultáneamente también en otros países de América Latina. Entre los estudiantes, existía una gran efervescencia por realizar actividades trascendentes, y ansias de compromiso y participación social. La comunidad científica de Argentina y Brasil comenzaba a articular estrategias comunes de desarrollo y cooperación.

Por otro lado, aun se estaba viviendo en plena Guerra Fría. Como se mencionó en los párrafos anteriores (ver gráfica 4) el mundo gastaba aproximadamente un billón y medio de dólares anuales (a valores equivalentes del año 2009) en emprendimientos militares. Entre un 10 y 15% de esos gastos estaban destinados exclusivamente a tareas de investigación y desarrollo con fines militares y casi un 20% al desarrollo exclusivo de nuevo armamento nuclear. Se estimaba que un 20% de los científicos del mundo, estaba directa o indirectamente involucrado con este tipo de investigaciones (Lemarchand 1988a).

Por aquel entonces, se gastaban miles de millones de dólares anuales en desarrollar –con gran resistencia de la comunidad científica- la llamada Iniciativa de Defensa Estratégica (vulgarmente conocida como *"Guerra de las galaxias"*). La misma pretendía desplegar una constelación de satélites en el espacio con sofisticadas armas láser y de energía dirigida, con la idea de destruir misiles balísticos intercontinentales en vuelo (Bloembergen *et al.* 1987). En forma independiente habían aparecido también los primeros modelos matemáticos y de simulación numérica que predecían cuáles serían las eventuales consecuencias climáticas globales que podrían devenir a raíz de una guerra nuclear, total o parcial (invierno nuclear).

La Argentina estaba participando activamente dentro de la llamada "Iniciativa de Paz de los Cinco Continentes" o "Grupo de los Seis" integrado por Raúl Alfonsín, Presidente de Argentina; Andreas Papandreu, Primer Ministro de Grecia; Rajiv Gandhi, Primer Ministro de la India; Miguel de la Madrid Hurtado, Presidente de México; Ingvar Carlsson, Primer Ministro de Suecia y Julius Nyerere, Primer Ministro de Tanzania. De las cumbres organizadas en Nueva Delhi (1985), México (1986) y Estocolmo (1988), no solo habían surgido pronunciamientos a favor de la paz, sino varias iniciativas concretas, de orden político y técnico, cuya aplicación intentaba allanar el camino de entendimiento entre las superpotencias de entonces. En cada Cumbre de mandatarios se destacó siempre la amplia participación de prominentes científicos, los

---

[6]    La Resolución de la Asamblea General le cambia el nombre original, reemplazando la palabra "científicos" por "ciencia" (ver Apéndice 3). En base a la resolución de la AG de la ONU, en el 2001, la Conferencia General de la UNESCO declara al 10 de noviembre como "Día Mundial de la Ciencia para la Paz y el Desarrollo" (ver Apéndice 5).

cuales tenían reuniones paralelas y emitían sus propias declaraciones.

Al mismo tiempo, según los datos publicados en el anuario del Instituto Internacional de Investigaciones para la Paz de Estocolmo (SIPRI), en 1987, la Argentina se encontraba encabezando la lista de gastos militares en América Latina, superando en esos años incluso a Australia, Corea del Sur, Israel y Sudáfrica[7].

Dentro de este contexto histórico, se enmarcaron los contenidos del *Simposio Internacional sobre los Científicos, la Paz y el Desarme* y también, la voluntad explícita de un importante grupo de jóvenes estudiantes de ciencias, para promover el uso pacífico del conocimiento científico, en beneficio de la humanidad y a favor de la paz.

De esta manera, los temas seleccionados para tratarse en la agenda incluían:

1. El análisis de diversos modelos de impacto climático de una guerra nuclear global o parcial y el fenómeno denominado "invierno nuclear" (Sagan 1984).

2. Las consecuencias biomédicas de un evento bélico que utilice armamento nuclear.

3. El estudio de los impactos económicos, ambientales, sociales y en vidas humanas del accidente de Chernobyl y el correspondiente manejo de información que realizaron los distintos gobiernos europeos acerca de los efectos reales de la radiación liberada sobre las personas y el medioambiente.

4. El análisis de los problemas técnicos y de control de armamentos que se derivan de los procesos de desarme nuclear.

5. Las consecuencias económicas y limitaciones para el desarrollo de los países que imponía la carrera armamentista.

6. El Tratado de Tlatelolco y la desnuclearización del Atlántico Sur.

7. El análisis de propuestas de cielos abiertos para el control del desarme nuclear.

8. La discusión sobre la militarización del espacio ultraterrestre y el desarrollo de la Iniciativa de Defensa Estratégica (IDE).

9. La dimensión ética, social, e individual, acerca de la responsabilidad de los científicos ante la carrera armamentista.

10. La propuesta de un Juramento Hipocrático para Científicos y de otras actividades para estimular la consciencia de los jóvenes científicos acerca del impacto social que se deriva de la labor científica.

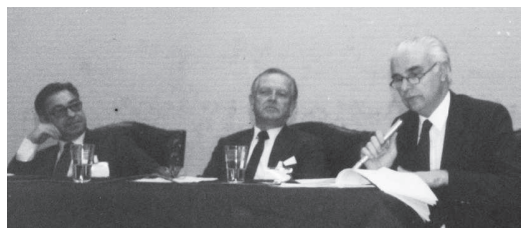

*Lectura de las conclusiones durante la mesa de cierre del Simposio Internacional sobre los Científicos, la Paz y el Desarme, celebrado en el aula magna de la Facultad de Ciencias Exactas y Naturales (FCEN) de la Universidad de Buenos Aires. De izquierda a derecha, Dr. Héctor Torres (Decano de FCEN), Dr. Gustavo Malek (Director de la Oficina Regional de Ciencia y Tecnología de la UNESCO para América Latina y el Caribe) y Dr. Carlos A. Mallmann (Director del Centro de Estudios Avanzados de la UBA). Foto: G. A. Lemarchand (c. 1988).*

Durante el Simposio, la Asociación Física Argentina y la Sociedad Brasileña de Física, dieron a conocer una declaración conjunta, en donde manifestaron su repudio a estudios y proyectos en relación con la construcción de armas nucleares y reiteraron la convicción de sus integrantes, de garantizar que los programas nucleares de ambas naciones se desarrollaran bajo un estricto control

---

civil. Asimismo, presentan una propuesta para promover las inspecciones y el control parlamentario de las instalaciones nucleares de los dos países, con la asistencia técnica de ambas sociedades. Entre estas visitas se coordinaron inspecciones mutuas a la planta de enriquecimiento de uranio de Pilcaniyeu (Argentina) y a la instalación de Iperó, relacionada con el llamado "plan nuclear autónomo" que era manejado estrictamente por los militares brasileños (Lemarchand y Pedace 1988: 377-278 y Bes 1988).

De las actividades mencionadas, la que recibió una mayor cobertura mediática e impacto, nacional e internacional, fue la propuesta de un Juramento Hipocrático para Científicos, conocida como "Juramento de Buenos Aires" (AAAS Committee on Scientific Freedom and Responsibility 1988; Bes 1988; Davis 1988, 1992; Harris 1988; Howard 1988; Lemarchand 1988b y 1988c; Méndez 1988; Morales 1988; Rydén 1990; SANA 1989; Soldatic 1989; Stone 1988; Sweet 1988; Tresch Fienberg 1988; entre otros).

El Simposio fue una reunión verdaderamente interesante, que movilizó tanto a destacados científicos internacionales como a jóvenes estudiantes de ciencias y cuyos frutos se siguen cosechando aun dos décadas después.

Luego de esta breve introducción histórica, en las próximas secciones, se describirá la importancia de la componente científica y tecnológica en la evolución de la capacidad destructiva militar mundial; la resistencia y el llamamiento de distintos grupos de científicos para detener la carrera de armamentos nucleares; se presentará un análisis de las distintas propuestas de compromisos éticos individuales de los científicos y tecnólogos y una descripción de la génesis del Juramento de Buenos Aires que cumplió más de dos décadas de uso durante las ceremonias de graduación de la Facultad de Ciencias Exactas y Naturales de la Universidad de Buenos Aires.

## 2. La evolución de la ciencia y la tecnología militar

Con el advenimiento de la Revolución Copernicana, la visión de la naturaleza, sus leyes, el concepto de la vida y el universo cambiaron drásticamente. Rápidamente, los nuevos descubrimientos facilitaron encontrar aplicaciones prácticas que permitieron que, en poco tiempo, se comenzara a vencer enfermedades, incrementar el rendimiento de las cosechas, revolucionar las comunicaciones y el transporte, comprender cómo se originó el universo y cómo apareció la vida en la Tierra, decodificar el genoma humano y hasta lograr que algunos representantes de nuestra especie *Homo sapiens*, tuvieran la posibilidad de caminar sobre la Luna.

Mientras que los avances de la ciencia, la tecnología y los procesos de innovación productiva, promovieron y generaron prodigiosos progresos en todos los ámbitos de la actividad humana, también han puesto a disposición del hombre un fabuloso poder destructivo[8].

El comienzo de la utilización del conocimiento científico de frontera en aplicaciones bélicas data de la época de la Magna Grecia. Son muy conocidas las historias del armamento desarrollado por Arquímedes (c. 287-212 a.C.), para evitar el avance de los romanos sobre Siracusa.

Diecisiete siglos después, durante el Renacimiento, Leonardo da Vinci (1452-1519) se

---

[8]   Sánchez Ron (2007: 707-869) hace una detallada y documentada descripción del proceso de militarización de la ciencia y de las actitudes de los científicos, desde 1939 hasta finales de la Guerra Fría. Otros textos interesantes sobre el mismo tema son los de S. Drell (1999), H. York (1995), L. Badash (2005), la selección de artículos publicados originalmente en Physics Today realizada por D. Hafemeister (1991) y de aquellos publicados en el Bulletin of the Atomic Scientists, compilados por L. Ackland y S. McGuire (1987). Además, existe una gran variedad de artículos disponibles en los portales de las Conferencias Pugwash sobre Ciencia y Asuntos Mundiales <www.pugwash.org>, de INES o Red Internacional de Científicos e Ingenieros por la Responsabilidad Global <www.inesglobal.com>, el sitio del Bulletin of the Atomic Scientists <www.thebulletin.metapress.com>, el portal del Union of Concern Scientists <www.ucsusa.org>, o en el sitio del Instituto Internacional de Estudios sobre la Paz de Estocolmo (SIPRI) <www.sipri.org>.

había asegurado un salario al usar sus habilidades en la construcción de fortificaciones para el Duque de Milán. Sin embargo, cuando percibió las potenciales aplicaciones militares de uno de sus inventos (el submarino), tomó la sabia decisión de deshacerse de todo el material vinculado con el diseño del mismo que ya había desarrollado. En uno de sus manuscritos se lee la siguiente justificación: *"en virtud de la naturaleza perversa de los hombres, quienes podrían utilizar desde el fondo del mar el invento, para asesinar a las tripulaciones completas de los barcos mediante la destrucción de las partes más bajas de las embarcaciones"*. Esta decisión de Leonardo retrasó 300 años el nefasto uso que visionariamente ya había pronosticado.

En los albores de la Revolución Científica, el filósofo y político inglés Francis Bacon (1521-1626), escribió una novela utópica *(La Nueva Atlántida),* en donde llegó a imaginarse la existencia de una civilización muy avanzada que estaba organizada y regulada únicamente por principios científicos. Los habitantes de esta verdadera República de la Ciencia, poseían estrictos mecanismos de auto-regulación que prohibían a sus científicos revelar conocimientos o invenciones potencialmente peligrosas para la sociedad. El siguiente párrafo extraído de dicha novela es más que elocuente:

*"Celebramos consultas para acordar cuáles son las invenciones y experiencias descubiertas que deben darse a conocer, y cuáles no; luego se toma a todos un juramento para guardar secreto respecto a las que consideramos que así conviene que se haga, y a veces unas las revelamos al Estado y otras no."*

Esta cita es la causa por la cual, en la literatura inglesa también se suele denominar a los "Juramentos Hipocráticos para Científicos" como "Juramentos Baconianos".

Durante la Guerra de Crimea el Gobierno Británico consultó al notable físico Michael Fa-

raday (1791-1867), sobre la posibilidad de atacar al enemigo con gases venenosos. Faraday respondió que era posible, pero que sería totalmente inhumano y se negó rotundamente a participar de alguna forma en ese plan.

Fue realmente la capacidad de invención e innovación -más que la ciencia- el verdadero motor detrás de la revolución industrial. Sin embargo, en esta época comenzaron a gestarse las primeras alianzas entre la investigación científica, el desarrollo tecnológico y la innovación productiva. Por entonces, el tiempo que transcurría entre un descubrimiento científico y su aplicación práctica podría llegar a extenderse decenas de décadas. Poco a poco, los nuevos conocimientos científicos y las tecnologías desarrolladas a partir de ellos, comenzaron a ser aplicados sistemáticamente, no solo al desarrollo industrial sino también al diseño de nuevo armamento.

A partir de 1939, los científicos ingresaron en número creciente a todos los "departamentos de guerra", se desarrolló rápidamente la investigación científica al servicio de lo militar y los métodos de la ciencia se aplicaron a la dirección efectiva de las operaciones en tierra, mar y aire. Una vez terminada la guerra ese proceso no se detuvo, sino que por el contrario continuó a un ritmo de crecimiento exponencial a lo largo de todo el siglo XX.

Para cuantificar el enorme impacto del conocimiento científico y tecnológico en el desarrollo de nuevo armamento, se representa en la gráfica 1–en escala semilogarítmica- el crecimiento de la capacidad destructiva de las armas disponibles a lo largo de los últimos 150 años. Aquí se muestra la evolución temporal del llamado *coeficiente de letalidad* del armamento más representativo de cada época. Este índice mide el número de bajas humanas que una determinada arma puede causar a lo largo de una hora de uso. En el cálculo se consideran las distintas características del armamento analizado, como por ejemplo, la tasa máxima de fuego, el número máximo de blancos, la eficiencia relativa, la precisión, la confiabilidad, el área afectada, etc.

Un somero análisis de la gráfica 1, no deja duda alguna que merced a la incorporación de conocimientos científicos y tecnológicos en el diseño del nuevo armamento, el coeficiente

de letalidad creció en forma exponencial a lo largo de todo el siglo XX (Lemarchand 2007). Se debe aclarar aquí, que en esta representación solo se ha considerado la evolución del armamento convencional y nuclear. No se ha incluido ni el armamento químico ni el biológico por carecer de datos certeros acerca del coeficiente de letalidad de éstos últimos.

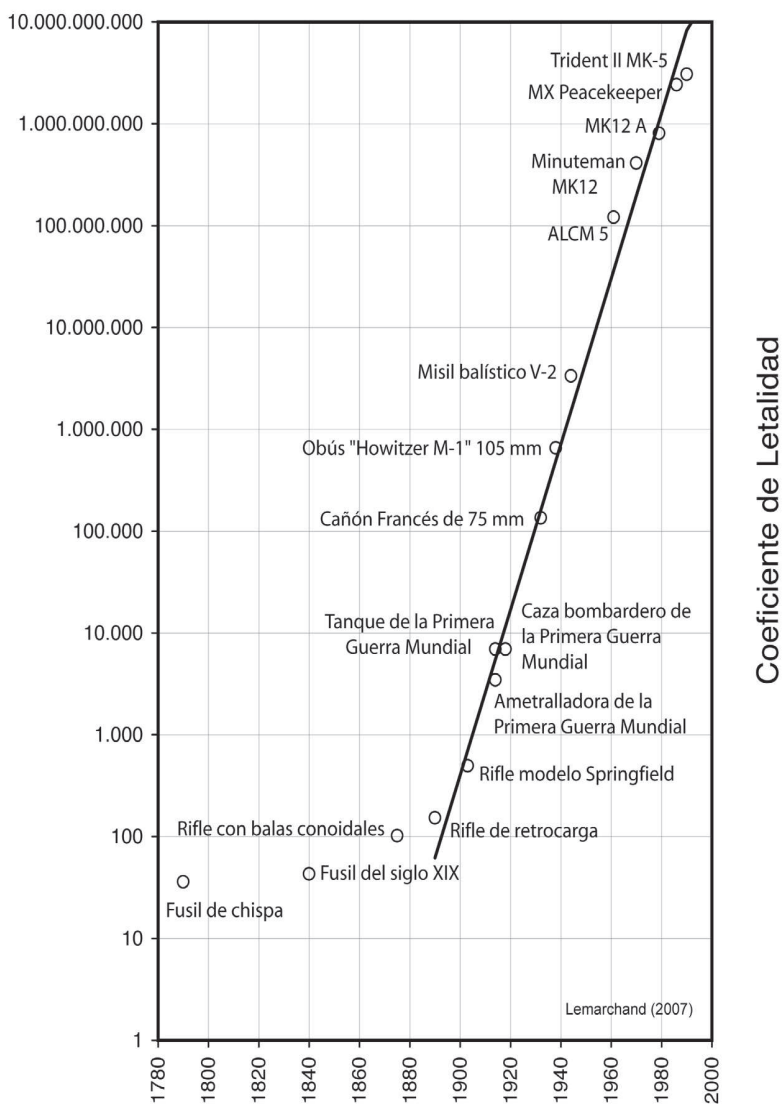

**Gráfica 1:** Representación en escala semilogarítmica de la evolución del coeficiente de letalidad del armamento entre 1780 y 2000, que mide el número teórico de bajas humanas por hora que cada arma puede originar. El índice toma en cuenta las propiedades del armamento como tasa máxima de fuego, número máximo de blancos, eficiencia relativa, precisión, confiabilidad, etc. Claramente, merced a la incorporación de conocimientos científicos y tecnológicos en el diseño de nuevo armamento, el coeficiente de letalidad creció exponencialmente en un factor 60.000.000 a lo largo del siglo XX. Fuente: Lemarchand (2007).

 

Si extendiéramos el estudio hacia atrás en el tiempo, se podría observar que desde la época de los primeros filósofos griegos (siglo V a.C.), hasta principios del siglo XIX, el coeficiente de letalidad del armamento disponible tenía un valor mínimo y máximo respectivamente de 10 y 50 personas por hora. O sea que se multiplicó solo en un factor 5 a lo largo de 2.300 años. Por otra parte, durante el siglo XIX se multiplicó en un factor 2, mientras que a lo largo del siglo XX este indicador llegó a multiplicarse en un factor 60.000.000. Esto implica que la tasa de crecimiento anual del coeficiente de la letalidad del armamento –disponible durante los últimos cien años- creció a un ritmo 276 millones de veces mayor que durante los 23 siglos anteriores al siglo XIX y 30 millones de veces mayor que durante el siglo XIX.

La gráfica 1 muestra que, con el armamento disponible, sería posible exterminar a la raza humana y a gran parte de la naturaleza en tan solo unas pocas horas. Este hecho determina que estemos atravesando un momento verdaderamente único dentro de la historia evolutiva de la especie. Desde hace solo unas pocas décadas, se dispone por primera vez de la tecnología capaz de garantizar la extinción total de la humanidad (Lemarchand 2010).

La gráfica 2 muestra, por otra parte, la distribución del armamento nuclear desplegado en el mundo, en función del tiempo desde el final de la Segunda Guerra Mundial (SGM) al presente. La figura señala que durante el año 1986 existían, emplazadas en el planeta, la aterradora cantidad de 70.586 ojivas nucleares. A veinte años de la caída del muro de Berlín y el desplome de la ex-Unión Soviética, aun en el año 2009, se cuenta todavía con aproximadamente 23.360 ojivas nucleares activas, localizadas en 111 sitios en 14 países diferentes.

Una noticia esperanzadora la constituye el hecho que aproximadamente la mitad de este arsenal nuclear, está a la espera de ser desmantelado, sumado al acuerdo firmado por EEUU y Rusia en el 2010 para reducir aún más el arsenal. Sin embargo, dada la obsolescencia del armamento nuclear aun en operación, se siguen presentando argumentos y desarrollando nuevos estudios para reemplazar el armamento nuclear actual por otros de nueva generación cuyo costo solo en EEUU sería del orden de USD 21.000 millones (Biello 2007).

En la tabla 1 se detalla la distribución geográfica de las ojivas nucleares que se encontraban aun en funcionamiento en el año 2009.

**Tabla 1:** Estimación del inventario del armamento nuclear mundial en 2009. Fuente: Adaptado de Norris y Kristensen (2009)

| País | Cantidad de ojivas nucleares |
|------|------------------------------|
| Federación de Rusia | 13.000* |
| Estados Unidos | 9.400** |
| Francia | 300 |
| China | 240 |
| Gran Bretaña | 180 |
| Israel | 80 -100 |
| Pakistán | 70–90 |
| India | 60–80 |
| Corea del Norte | ¿? |
| Total | ≈ 23.360 |

**Notas de la Tabla 1:**

* Aproximadamente 4.850 ojivas en Rusia están operacionales y activas. El estatus de las 8.150 ojivas restantes es incierto. Una fracción estaría en reserva esperando ser desmantelada.

** Aproximadamente 5.200 ojivas de EEUU se encuentran dentro del arsenal militar. Unas 2.700 están desplegadas otro tanto en reserva; 4.200 están esperando ser desmanteladas.

Otra manera de percibir el peligro real que el armamento nuclear representa para la vida en el planeta es mediante la estimación de la capacidad destructiva equivalente del arsenal atómico. En general, la unidad de medida que

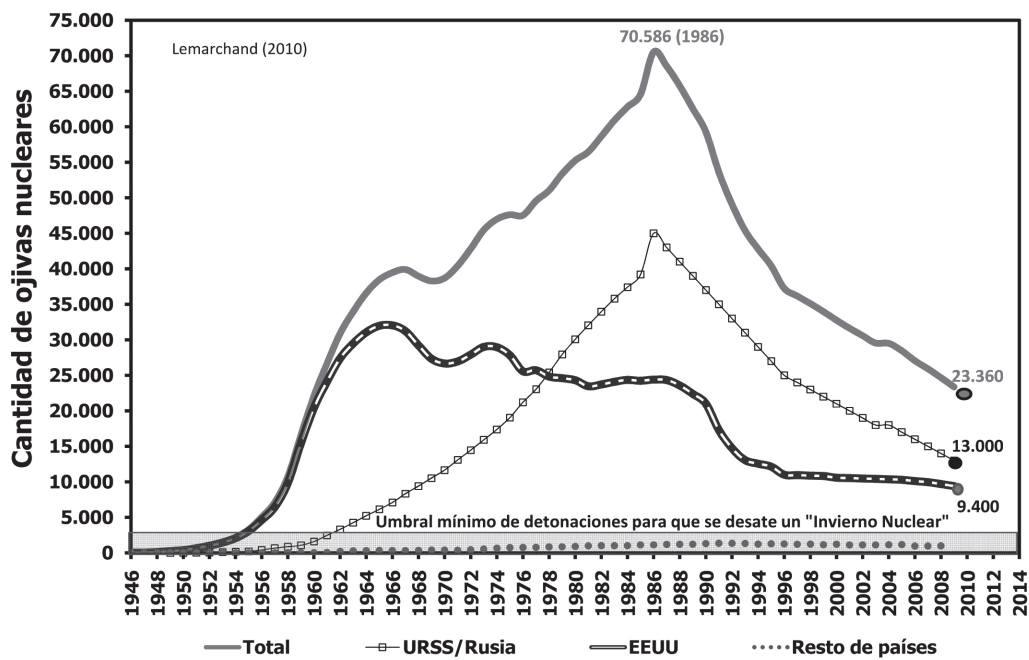

**Gráfica 2:** Evolución de la distribución del número de ojivas nucleares existentes en el planeta (1946-2009). En el año 1986 se llegó a tener 70.586 ojivas nucleares desplegadas. En la actualidad hay más de 23.360, lo que equivale a casi once veces el umbral mínimo de 2.000 ojivas -según los modelos originales de los años ochenta- necesario para desatar una catástrofe climática global denominada "invierno nuclear", que precipitaría una extinción en masa de todas las formas de vida en el planeta. Sin embargo, recientes estudios de Robok *et al.* (2007), Toon *et al.* (2008) y Mills *et al.* (2008) con modelos más realistas de la atmósfera terrestre demostraron que una pequeña guerra regional con solo 100 bombas nucleares podría generar una catástrofe global. Fuente: Elaboración propia en base a datos fuente publicados a lo largo de los años en diversos números del *Bulletin of the Atomic Scientists.*

se suele utilizar en estos casos para determinar el poder de destrucción de las bombas nucleares es el "megatón", que equivale a un millón de toneladas de TNT (dinamita). Estos son números tan indecentemente grandes, que su valor real escapa a toda comprensión basada en la experiencia humana. Para llevar el análisis a una unidad de medida accesible, en la gráfica 3, se representa la evolución anual del poder destructivo en megatones del arsenal nuclear global, dividido por la población mundial en cada año. De esta manera se representa la evolución temporal (1954-2008) del número de toneladas equivalentes de TNT "por persona".

Una rápida mirada muestra que en los años 1960 y 1970 se llegó a disponer de unos

7.000 kg de dinamita por cada ser humano, por cada anciano, cada niño, cada mujer y hombre. Merced a los tratados de desarme, especialmente luego del final de la Guerra Fría y también debido al efecto del crecimiento demográfico, en el presente, ese número bajó a unos 1.200 kg de dinamita por persona. Sin embargo, si se quisiera ser más riguroso, los escandalosos números anteriores deberían ser considerados solo como umbrales mínimos, ya que para poder estimar las proporciones reales es imprescindible añadirles los valores del poder destructivo del armamento convencional, químico y biológico que existe emplazado hoy en el mundo.

Debido al envejecimiento de los sistemas de alerta temprana de ataque nuclear, a las fa-

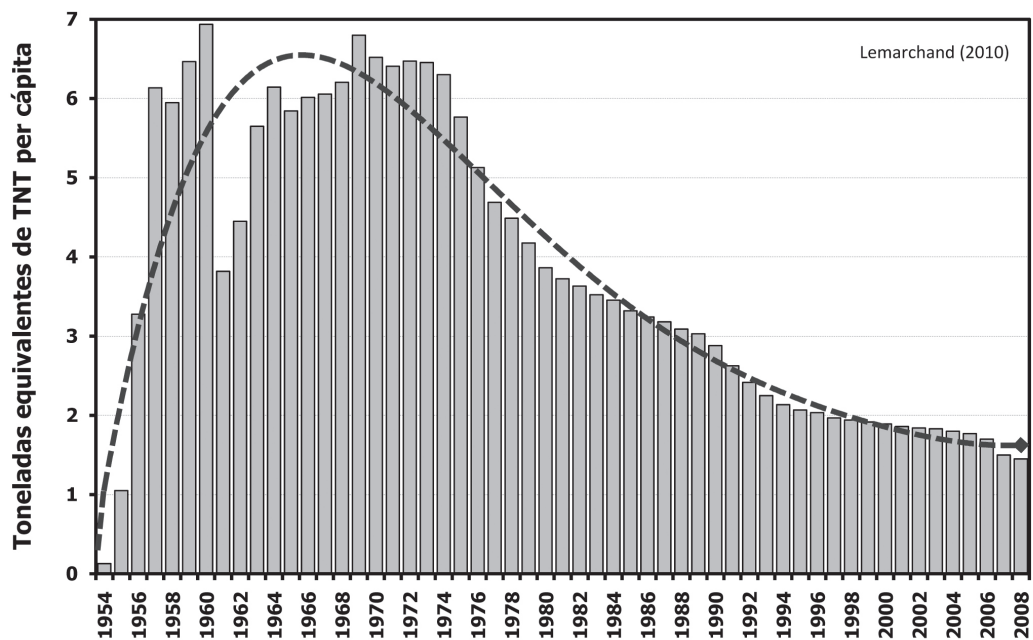

**Gráfica 3:** Evolución de la capacidad destructiva equivalente del armamento nuclear desplegado por año dividido la población mundial de ese mismo año (1954-2008). La unidad resultante de medida es la cantidad de toneladas de TNT equivalentes por persona en cada año. Fuente: Elaboración propia basado en Lemarchand (2010).

llas humanas y a los errores no detectados en sofisticados sistemas de software, las estadísticas publicadas muestran que durante la Guerra Fría existió una altísima tasa de falsas alarmas de ataques nucleares que por fortuna pudieron ser detectados a tiempo y evitar de esa manera que se desate una guerra total por error[9] (Wallace *et al.* 1986).

Aplicando la teoría matemática de colas se puede determinar, estadísticamente, cuánto tiempo se debe esperar para que dos o más sistemas de alerta temprana de ataque nu-clear indiquen simultáneamente una falsa alarma. Usando los niveles de tasas semanales de falsas alarmas publicados oficialmente, a principios de la década del ochenta, los cálculos realizados por Wallace *et al.* (1986) y refinados luego por Blair (1993) muestran que aproximadamente cada 7 años existe la probabilidad de que un conjunto de sistemas indiquen simultáneamente, por error, que existe un ataque nuclear masivo desatado por el bando contrario.

Considerando que los tiempos de vuelo son de aproximadamente 30 minutos para los misiles balísticos intercontinentales, y de solo 10 minutos para aquellos misiles lanzados desde submarinos, los responsables tienen solo la mitad de ese tiempo para verificar la autenticidad de dicho ataque y eventualmente decidir tomar represalias. Los modelos desarrollados muestran también cómo las circunstancias del clima político internacional aligeran (estado actual) o presionan (por ejemplo durante la crisis de los misiles de Cuba) los procesos de toma de decisión, disminuyendo o aumentando respectivamente la probabilidad de que se desate una guerra nuclear por error. Debido al impacto que se generó ante la opinión pública, estas estadísticas se dejaron de dar a conocer a mediados de la década del ochenta.

---

[9]     *Por ejemplo, el 5 de octubre de 1960, el sistema de alerta temprana del North American Aerospace Defense Command (NORAD) indicaba que EEUU estaba bajo un ataque nuclear con una certeza del 99,9%, cuando en realidad la señal estaba siendo generada por el eco en la Luna de un radar en Groenlandia. El 9 de noviembre de 1979, una cinta con datos de simulación de un ataque nuclear fue conectada por error del operador al sistema del NORAD. Durante los 6 minutos del alerta, 10 bombarderos despegaron de las bases del norte de EEUU. El 3 de junio de 1980, una falla en un chip del sistema generaba números aleatorios de misiles en dirección de EEUU. Para dar un ejemplo de la frecuencia de este tipo de eventos, entre 1977 y 1984 se generaron 1.152 falsas alarmas de ataque nuclear. Esto representa una tasa de 3 falsas alarmas por semana.*

En el presente hay 9 países que tienen armamento nuclear y existe una controversia acerca de la posibilidad real de que Irán se una al club. Debido a que algunos de esos países también disponen de misiles balísticos intercontinentales (Estados Unidos, Rusia y China) éstos poseen la capacidad técnica de bombardear con armamento nuclear prácticamente cualquier región del planeta. Asimismo, se debe añadir a la lista anterior aquellos países que tienen la disponibilidad de lanzar armamento nuclear desde submarinos (Francia y Gran Bretaña). Con este grupo de actores, Fischetti (2007) realizó un conjunto de simulaciones numéricas para demostrar, tanto el rango de alcance de ataque de cada miembro del club nuclear, como así también las consecuencias inmediatas -en vidas humanas- del estallido a gran altura de bombas de hidrógeno de un megatón en algunas de las grandes ciudades del mundo. Por ejemplo, en el caso de Nueva York morirían a las pocas horas unos 4,2 millones de personas; en Londres 2,8 millones; en Nueva Delhi 8,5 millones y en Beijing 4,6 millones. A estos umbrales mínimos le debemos sumar las muertes que devienen del colapso de los sistemas sanitarios, el exceso de radiación y otras catástrofes que se desatarían a partir de una tragedia de semejante magnitud.

Se debe destacar aquí, que los conservadores cálculos de los primeros modelos de la década de los ochenta (Sagan 1984) mostraban que con tan solo unas 2.000 ojivas nucleares se podría desatar un invierno nuclear de características globales. Este evento climático planetario podría generar la extinción masiva de la especie humana y también un porcentaje significativo de toda la naturaleza. La gráfica 2 mostraba que aun hoy se está once veces por encima de ese umbral estimado hace 25 años.

Sin embargo, estudios recientes, utilizando modelos más sofisticados y realistas sobre el comportamiento de la atmósfera terrestre, llegaron a la conclusión que esos umbrales originales estaban severamente subestimados (Robok *et al.* 2007; Toon *et al.* 2008, Robok y Toon 2010).

Se demostró que un intercambio nuclear a escala regional podría desatar consecuencias ambientales a escala global. Ahora se sabe que si tan solo 100 bombas del tamaño de la de Hiroshima (que representan el 0,4% del arsenal mundial de 2009) fueran lanzadas sobre ciudades y zonas industriales, se podría inyectar tanto humo y polvo a la atmósfera que sería suficiente para aniquilar completamente el sistema agrícola mundial. Una guerra regional podría causar la pérdida generalizada de vidas humanas incluso en países alejados del conflicto.

Por ejemplo, si se desatara una hipotética guerra nuclear entre la India y Pakistán empleando el arsenal que estos países ya disponen (Tabla 1), no solo se masacraría a la población de estos países, sino que se daría lugar también a un colapso climático global (Robok y Toon 2010). Veinte millones de personas en la región podrían morir instantáneamente a causa de las explosiones de las bombas, los incendios y la radiación. Luego mil millones de personas en todo el mundo podrían morir de hambre debido al colapso agrícola resultante.

Por otra parte, Mills *et al.* (2008), determinaron cuáles serían las consecuencias para la capa de ozono atmosférico. Aplicando el mismo escenario de conflicto regional planteado en el párrafo anterior, encontraron pérdidas de más del 20% del ozono total a nivel mundial, 25-45% del ozono en las latitudes medias, y 50-70% en latitudes septentrionales altas. Los estudios muestran que estas bajas en los niveles de ozono persistirían al menos por unos 5 años. Como resultado se obtendrían aumentos de la radiación UV que podrían afectar significativamente la biosfera, incluidas graves consecuencias para la salud humana. La causa primaria para esta dramática y persistente disminución del ozono atmosférico, sería el calentamiento que sufriría la estratosfera por acción del humo, que absorbe fuertemente la radiación solar.

Las pérdidas de ozono predichas por los nuevos modelos son significativamente mayores que las estimadas en los primeros cálculos de invierno nuclear. A la distancia, los modelos originales se asemejan a lo que ahora sería considerada una verdadera "primavera nuclear". En particular si se considera específicamente el agotamiento de los niveles de ozono y el incremento de los niveles de radiación UV que llegarían a la superficie terrestre que los cálculos modernos muestran.

Los avances científico-tecnológicos de los últimos cien años no solo han sido aplicados al desarrollo del costoso y peligroso armamento nuclear, otras áreas, como la química y la biológica, resultaron tener aplicaciones militares tan letales como la primera pero mucho más económicas.

Las armas químicas fueron ya utilizadas durante la Primera Guerra Mundial. Su acción indiscriminada generó aproximadamente 100.000 muertes. Pese a la vigencia de tratados y convenciones internacionales, durante la Guerra Fría se desarrollaron y acumularon una gran variedad de armamento químico de varios tipos (agentes nerviosos, agentes vesicantes, agentes asfixiantes y agentes incapacitantes). No todos los países han declarado su verdadero inventario de armamento químico (Gillis 2009). Se sabe que hasta julio del 2009 solo se había desactivado el 44% de las 70.000 toneladas métricas de armamento químico declaradas oficialmente.

Formalmente, tampoco ningún Estado reconoce poseer armamento biológico. Sin embargo, en los últimos años ha habido un creciente interés y preocupación sobre el tema ya que este tipo de armamento -por su bajo costo de producción y alta letalidad- se adecua a su posible uso en ataques terroristas. La Convención sobre Armas Bacteriológicas y Tóxicas (CABT) fue firmada en 1972 y entró en vigor en 1975. Hasta junio de 2009, la Convención reunía a 163 Estados Parte y 13 Estados Signatarios (www.unog.ch).

Se suele definir a la biotecnología como toda aplicación tecnológica que utiliza sistemas biológicos y organismos vivos o sus derivados para la creación o modificación de productos o procesos destinados a usos específicos. Durante las últimas tres décadas, merced a los prodigiosos avances de la biología molecular e ingeniería genética, se produjo una verdadera revolución en las técnicas biotecnológicas. Éstas han tenido y siguen teniendo un gran impacto en campos tan variados como la medicina y en el control de calidad y seguridad de los alimentos.

La CABT prohíbe el desarrollo, producción y almacenamiento de armas biológicas, pero permite a los Estados Parte realizar actividades de investigación, desarrollo y producción con fines pacíficos o para la defensa y protección contra agentes de guerra biológica. Por otra parte, las técnicas utilizadas para mejorar la salud o para proteger a los soldados y civiles de las peores consecuencias de la guerra biológica podrían también ser aplicadas para diseñar una nueva generación de armas biológicas. Esta dualidad entre fines permitidos y prohibidos por la CABT crea dilemas éticos y vínculos inciertos para los científicos y técnicos involucrados en este tipo de investigaciones. Dado el rápido desarrollo de esta área (las capacidades tecnológicas se duplican cada año), los científicos se están moviendo sobre terreno poco firme para poder darse cuenta si sus actividades de investigación y desarrollo podrían tener o no aplicación directa al diseño de nuevo armamento biológico.

Nixdorff y Bender (2002) analizaron detalladamente las implicaciones éticas que se derivan de las tareas de I+D en biotecnología que podrían tener impacto en el desarrollo de nuevo armamento bacteriológico y toxínico. Desde el advenimiento de la ingeniería genética, identificaron cuatro categorías de manipulaciones o modificaciones de los microorganismos y sus productos que han sido objeto de debate por sus posibles aplicaciones militares: (1) la transferencia a los microorganismos de la resistencia a los antibióticos, (2) la modificación de las propiedades antigénicas de microorganismos; (3) la modificación de la estabilidad del microorganismo hacia el medio ambiente, y (4) la transferencia de propiedades patógenas a los microorganismos. Está claro que estos cuatro tipos de manipulaciones pueden tener también objetivos pacíficos legítimos y eventualmente también contribuir al fortalecimiento del CABT. Sin embargo, no existe duda alguna que también podrían ser utilizados para desarrollar nuevos armamentos bacteriológicos y toxínicos (por ej. manipulación del *Bacillus anthracis*).

En este punto, es obvio que ninguno de los programas de armamento biológico pudo haberse desarrollado sin el liderazgo, participación y estrecha colaboración de científicos pertenecientes a la comunidad biológica y médica. Este hecho desata el siguiente interrogante ¿cómo científicos educados para ayudar a la humanidad utilizan su conocimiento

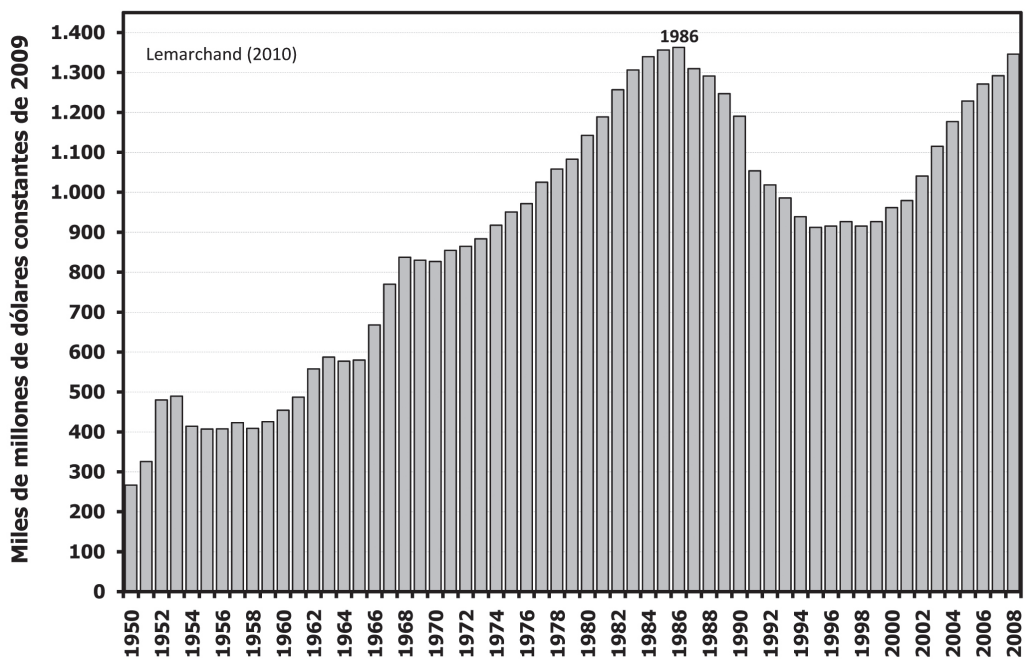

**Gráfica 4:** Evolución de los gastos militares mundiales (1950-2008) expresados en dólares constantes del año 2009. El pico máximo se encuentra en el año 1986. Fuente: Elaboración y cálculos de normalización propios en función de datos originales en moneda corriente publicados en SIPRI Yearbook of World Armaments and Disarmaments (1966, 1972, 1979, 1981, 1986, 1992, 1996, 1999, 2003, 2006, 2009) utilizando el deflactor de PBI de EEUU.

especializado para crear armamento que asesine en masa tanto a militares como civiles? (Guillemin 2006).

Dado el pronunciado carácter de "doble-uso" de estas biotecnologías, resulta sumamente complicado el diseño y establecimiento de regulaciones efectivas para este tipo de tareas de I+D. Se hace necesario, dentro del ámbito académico y político, profundizar los estudios y transparentar aquellos mecanismos que garanticen que estas poderosas herramientas científico-tecnológicas sean utilizadas únicamente en aplicaciones pacíficas.

La gráfica 4 muestra la evolución temporal de los gastos militares mundiales. Se ve claramente un crecimiento constante desde la SGM, llegando a un máximo de USD 1.482.000.000.000 (dólares constantes de 2009) en el año 1986, para disminuir a USD 996.000.000.000 en 1998 y luego del ataque a las Torres Gemelas, volver a aumentar hasta la alarmante cifra de USD 1.440.000.000.000 en el 2008. Estos números, de la misma manera a lo descripto en el caso de la capacidad destructiva del armamento nuclear, están por fuera de la capacidad de percepción del ser humano común.

**Tabla 2:** El costo de alcanzar los Objetivos de Desarrollo del Milenio de las Naciones Unidas como porcentaje del gasto militar mundial

| |
|---|
| **Objetivo:** Erradicar la Pobreza Extrema y el hambre para el 2015<br>Llevar a la mitad la proporción de personas que viven con menos de 1 dólar diario y sufren hambre<br>**Costo:** USD 39.000 a 54.000 millones<br>**Porcentaje del Gasto Militar Global:** 2,6% a 3,7% |
| **Objetivo:** Promover la educación universal y el equilibrio de género para el 2015<br>Alcanzar la educación universal y eliminar la disparidad de género en la educación<br>**Costo:** USD 10.000 a 30.000 millones<br>**Porcentaje del Gasto Militar Global:** 0,7% a 2,0% |
| **Objetivo:** Mejorar la salud para el 2015<br>Reducir en 2/3 la tasa de mortalidad infantil antes de los 5 años, reducir en 3 /4 la tasa de mortalidad materna y revertir la difusión del HIV/SIDA<br>**Costo:** USD 20.000 a 25.000 millones<br>**Porcentaje del Gasto Militar Global:** 1,4% a 1,7% |
| **Objetivo:** Medio ambiente sostenible para el 2015<br>Llevar a la mitad el número de personas sin acceso al agua potable, mejorar las condiciones de vida de más de 100 millones de personas que habitan en villas miserias<br>**Costo:** USD 5.000 a 21.000 millones<br>**Porcentaje del Gasto Militar Global:** 0,3% a 1,4% |
| **Los gastos militares globales utilizados solo en el año 2008, equivalen a los costos totales de las Naciones Unidas durante 732 años de funcionamiento o a los de la UNESCO durante 4.364 años...** |

**Fuente:** Adaptado de Gillis (2009: 13) usando datos tomados de la gráfica 4 y de la publicación: *The Costs of Attaining the Millenium Development Goals*, The World Bank, Washington. Accesible en: http://www.worldbank.org/html/extdr/mdgassessment.pdf

**Nota de la Tabla 2:** La metodología utilizada por el Banco Mundial para estimar los costos de los Objetivos del Milenio, asume que -debido a la superposición de tareas- resulta muchísimo más económico agrupar distintos Objetivos del Milenio entre sí y estimar el costo agregado de lograr dichas metas. Si se optara por estimar el costo en forma individual de cada uno de los 8 objetivos previstos en el programa, la suma total sería muy superior. Por esta razón, en esta tabla se presentan los costos de los objetivos en forma agregada, tal cual se presenta en la estimación del Banco Mundial.

La Tabla 2 tiene por objeto mostrar que con menos del 9 % de los fondos que se emplean durante un solo año en gastos militares globales, sería posible financiar totalmente los programas que permitirían alcanzar los Objetivos del Milenio planteados por las Naciones Unidas.

El 16 de noviembre de 1945 se fundó la Organización de las Naciones Unidas para la Educación, la Ciencia y la Cultura (UNESCO), con el fin alcanzar gradualmente, mediante la cooperación de las naciones del mundo en las esferas de la educación de la ciencia y de la cultura, los objetivos de paz internacional y de bienestar general de la humanidad. La UNESCO obra por crear condiciones propicias para un diálogo entre las civilizaciones, las culturas y los pueblos, fundado en el respeto de los valores comunes. Propone, por medio de este diálogo, forjar en el mundo concepciones de un desarrollo sostenible que suponga la observancia de los derechos humanos, el respeto mutuo y la reducción de la pobreza. Como dice la célebre frase de su preámbulo *"Puesto que las guerras nacen en la mente de los hombres es en la mente de los hombres donde deben erigirse los baluartes de la paz"*.

Si se contrasta el presupuesto asignado por los Estados Miembros a la UNESCO a lo largo de toda su historia, con los gastos militares mundiales acumulados durante el mismo período (1945-2008) se llega a la conclusión que el mundo gastó, durante los últimos 65 años, 310.000 veces más fondos en prepararse para la guerra que en asegurarse la paz a través de la cooperación internacional en temas de educación, ciencia y cultura.

Por otra parte, cuando se analiza la inversión mundial en tareas de investigación y desarrollo militar, en el 2008 se invirtió el equivalente al 10% de los gastos militares mundiales (≈ USD 140.000.000.000). Esta cifra también representa el 15% de todos los gastos mundiales –públicos y privados- en tareas de I+D en todas las áreas del conocimiento (sector farmacéutico, aeroespacial, nuclear, energético, TIC, medicina, ciencias exactas, naturales, sociales, humanas, etc.). Estudios más detallados de cómo se distribuyen las inversiones en tareas de I+D militar pueden encontrarse en los trabajos de Hartley (2006), Setter y Tishler (2006) y Trajtenberg (2006).

El sector de defensa siempre ha estado a la vanguardia identificando aplicaciones militares para las tecnologías emergentes que aparecen constantemente merced a los descubrimientos científicos de frontera. Por ejemplo, el conjunto de tecnologías genéricas englobadas dentro de las llamadas nanotecnologías (NT), resultó ser una de las áreas temáticas que más recursos económicos y humanos en CyT está siendo demandada desde el sector militar. En diez o veinte años, las aplicaciones militares de las NT pueden llegar a ser empleadas en el desarrollo de diminutas computadoras, robots moleculares, nuevas tecnologías de misiles, satélites, lanzadores y sensores. También pueden proporcionar materiales más ligeros y fuertes para la nueva generación de vehículos de transporte y nuevas armas, sofisticados sistemas de vigilancia y control de personas, implantes en los cuerpos de los soldados, armas de fuego libres de metales, sistemas autónomos de lucha, y nuevas generaciones de armas químicas y biológicas mucho más fáciles de transportar (Altmann 2006; Sparrow 2009; Wang y Dotmans 2004).

No existe duda alguna que estas potenciales aplicaciones militares de las NT pueden llegar a plantear graves problemas. Será necesario establecer criterios innovadores para el control de los nuevos armamentos y definir nuevas normativas del derecho internacional, sortear los potenciales peligros de carreras armamentistas y fenómenos de proliferación, y atender las insospechadas consecuencias para los seres humanos y la sociedad.

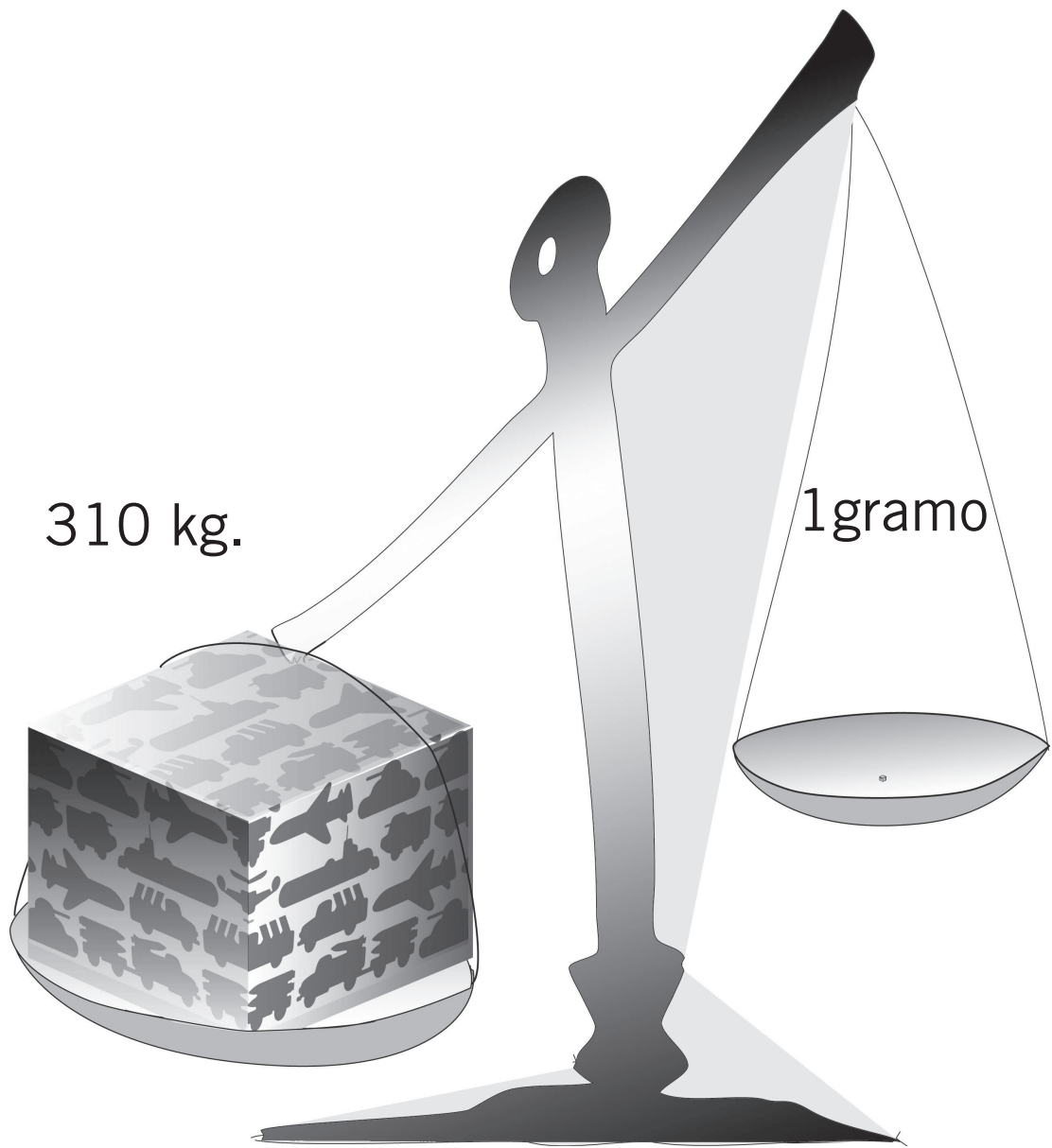

310 kg.

1gramo

Comparación entre la asignación total de recursos (1945-2008) por parte de sus Estados Miembros a la UNESCO, una organización creada para fortalecer la paz a través de la educación, la ciencia y la cultura y la asignación total de recursos que los países del mundo destinaron durante ese mismo período (1945-2008) a gastos militares. Los últimos resultaron ser 310.000 veces más cuantiosos. Ilustración: María Noel Pereyra (UNESCO, 2010).

Algunas de las aplicaciones militares de la NT, tales como nuevos sistemas de computación, estarán también muy cerca de sus posibles aplicaciones civiles. Otras, como los sensores para los agentes de guerra bacteriológica, pueden contribuir a lograr una mayor protección contra los ataques terroristas y ayudar a un mejor control del cumplimiento de los tratados de desarme. Estas aplicaciones de "doble-uso" requieren del desarrollo de nuevos códigos de conducta y procedimientos que pongan el énfasis en determinar adecuadamente cuál será el uso final de cada nuevo desarrollo en NT (Selgelid 2009).

Las llamadas neurociencias, constituyen otra de las áreas que recientemente han sido incorporadas con gran ímpetu a los proyectos de I+D militar. Las implicaciones éticas y filosóficas, que están emergiendo dentro de esta área del conocimiento científico de frontera, son tan sutiles y profundas que una nueva especialidad denominada "neuroética", se instaló ya en el centro de discusión de las revistas especializadas.

Jonathan Moreno (2006), un destacado experto en temas bioéticos, describe algunos de los programas en neurociencias que ya han sido financiados por la Agencia de Proyectos de Investigación Avanzada de Defensa (DARPA) de los Estados Unidos: (1) Interfaces cerebro-máquina ("prótesis neuronales") que permitirían a los pilotos y soldados controlar armas de alta tecnología a través del pensamiento; (2) Desarrollo de cascos de retroalimentación cognitiva para implementar la percepción remota del estado mental y anímico de los soldados; (3) Tecnologías de resonancia magnética *("brain finger printing")* para su uso durante los interrogatorios, inspecciones en los aeropuertos o en la identificación de terroristas[10]; (4) Desarrollo de armas neuro-disruptoras que podrían hacer estragos dentro del cerebro de los soldados enemigos; (5) Desarrollo de agentes biológicos para excitar la liberación de neurotoxinas; (6) Desarrollo de nuevas drogas que permitirían a los soldados dejar de dormir durante días, suprimir recuerdos traumáticos, eliminar el miedo, o reprimir las inhibiciones psicológicas para matar.

Estas temáticas que parecen extraídas de la literatura de ciencia ficción han sido recientemente motivo de un exhaustivo debate entre los científicos que están trabajando en dichos proyectos y destacadas autoridades en el campo de las neurociencias (Canli *et al.* 2007; Resnik 2007; Rosenberg y Gehrie 2007, Marchant y Gulley 2010).

A lo largo de esta sección, se mostraron los intensos vínculos que se han construido -principalmente después de la SGM- entre la comunidad científico-tecnológica y el sector militar, destinados al desarrollo de armamento cada vez más sofisticado. Durante este período se han abierto puertas que podrían conducir a la extinción masiva de la especie humana y otras que despiertan nuevos dilemas éticos.

En 1922, décadas antes de la invención de la bomba atómica, el padre de la geoquímica e inventor del concepto de biosfera, Vladimir Vernadsky (1863-1945) escribía: *"No está lejano el día en que el hombre llegue a adueñarse de la energía atómica, fuente de poder que le permitirá edificar su vida a su gusto ¿Será capaz de utilizar esta fuerza y dirigirla hacia el bien, o por el contrario, la dedicará a su autodestrucción? ¿Está ya lo suficientemente maduro como para saber emplear el poder que la ciencia ha de otorgarle inevitablemente?"*

En este punto, nada parece más apropiado para sintetizar la presentación de esta sección, que finalizar parafraseando a Louis de Broglie (1892-1987). Casi contestándole a Vernadsky, el padre de la mecánica ondulatoria sostenía que para poder sobrevivir al propio progreso de sus conocimientos, los seres humanos deben encontrar en la elevación de su ideal moral, la sabiduría de no abusar de sus fuerzas acrecentadas. Este resulta ser un emprendimiento, que en los albores del siglo XXI, aun dista mucho de ser alcanzado por nuestra especie humana.

---

[10]   Este tipo de aplicaciones tecnológicas abren fuertes interrogantes sobre los posibles errores instrumentales y metodológicos durante los procesos de medición y evaluación, y también acerca de las cuestiones legales de autoincriminación involuntaria.

## 3. La introducción de los temas de ética, responsabilidad social y carrera armamentista en la agenda de los científicos, ingenieros y tecnólogos

Los eventos sucedidos en Hiroshima y Nagasaki despertaron la conciencia acerca de la necesidad de encauzar los resultados de la ciencia y la tecnología únicamente a favor de la paz y en beneficio de toda la humanidad, por parte de intelectuales, filósofos, científicos, humanistas, educadores y en mayor o menor medida de cada ser humano sobre la faz de la Tierra.

Algunos ejemplos históricos muestran el fuerte temor desatado a nivel mundial por las enormes consecuencias que se derivarían de una guerra nuclear. Por ejemplo, en noviembre de 1945, durante el discurso de bienvenida a los participantes que fundarían la UNESCO, Ellen Wilkinson (1891-1947), Ministra de Educación de Gran Bretaña y Presidente de la Conferencia declaró: *"Aunque en el nombre original de la Organización no figura la ciencia, la delegación británica presentará una propuesta para que se la incluya, de modo que el nombre sea "Organización de las Naciones Unidas para la Educación, la Ciencia y la Cultura". En esta época, cuando todos nos preguntamos, quizá con miedo, qué más van a hacer los científicos, importa que éstos se mantengan estrechamente relacionados con las humanidades y tengan conciencia de su responsabilidad para con la humanidad por el resultado de sus trabajos. No creo que, tras la catástrofe mundial, exista algún científico que pueda sostener todavía que no le intere-*

*sa en modo alguno las consecuencias de sus descubrimientos."*

Simultáneamente, otras voces comenzaron a surgir en los Estados Unidos. A instancias de Hyman Goldsmith (1898-1949) y Eugene Rabinowitch (1901-1973), dos destacados físicos que habían trabajado en el Proyecto Manhattan (PM), en 1945 se funda el *Bulletin of the Atomic Scientists* (BAS). Desde entonces, este boletín, se transformó en uno de los más importantes medios de comunicación dedicado al debate sobre los problemas de la carrera armamentista y el papel que en ella cumplen los hombres de ciencia. A los pocos meses, otro grupo de científicos del PM, reconocieron que la ciencia se había convertido en tema central de muchas cuestiones fundamentales de la política pública, y por esa razón fundaron la Federación de Científicos Americanos (FAS). La misión de FAS asume que los científicos tienen la responsabilidad única de advertir tanto al público como a los líderes políticos acerca de los peligros potenciales de los avances científicos y técnicos y demostrar cómo los nuevos conocimientos científicos pueden contribuir a mejorar la calidad de vida de los habitantes si se aplica una política pública adecuada.

Resulta complejo, tras 65 años, describir en forma precisa el grado de consternación y cierta aprensión hacia los científicos, que desató el uso de las bombas atómicas. Cuando se hizo el anuncio de las consecuencias generadas sobre Hiroshima y Nagasaki, el prominente dramaturgo alemán Bertolt Brecht (1898-1956) se encontraba justamente trabajando en la edición de la segunda versión de "La vida de Galileo" (1946). Tal vez, nadie mejor que él para traducir en palabras el impacto social y la impresión que la gente común tuvo acerca de los científicos en esa época.

Inmediatamente después de la aterradora noticia, Brecht decidió introducir un cambio decisivo en la última escena de su famosa obra teatral. Había percibido que la bomba atómica sólo había impresionado a la gente común como algo terrible. Sin embargo, al igual que el propio Einstein, se había dado cuenta de algo que no todos veían: *se había ganado la guerra, pero no la paz.* Diez días después de la noticia escribe en su diario "La bomba atómica ha convertido realmente las relaciones entre sociedad y ciencia en un problema de vida o muerte" (Fernández Buey, 2010).

El nuevo desenlace de la obra teatral empieza a tener a partir de allí una dimensión trágica. Hay un paso intermedio que ayuda a entender lo que será su final: el diálogo de Galileo con su discípulo y el monólogo del científico con que concluiría la obra.

Brecht empieza a pensar en una inversión casi paródica del mito de Prometeo. Imagina un Prometeo que descubre el fuego por sí mismo y en un acto delictivo se lo entrega a los dioses, que son ignorantes y malignos, que explotan a los hombres y que viven de las riquezas que producen éstos. Prometeo, encadenado por los dioses para que no revele el secreto del fuego a los hombres, descubre un día un resplandor rojizo en el horizonte y sabe así que los dioses han utilizado el fuego para extorsionar a los hombres (Fernández Buey, 2010). Con gran maestría pone, entonces, en palabras de Galileo el siguiente texto que manifiesta cabalmente la angustia de muchos científicos de la época:

*"Como científico, tuve una posibilidad excepcional. En mi época, la Astronomía llegó a la plaza pública. En esas condiciones muy especiales, la firmeza de un hombre hubiera podido provocar grandes conmociones. Si yo hubiera resistido, los hombres dedicados a las ciencias naturales hubieran podido desarrollar algo así como el Juramento de Hipócrates de los médicos: ¡la promesa de utilizar la Ciencia únicamente en beneficio de la Humanidad! Tal como están las cosas, lo más que se puede esperar es una estirpe de enanos inventores,* que podrán alquilarse para todo. Además, he llegado al convencimiento, Sarti, de que nunca estuve verdaderamente en peligro. Durante algunos años fui tan fuerte como la autoridad. Y entregué mi saber a los poderosos para que lo usaran, no lo usaran o abusaran de él, según conviniera mejor a sus fines. He traicionado a mi profesión. Un hombre que hace lo que yo he hecho no puede ser tolerado en las filas de la Ciencia".

Mientras Brecht se imaginaba un Galileo implorando un Juramento Hipocrático en donde los científicos usaran su conocimiento para beneficio de la humanidad, del otro lado del Atlántico, la antropóloga Gene Weltfish (1902-1980), quien por entonces era la vice presidenta de la Federación Internacional de Mujeres Democráticas, propone en un pequeño artículo en *Scientific Monthly* (sept. 1945), el texto de un nuevo Juramento Hipocrático para Científicos en la Era Nuclear. Unos meses después, el novelista Aldous Huxley[11] (1894-1963), consternado por las consecuencias de la detonación de las dos bombas atómicas, publica en 1946 un ensayo titulado *"Science, Liberty and Peace"* donde sugiere que los científicos deberían tener su propio Juramento Hipocrático. En su libro reproduce el texto propuesto por Weltfish.

Este momento histórico, marcó un verdadero punto de inflexión y de división de aguas dentro de la comunidad científica y de la sociedad en general. De la misma manera que los gobiernos de los países más avanzados comenzaron a invertir escandalosas sumas de dinero en tareas de I+D con fines militares, también comenzaron a surgir voces desde dentro de la comunidad científica, señalando

---

[11]  Autor de "Un Mundo Feliz" (1932), una irónica utopía en donde en el futuro la sociedad había alcanzado la felicidad y eliminado la guerra y la pobreza a expensas de haber exterminado la familia, la diversidad cultural, el arte, la ciencia, la literatura, la religión y la filosofía. Aldous Huxley era hermano del prominente biólogo, escritor y humanista británico Julian Huxley (1887-1975) quien fue el primer Director General de la UNESCO y nieto de Thomas H. Huxley (famoso por ser uno de mayores defensores de las ideas de Charles Darwin, del cual fue amigo y colega).

y reflexionando acerca del peligroso camino que la humanidad estaba tomando.

Por ejemplo, al finalizar la SGM, Hugo R. Kruyt (1882-1959) en su discurso presidencial ante la Primera Asamblea General del Consejo Internacional de la Ciencia (ICSU), afirmaba lo siguiente: "*Con más claridad que nunca comprendemos que el conocimiento no lo es todo. Necesitamos de la moral y de la fraternidad para evitar que la ciencia se convierta en una maldición*".

A los pocos meses, en 1946, se funda la Federación Mundial de Trabajadores Científicos (WFSW), que permitió por primera vez que los científicos elaboraran y expresaran esas preocupaciones de modo colectivo y a nivel internacional. En 1948 adoptaron la Carta de Trabajadores Científicos, en la que se bosquejaron las bases de una relación fructífera, responsable y armoniosa entre los trabajadores científicos y la comunidad más amplia. En la Carta se reconoce que "la profesión científica entraña responsabilidades especiales más allá y por encima de los deberes ordinarios de la ciudadanía" describiéndose allí también estas responsabilidades en relación con la ciencia, la comunidad y el mundo en general.

De alguna manera, a medida que el armamento nuclear y la carrera armamentista se expandían con un crecimiento acelerado de dimensiones inusitadas, cada vez con mayor frecuencia surgieron voces -dentro de la comunidad científica internacional- que señalaron los peligros de aplicar el conocimiento científico y tecnológico a fines bélicos y acerca de la necesidad de encauzarlo hacia objetivos totalmente pacíficos. La Tabla 3 da cuenta de algunos ejemplos de este tipo de declaraciones o llamamientos. En los Apéndices 1 a 5, se pueden encontrar los textos completos de aquellas resoluciones y declaraciones que surgieron del ámbito de las Naciones Unidas.

Después de las bombas de Hiroshima y Nagasaki, personajes de la talla de Albert Einstein (1879-1955), Bertrand Russell (1872-1970), Linus Pauling (1901-1994), Joseph Rotblat (1908-2005), entre un enorme grupo de científicos e ingenieros de todas partes del mundo, dedicaron gran parte de sus vidas a educar al público en general y a los políticos acerca de los peligros de la carrera armamentista nuclear. El Manifiesto Russell-Einstein (Apéndice 6) es posiblemente uno de los documentos más destacados y referenciados que se ha utilizado como ejemplo del reclamo realizado por destacadísimos científicos para construir un mundo mejor y garantizar la continuidad de la civilización humana.

Al respecto, también resulta muy interesante tanto la carta que Albert Einstein escribiera en ocasión de su ingreso a la Sociedad para la Responsabilidad Social de la Ciencia (Recuadro 1), como la que Bertrand Russell le escribiera al matemático Mischa Cotlar (1912-2007) a principios de la década del sesenta (Recuadro 2). Estos documentos muestran la firmeza y convicción de grandes personajes de la historia, acerca de cuáles deben ser las actitudes de los científicos con respecto al desarrollo de tareas de I+D con objetivos militares.

Una de las consecuencias directas del Manifiesto Russell-Einstein, fue el nacimiento, a partir de 1957, de las Conferencias Pugwash sobre Ciencia y Asuntos Mundiales. El grupo de científicos, tecnólogos y diplomáticos así constituido, se ha transformado desde entonces, en uno de los más importantes movimientos internacionales de personalidades orientados a garantizar la paz en nuestro frágil planeta azul. Durante la Guerra Fría, estas conferencias -que reunían a científicos y diplomáticos de los dos bloques en pugna- sirvieron de lugar de encuentro para dialogar acerca de los términos de referencia de la gran mayoría de los acuerdos internacionales de desarme (Rotblat 1972; Lemarchand 1991). En el año 1995, el Movimiento Pugwash compartió el

Premio Nobel de la Paz junto con su fundador
y líder, Joseph Roblat.

**Tabla 3**: Ejemplos de Declaraciones de Organismos Internacionales y grupos de académicos sobre la responsabilidad social de los científicos, la carrera armamentista y temas afines. Fuente: Elaboración Propia.

| Fecha | Nombre del Documento | Endosado por: |
|---|---|---|
| 1945 | Reporte del Comité Franck | James Franck , Donald J. Hughes, J. J. Nickson, Eugene Rabinowitch, Glenn T. Seaborg, J. C. Stearns, y Leo Szilard. |
| 1945 | Declaración del Comité de Emergencia de los Científicos Atómicos | Albert Einstein, Hans Bethe, Linus Pauling, Leo Szilard. Harold Urey, Victor Weisskopf |
| 1950 | Carta Abierta a las Naciones Unidas | Neils Bohr |
| 1955 | Manifiesto Russell-Einstein | Max Born, Percy W. Bridgman, Albert Einstein, Leopold Infeld, Frederic Joliot-Curie, Herman J. Muller, Linus Pauling, Cecil F. Powell, Joseph Rotblat, Bertrand Russell, Hedeki Yukawa |
| 1957 | Declaración de consciencia | Albert Schweitzer |
| 1958 | Petición de los científicos a las Naciones Unidas para el cese de las pruebas nucleares | Linus Pauling y más de 13.000 firmas de científicos, enero de 1958 |
| 1958 | Declaración de Viena sobre la Responsabilidad de los Científicos | III Conferencia Pugwash |
| 1969 | Declaración de Responsabilidad Profesional de los Especialistas de América Latina | Bulletin of Peace Proposals (vol.2: 15-16, 1969) |
| 1974 | Declaración de Monte Carmelo sobre la Tecnología y la Responsabilidad Moral | Haifa y Jerusalem, Israel, 25 de diciembre de 1974 |
| 1974 | Recomendación relativa a los investigadores científicos | 18 Conferencia General de la UNESCO |
| 1975 | Declaración sobre la utilización del progreso científico y tecnológico en interés de la paz y en beneficio de la humanidad | Declaración 3384 (XXX) de la Asamblea General de las Naciones Unidas, 10 de Noviembre de 1975 |
| 1978 | Declaración de Principios de Política Científica y Tecnológica | Quinta Conferencia permanente de dirigentes de consejos nacionales de política científica y de investigación de los Estados Miembros de UNESCO en América Latina y el Caribe, Quito, Ecuador, 13-18 de marzo de 1978 |
| 1980 | Llamamiento al "Parlamento Mundial de los Pueblos por la Paz" | Federación Mundial de Trabajadores Científicos (WFSW) |
| 1981 | Declaración del Simposio de Bucarest: "Científicos y paz" | 68 científicos representando a 38 países, 4-5 de septiembre de 1981 |
| 1982 | Recomendaciones del Simposio UNESCO/Pugwash: "Científicos, carrera armamentista y desarme" | Ajaccio, Francia, 19-23 de febrero de 1982 |
| 1982 | Manifiesto de Erice | Paul Dirac, Piotr Kapitza y Antonino Zichichi y más de 10.000 firmas |
| 1982 | Declaración sobre la Prevención de la Guerra Nuclear | Asamblea de Presidentes de Academias Nacionales de Ciencia, Vaticano, 23-24 de septiembre de 1982 |
| 1983 | Llamamiento a los científicos del mundo | Científicos Soviéticos, 10 de abril de 1983 |
| 1983 | Llamamiento de los médicos internacionales para la finalización de la carrera armamentista nuclear | Tercer Congreso del IPPNW, La Haya, 17-21 de Junio de 1983 |

| Fecha | Nombre del Documento | Endosado por: |
|-------|----------------------|---------------|
| 1984 | Declaración de alerta sobre el Invierno Nuclear | Asamblea de Presidentes de Academias Nacionales de Ciencia, Vaticano, 23-25 de enero de 1984 |
| 1985 | Declaración de apoyo a la Iniciativa de Paz de los 5 Continentes: Por las especies y el planeta | Declaración firmada por cientos de científicos, entre los cuales se encontraban 40 que habían obtenido el Premio Nobel |
| 1985 | Propuesta de Carta de las Naciones Unidas sobre la Responsabilidad de los Científicos en el desarrollo de Armamento Nuclear | Christopher G. Weeramantry (1987) |
| 1988 | Semana Internacional de la Ciencia y la Paz | Resolución de la Asamblea General de las Naciones Unidas |
| 1991 | Resolución de Toronto | Universidad de Toronto – Science for Peace |
| 1999 | Declaración sobre la Ciencia y el Uso del Saber Científico y Programa en Pro de la Ciencia, Marco General de Acción | Conferencia Mundial de la Ciencia de Budapest (UNESCO/ICSU) y 30 Conferencia General de la UNESCO |
| 2001 | Día Mundial de la Ciencia para la Paz y el Desarrollo | 31 Conferencia General de la UNESCO |

## Recuadro 1:
## Carta de Albert Einstein en ocasión de su ingreso a la Sociedad para la Responsabilidad Social en Ciencia.

Estimados colegas:

La cuestión de cómo debe actuar un individuo cuando su gobierno le ordena proceder de cierta manera o cuando la sociedad espera de ella o él una actitud contraria a su propia conciencia, es realmente un problema de larga data. Parece fácil afirmar que no es posible responsabilizar a un individuo de lo realizado en una situación de coacción que no se puede resistir, porque el individuo depende totalmente de la sociedad en la que vive, por lo que debe aceptar sus reglas. Pero la sola formulación de esta idea evidencia hasta qué punto este concepto contradice nuestro sentido de la justicia.

La imposición externa puede, hasta cierto punto, reducir la responsabilidad del individuo, pero nunca puede eliminarla. En los juicios de Nüremberg esta idea se consideró indiscutible. Los orígenes de aquello a lo que le damos importancia moral en nuestras constituciones, leyes y costumbres se encuentran en la interpretación que innumerables individuos poseen del sentido de la justicia. En términos morales, a menos que se apoyen en un sentido de responsabilidad de los individuos vivos, las instituciones son impotentes. Los esfuerzos dirigidos a generar y fortalecer este sentido de responsabilidad individual representan un importante aporte para la humanidad

Hoy en día los ingenieros y científicos cargan con una responsabilidad particular, ya que dentro de su esfera de actividad se incluye el desarrollo de medios militares de destrucción masiva. Por ello, creo que la creación de la *Sociedad para la Responsabilidad Social en Ciencia* satisface una verdadera necesidad. A través de la discusión de los problemas que le competen, esta sociedad ayudará al individuo a aclarar su mente y llegar a una definición clara de su posición. La ayuda mutua es esencial para quienes enfrentan dificultades por actuar de acuerdo a lo que les dicta su conciencia.

Atentamente,
Albert Einstein

*Publicada en* <u>Science</u>, *vol.112: 760-761, 1950*
*Traducida al español por Victoria De Negri.*

## Recuadro 2:
## Extracto de la carta de Bertrand Russell dirigida al matemático Mischa Cotlar (*)

Londres, 6 de julio de 1962

Estimado Dr. Cotlar:

...me gustaría que usted leyera el siguiente mensaje ante la Conferencia Internacional de Matemática:

Aquellos de ustedes, que hayan buscado aportar precisión y claridad al pensamiento humano, estarán profundamente afligidos por lo que han hecho los hombres del poder con nuestro esfuerzo creativo. La más imparcial y teórica de las obras de la matemática suministra hoy las bases para la concepción de minuciosos e ingeniosos medios de ocasionar el sufrimiento y la muerte de cientos de millones de nuestros congéneres. Creo que, si no existen cambios radicales en nuestras políticas y en el curso de los acontecimientos actuales, la posibilidad de que una guerra nuclear accidental ocurra es un asunto de certeza estadística. La tecnología nuclear es poco fiable y las personas a cargo de su operación serán invocadas en cuanto sus semillas de maldad se dispersen por el mundo. Por ello, clarificar los hechos de esta gran amenaza a la civilización humana es un deber moral de la conciencia de cada uno de nosotros. Nuestra responsabilidad es especial, pues sin nosotros esas armas de guerra de destrucción masiva nunca se hubieran podido fabricar.

Los insto a considerar que no basta con pronunciamientos ajenos a la acción. El trabajo de las Conferencias Pugwash sobre Ciencia y Asuntos Mundiales, por ejemplo, ha sido de gran valor para suministrar información científica sobre la naturaleza del peligro al que nos enfrentamos y los medios por los cuales el conflicto podría resolverse. Sin embargo, el problema reside no tanto en la formulación de esquemas inteligentes, o incluso en la delimitación de hechos científicos, sino en el suministro de aquellos medios por los cuales los gobiernos dementes se verán obligados a modificar su forma de actuar.

Hago este llamado a ustedes como individuos, y matemáticos, para que consideren recomendar el cese de sus servicios ante cualquier gobierno o institución privada que trabaje para programas de armamento o tecnología aplicada relacionada con este. Soy consciente del hecho de que gran parte del trabajo teórico puede utilizarse sin el consentimiento de sus autores. Sin embargo, en la medida en que seamos capaces, debemos condenar la prostitución de nuestro trabajo y buscar revelar la verdad de la amenaza a la vida y al mundo de las mentes...

Bertrand Russell

*(*) De la copia del original entregada por Mischa Cotlar a GAL. Traducción al español realizada por Paula Santos.*

A lo largo de más de 50 años, el éxito de las Conferencias Pugwash, ha sido el resultado del decidido esfuerzo de un grupo de científicos resueltos a mantener una posición independiente e imparcial, deseosos de construir y desarrollar la comprensión y la cooperación entre las naciones. Las conferencias demostraron que es posible aplicar la óptica científica a otros problemas que se encuentran fuera de las esferas tradicionales de la ciencia. Aun tratándose de cuestiones controvertidas, se demostró que es posible tratar los problemas más acuciantes de la humanidad, sin perder la objetividad científica y el respeto a las ideas distintas.

La gráfica 5 muestra cómo el punto de inflexión acerca de la reflexión sobre la responsabilidad social de los científicos se sitúa claramente a finales de la SGM. En la figura se representa -en escala semilogarítmica- el número de artículos científicos de corriente principal listados en las bases internacionales *Science Citation Index (SCI)*, *Social Science Citation Index (SSCI)* y *Arts & Humanities Citation Index (A&HCI)* dedicados específicamente a temas de ética en ciencia y tecnología, publicados anualmente entre 1895 y 2009 (114 años). Se puede apreciar, que antes de la SGM se publicaban en esta temática entre 10 y 30 artículos por año sin ningún patrón de crecimiento definido. Sin embargo, a partir de la SGM el número de artículos publicados en las revistas de corriente principal de todas las áreas de las ciencias, muestra un crecimiento tipo exponencial. En el presente se llegan a publicar anualmente del orden de 4.000 artículos científicos en temáticas que vinculan la ética con la ciencia y la tecnología.

En 1974, los Estados Miembros de la UNESCO, durante su 18ª Conferencia General celebrada en París, aprobaron sin ningún voto en contra, la "Recomendación relativa a la situación de los investigadores científicos" (Apéndice 1). Allí se reconoce que "*que los descubrimientos científicos y los adelantos y aplicaciones tecnológicas conexas abren vastas perspectivas al progreso que provienen en particular de utilizar con la máxima eficacia la ciencia y los métodos científicos en beneficio de la humanidad y para contribuir a preservar*

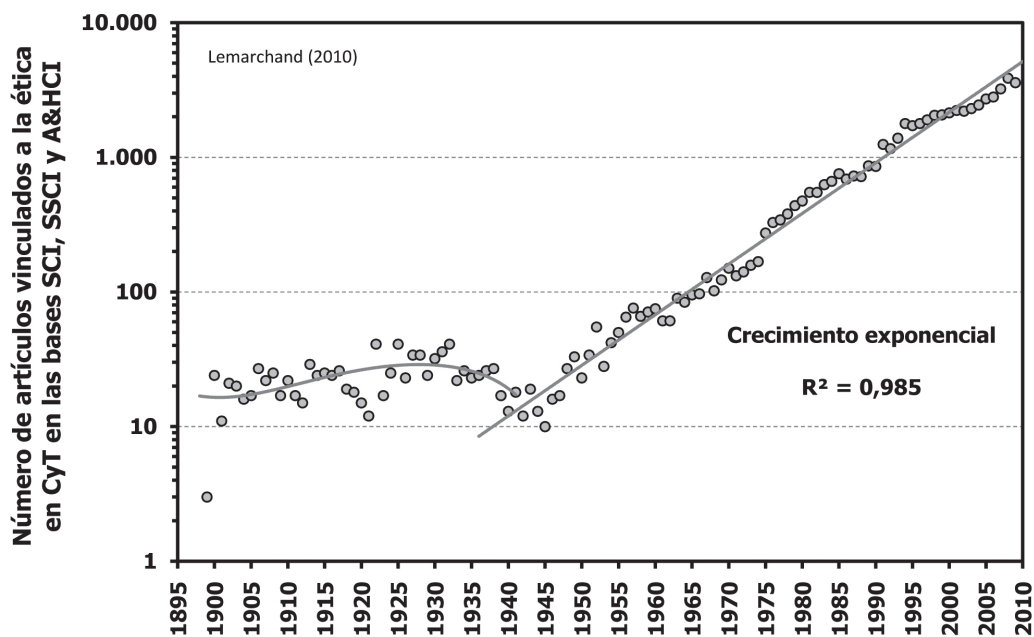

**Gráfica 5:** Evolución temporal (1895-2009) del número de publicaciones científicas listadas en el Science Citation Index (SCI), *Social Science Citation Index* (SSCI) y *Arts & Humanities Citation Index* (A&HCI) dedicadas a temas de ética en ciencia y tecnología. La representación se encuentra en escala semilogarítmica. Se puede observar que hasta la Segunda Guerra Mundial (SGM) hay un comportamiento errático que muestra entre 10 y 20 publicaciones por año. Sin embargo, a partir de la SGM el número de artículos científicos vinculados a la ética en la ciencia y la tecnología creció en forma exponencial, con un coeficiente de Pearson $R^2 = 0,985$. Fuente: Elaboración propia.

*la paz y reducir las tensiones internacionales, pero que, al mismo tiempo, entrañan ciertos peligros, que constituyen una amenaza, sobre todo en el caso de que los resultados de las investigaciones científicas se utilicen contra los intereses vitales de la humanidad para la preparación de guerras de destrucción masiva o para la explotación de una nación por otra y que, en todo caso, plantean complejos problemas éticos y jurídicos."*

Al año siguiente, la Asamblea General de las Naciones Unidas aprueba la Resolución 3384 (XXX) proclamando la "Declaración sobre el uso del progreso científico y tecnológico en interés de la paz y en beneficio de la humanidad" (Apéndice 2). Este es uno de los documentos más contundentes emitidos por las Naciones Unidas en donde se insta a sus Estados Miembros a promover la cooperación internacional con objeto de garantizar que los resultados del progreso científico y tecnológico se usen en pro del fortalecimiento de la paz, e insta a abstenerse de todo acto que entrañe la utilización de los logros científicos y tecnológicos para violar la soberanía y la integridad territorial de otros Estados, intervenir en sus asuntos internos, hacer guerras de agresión, sofocar movimientos de liberación nacional o seguir políticas de discriminación racial. Considera que estos actos no sólo constituyen una patente violación de la Carta de las Naciones Unidas y de los principios del derecho internacional, sino que además representan una aberración inadmisible de los propósitos que deben orientar al progreso científico y tecnológico en beneficio de la humanidad.

En 1978, durante la "Quinta conferencia permanente de dirigentes de los consejos nacionales de política científica y de investigación de los Estados Miembros de la UNESCO de América Latina y el Caribe" celebrada en la ciudad de Quito, dentro de la "Declaración de Principios de Política Científica y Tecnológica" se incluye en su punto 8 lo siguiente: *"Que tanto a nivel nacional como internacional, la*

*política científica y tecnológica debe dirigirse a crear o fortalecer la capacidad de los países para generar y adaptar los conocimientos y tecnologías más adecuados a sus necesidades y sus recursos, acorde con los principios de endogeneidad y autodeterminación, así como evitar que resultados de la investigación científica y tecnológica sean utilizados a los fines de desarrollar y perfeccionar medios bélicos de exterminación/agresión o sirvan para elaborar medios de presión política de unos estados respecto a otros".*

El 18 de diciembre de 1982, la Asamblea General de las Naciones Unidas aprueba una nueva Resolución (37/189A) en donde hace un llamamiento a los Estados Miembros, a las agencias especializadas, a las organizaciones intergubernamentales y no gubernamentales para que se tomen las medidas necesarias que aseguren que los resultados del progreso científico y tecnológico sean utilizados exclusivamente para asegurar la paz internacional, en beneficio de la humanidad, y para promover y fomentar el respeto de los derechos humanos y las libertades fundamentales.

En 1984 el Comité de Derechos Humanos de las Naciones Unidas, durante su 563[ava] reunión declaró que la producción, prueba, almacenamiento y despliegue de armas nucleares deberían ser prohibidos y reconocidos como crímenes de lesa humanidad.

En virtud de la exitosa iniciativa del Prof. Hendrik Bramhoff de la Universidad de Hamburgo, quien desde 1986 convocó regularmente a los científicos de todo el mundo a participar de la "Semana Internacional de los Científicos y la Paz", mediante la organización eventos simultáneos en los principales centros de producción científica del mundo, en 1988, la Asamblea General emite una Resolución en donde invita a sus Estados Miembros a celebrar anualmente la "Semana Internacional de la Ciencia y la Paz" (Apéndice 3).

En 1985, el Prof. Christopher G. Weeramantry, vicepresidente de la Corte Internacional de Justicia de La Haya, y ganador del Premio UNESCO de Educación para la Paz en el año 2006, propuso un código de ética para científicos. El mismo fue incluido dentro de un proyecto de resolución de la Asamblea General de las Naciones Unidas que se reproduce en el Recuadro 3. Apoyándose en la jurisprudencia internacional, presenta argumentos para fundamentar que la participación en la investigación científica y tecnológica vinculada al armamento nuclear sería opuesta al derecho internacional vigente y debería ser considerada como un crimen de lesa humanidad. Considera además que aquellos que conscientemente participen en la fabricación de armas nucleares y en investigación sobre armas nucleares son personalmente culpables de violación del derecho internacional y del delito de lesa humanidad y/o de complicidad en tales actos.

## Recuadro 3:
## Propuesta de Declaración de las Naciones Unidas sobre la Responsabilidad Científica con Relación al Armamento Nuclear (*)

Preámbulo

LA ASAMBLEA GENERAL

*Reconociendo* que en una era en la que la ciencia y la tecnología prevalecen, es esencial que estas se encuentren al servicio de la humanidad

*Profundamente preocupada* porque el desarrollo y la producción de armas nucleares y por la carrera armamentista nuclear están poniendo en peligro el futuro de la humanidad y, en efecto, la vida en el planeta

*Consciente* de que las más recientes y meticulosas investigaciones científicas establecen la probabilidad de un invierno nuclear con consecuencias desastrosas para la humanidad y nuestro planeta en caso de un enfrentamiento nuclear

*Percatándose* de que la carrera armamentista nuclear sería insostenible sin la activa cooperación de científicos y tecnólogos.

*Teniendo en cuenta* el hecho de que los principios generales del derecho internacional contenidos en:

a. la costumbre internacional

b. los principios legales generales reconocidos por las naciones civilizadas

c. las decisiones judiciales y la formación de juristas

d. las convenciones internacionales

Exime de todo cuestionamiento a la ilegalidad del uso de armamento nuclear, por estar relacionado con la violación de los principios de proporcionalidad, la discriminación, la agravación del dolor y el sufrimiento, la nulidad de un retorno a la paz y la inviolabilidad de estados neutrales, entre otros

*Consciente de que* el uso de armamento nuclear resultaría indudablemente en ecocidio, genocidio, y si hubiera sobrevivientes, en daño intergeneracional masivo

*Convencida* de que el concepto de una guerra nuclear limitada es irreal y que una vez que una guerra nuclear comience será poco probable contenerla

*Persuadida* de que los conceptos de legítima defensa y la disuasión han perdido sentido en el contexto del armamento nuclear y por consiguiente, no ofrecen justificación alguna para su producción, posesión, ensayo o despliegue.

*Consciente* de que el uso, la producción, el ensayo, la posesión y el despliegue de armas nucleares constituyen por lo tanto una violación del derecho internacional y un crimen contra la humanidad

*Recordando* que esta Asamblea en su Resolución 3384 (XXX), de 10 de noviembre 1975 proclamó la Declaración sobre la utilización del progreso científico y tecnológico en interés de la paz y en beneficio de la humanidad y desde entonces ha tomado numerosas medidas para la aplicación de la presente resolución, incluida la aprobación de la Resolución 37/189A de 18

de diciembre 1982 instando a todos los Estados, organismos especializados, organizaciones intergubernamentales y no gubernamentales a adoptar las medidas necesarias para garantizar que los resultados del progreso científico y tecnológico se utilicen exclusivamente en favor de la paz internacional, en beneficio de la humanidad y para la promoción y estímulo del respeto de los derechos humanos y las libertades fundamentales

*Observando* que el Comité de Derechos Humanos de las Naciones Unidas en su 563a reunión (23ª sesión), celebrada en noviembre de 1984 en su comentario general 14 (23) / c (artículo 6) declaró que la producción, ensayo, posesión, despliegue y utilización de armas nucleares deben prohibirse y reconocerse como un crimen contra la humanidad

*Observando* también que dicho Comité en su comentario general, realizó un llamamiento a todos los Estados, sean Partes en la Convención o no, para que adopten medidas urgentes unilateralmente y mediante acuerdos para liberar al mundo de esta amenaza

*Persuadidos* de que la responsabilidad legal y moral de los científicos que participan en dichas actividades es hoy infinitamente mayor que en el momento de creación de las primeras armas nucleares debido, entre otras cosas, al mayor conocimiento disponible actualmente sobre los catastróficos impactos atmosféricos, agrícolas, médicos y sociales del uso de armamento nuclear, la posibilidad de una represalia nuclear, el enormemente mejorado poder destructivo del armamento nuclear actual y los vastos arsenales nucleares disponibles en caso de guerra nuclear

*Profundamente conmovida* por la consideración de que el poder de la ciencia es tal, en palabras del Manifiesto Russell-Einstein, que conducirá hacia un nuevo paraíso o al riesgo de una muerte universal

*Estimando* que la participación de científicos y tecnólogos es crucial para la determinación de la elección entre estas alternativas

*Convencida* de que el principio de responsabilidad individual por crímenes contra la humanidad se encuentra plenamente establecido en el derecho internacional

*Convencida* también de que las órdenes superiores no constituyen una defensa en derecho internacional en lo que respecta a los crímenes contra la humanidad

*y Convencida* de que a la luz de las circunstancias anteriormente mencionadas la comunidad internacional no deberá dilatar el análisis de la responsabilidad de los científicos y tecnólogos dedicados a la empresa de las armas nucleares

Esta Asamblea reafirma los principios de que:

a. el uso, la producción, la posesión, el ensayo y el despliegue de armas nucle-

ares se oponen al derecho internacional y constituyen un crimen contra la humanidad

b. la participación en la investigación científica y tecnológica en este ámbito se opone al derecho internacional y es un crimen de lesa humanidad

c. aquellos que conscientemente participen en la fabricación de armas nucleares y en investigación sobre armas nucleares son personalmente culpables de violación del derecho internacional y delito de lesa humanidad y/ o de complicidad en tales actos

d. ese tipo de actividad es incompatible con el principio dominante que subyace a toda la actividad científica, a saber, servicio a la humanidad, y por tanto es inmoral y se opone a las Declaraciones expresas de la presente Asamblea

*y exhorta* a todos los científicos y tecnólogos de todo el mundo a cumplir con las obligaciones legales y éticas establecidas en este documento y abstenerse de cualquier actividad que implique el desarrollo, la producción, el ensayo, la posesión, el despliegue o la utilización de armas nucleares.

(*) Texto propuesto por el Vicepresidente de la Corte Internacional de La Haya Christopher G. Weeramantry (1987). Traducción del original en inglés por Paula Santos.

Este controversial proyecto fue circulado por su autor a través de diversos canales diplomáticos, entre todos los Estados Miembros de las Naciones Unidas y sociedades científicas de varios países. El documento generó posturas antagónicas, desde la oposición absoluta hasta el apoyo más entusiasta. El entonces Primer Ministro de Suecia, Olof Palme (1927-1986), consideraba que si bien la propuesta tenía elementos muy positivos, le correspondía a las sociedades científicas y no a las Naciones Unidas establecer ese tipo de código de conducta para los científicos (Weeramantry 1987: 179-180). Finalmente, la propuesta fue desestimada ya que ningún Estado Miembro de la ONU tomó la iniciativa de presentarlo formalmente.

En 1999 el Consejo Internacional para la Ciencia (ICSU) y la UNESCO organizaron la Conferencia Mundial de la Ciencia en Budapest, con el objetivo de contribuir a reforzar el compromiso de los Estados Miembros de la UNESCO y otros interesados principales en aquellas temáticas relacionadas con la educación científica y las actividades en materia de investigación y desarrollo, y para definir una estrategia global gracias a la cual la ciencia corresponda mejor a las necesidades y aspiraciones de la sociedad en el siglo XXI. Durante la misma se emitió una "*Declaración sobre la ciencia y el uso del saber científico*" y un "*Programa en pro de la ciencia: Marco general de acción*" (Apéndice 4). Ambos documentos fueron adoptados en 1999 por los Estados Miembros de la UNESCO en su 30ª Conferencia General celebrada en París el 18 de agosto de 1999 (Doc. 30/C15) y por el ICSU en su XXVI Asamblea General celebrada en El Cairo entre el 28 y 30 de septiembre de 1999.

Los documentos anteriores consideran que la cooperación mundial entre científicos es una contribución valiosa y constructiva a la seguridad mundial y al desarrollo de relaciones pacíficas entre países, sociedades y culturas diferentes. Los principios fundamentales de la paz y la coexistencia son considerados parte integrante de la enseñanza en todos los niveles. También insta a los Estados Miembros a lograr que los estudiantes de carreras científicas y tecnológicas cobren conciencia de su deber de no utilizar sus competencias y conocimientos científicos para actividades que hagan peligrar la paz y la seguridad. Se promueve un diálogo entre representantes del gobierno, de la sociedad civil y de los científicos para tratar de reducir el gasto militar y lograr que la ciencia se oriente menos hacia las aplicaciones militares.

El Programa en Pro de la Ciencia, considera que la ética y la responsabilidad de la ciencia deberían ser parte integrante de la educación y formación que se imparte a todos los científicos, ingenieros y tecnólogos. Se reconoce que es importante infundir en los estudiantes una actitud positiva de reflexión, vigilancia y sensibilidad respecto de los problemas éticos con los que pueden tropezar en su vida profesional. Recomienda que los científicos jóvenes sean estimulados a respetar y observar los principios de ética y responsabilidad de la ciencia. Le otorga a la Comisión Mundial de Ética del Conocimiento Científico y la Tecnología (COMEST) de la UNESCO la responsabilidad especial de realizar el seguimiento de esta cuestión, en cooperación con el Comité Permanente sobre Responsabilidad y Ética Científicas (SCRES) del ICSU.

Los documentos de la Conferencia Mundial de la Ciencia encomiendan a los gobiernos, las ONG, y más concretamente a las asociaciones científicas y eruditas, organizar debates –incluso públicos– sobre las consecuencias éticas del trabajo científico. Sugieren también, que los científicos, ingenieros y tecnólogos, sus organizaciones y las sociedades eruditas, deberían estar representadas convenientemente en los organismos competentes de reglamentación y adopción de decisiones. Esas actividades deberían ser fomentadas en el plano institucional y reconocidas como parte de la labor y responsabilidad de los científicos. Al respecto *recomiendan a las asociaciones científicas que adopten un código deontológico para sus miembros.*

La propuesta realizada por Rotblat (1999, 2000) durante la Conferencia Mundial de la Ciencia, para que los científicos adoptaran un juramento o compromiso ético del estilo del Juramento Hipocrático de los médicos, durante las ceremonias de graduación, se difundió rápidamente dentro y fuera de la comunidad científica. Sin embargo, esta idea distaba mucho de ser original y como se verá en la próxima sección a la fecha de esta propuesta ya existían unas 70 propuestas similares, muchas de las cuales (entre ellas el "Juramento de Buenos Aires") ya habían sido implementadas en distintas universidades del mundo.

En el 2001, durante la 31ª Conferencia General de la UNESCO, se adoptó la Resolución 20, en donde en virtud de las Recomendaciones de la Conferencia Mundial de la Ciencia y la Resolución 43/61 de la 71a. Sesión Plenaria de la Asamblea General de las Naciones Unidas del 6 de diciembre de 1988, la UNESCO proclamó al día 10 de noviembre de cada año, como el Día Mundial de la Ciencia para la Paz y el Desarrollo (Apéndice 5).

## 4. Del Juramento Pitagórico a las modernas versiones de Juramentos Hipocráticos para Científicos

Los descubrimientos científicos y los nuevos desarrollos tecnológicos abren posibilidades cuyas aplicaciones generan consecuencias que no necesariamente están previstas en el momento en que se realizan los nuevos descubrimientos. En virtud de ello, individuos, sociedades profesionales y estados nacionales han intentado desarrollar e implementar diversos tipos de normas éticas para guiar la conducta de los científicos, ingenieros y tecnólogos.

Un somero análisis de las distintas propuestas, permite reconocer la existencia de diversas categorías de normas éticas, entre las que se destacan las siguientes: *ethos*, las promesas, los juramentos, conjuntos de principios o directrices, códigos, recursos, recomendaciones, manifiestos, declaraciones, resoluciones, convenciones, cartas, y leyes.

A continuación se enumeran las propiedades más representativas de aquellas categorías de normas éticas más frecuentes (Evers 2004):

**(1) Ethos:** La palabra *ethos* es de origen griego y significa -en la acepción vinculada a este ensayo- carácter, costumbre, moral. Se trata de una creación genuina y necesaria del hombre, pues éste desde el momento en que se organiza en sociedad, siente la necesidad imperiosa de crear reglas para regular su comportamiento y permitir modelar así su carácter.

Dentro de la perspectiva de la sociología de la ciencia, en 1942, Robert K. Merton (1910-2003) sugirió que el comportamiento de los investigadores científicos podría ser descripto por un *ethos científico* que enfatiza las normas de trabajo del investigador agrupándolas bajo el acrónimo en inglés de CUDAS (comunalismo, universalismo, desinterés y escepticismo organizado). Este tipo de análisis no incluye ningún área que evalúe, por ejemplo, si es ético o no trabajar en el diseño de armamento de destrucción masiva (Kuipers 2010).

A mediados del siglo XX se formularon las primeras iniciativas de códigos universales de ética para científicos (Conant 1948; Pigman y Carmichael 1950; Leys 1952; Glass 1965; Cournand y Meyer 1976; Cournand 1977). En este sentido resulta interesante comprobar que las mismas se centraron casi exclusivamente dentro del marco teórico del *ethos mertoniano.* En ellas el concepto de responsa-bilidad social del científico no era considerado como parte del ethos.

**(2) Juramentos y promesas:** la ética de responsabilidad puede manifestarse concretamente en, por ejemplo, a través de la expresión de un juramento, promesa o compromiso. Cualquiera de ellos puede considerarse como una manifestación concreta de una ética abstracta subyacente. Las nociones anteriores (juramento versus promesa) no son idénticas en sí, pero pueden ser tratadas como expresiones equivalentes sin negar que una se puede diferenciar de la otra. Ambas comparten las características esenciales de mantener los elementos importantes del testimonio, la promesa, la palabra de honor o la garantía. Puesto que estos términos son de uso general, los juramentos y las promesas suelen ser afirmaciones públicas de un compromiso a mantener ciertos principios específicos o responsabilidades.

**(3) Códigos y directrices:** La palabra 'código' proviene del vocablo latino "códice", que puede significar tronco de árbol, o libro. Originalmente, un códice fue un libro hecho con tapas de madera cubiertas de cera. En su sentido moderno un código es un conjunto de leyes, y reglamentos y un texto escrito que ofrece pautas - por ejemplo, reglas, directrices o principios de conducta moral. La idea moderna de los códigos, se deriva del ideal renacentista de la racionalización en el derecho romano, poniendo las diversas partes en orden e indicando brevemente y con claridad cuál es la esencia de la norma. En consecuencia, el código puede ser descrito como una colección ordenada que guía algunos campos específicos. Por otro lado se pude concebir a la directriz, como un código que se expresa a través de un juramento.

En esta sección se muestra cómo el concepto de Juramento Hipocrático utilizado por más de 2500 años en la tradición médica, ha sido propuesto decenas de veces, desde la Revolución Científica (por ej. con Francis Bacon), como un instrumento para que los científicos se comprometan a utilizar sus conocimientos en beneficio de la humanidad, y para que asuman las pertinentes responsabilidades individuales y colectivas destinadas a evitar daños a la humanidad y a la naturaleza.

¿Cómo se inició la tradición médica del Juramento Hipocrático? De la vida Hipócrates de Cos, fundador de la más famosa escuela médica de la antigüedad, son pocos los datos que han podido ser corroborados históricamente. Se sabe que nació cerca del 460 A.C. y falleció aproximadamente en el 370 A.C. Acerca de los posibles escritos de Hipócrates existe la misma incertidumbre de autenticidad que en la mayor parte de los textos que componen el llamado *corpus* hipocrático. Se sabe con cierta certeza que el juramento que lleva su nombre es anterior al propio Hipócrates, aunque su adopción por parte de la Escuela Hipocrática es signo claro de la exigencia ética de su maestro y seguidores en relación con el respeto por la vida y la profesión médica.

Tal vez uno de los trabajos más profundos, detallados y mejor documentados acerca del origen y significado trascendental del llamado "Juramento Hipocrático" sea el estudio realizado por Ludwig Edelstein (1943). En esta monografía, se fundamenta abrumadoramente –casi palabra por palabra- la tesis de que el verdadero origen del juramento se remonta a la Escuela Pitagórica. Para Edelstein el Juramento de Hipócrates es un documento uniformemente concebido y atravesado profundamente por la filosofía pitagórica. En su espíritu, forma, texto y contenido, el Juramento es un verdadero manifiesto pitagórico. En este punto es interesante señalar que Pitágoras también fue la primera persona en la historia en percatarse acerca de la conexión existente entre la matemática y un fenómeno físico, cuando descubrió que las notas musicales pueden ser descriptas a través de proporciones numéricas sencillas. En su concepción del mundo, la esencia del universo estaba regulada por la matemática. Por eso resulta más que significativo que la idea de un compromiso ético para el manejo responsable del conocimiento se originara en una escuela de pensamiento mucho más amplia que la medicina.

En forma independiente, la tradición india, reconoce un texto similar publicado en sánscrito en el *Charaka Samhita* que data al menos del siglo III A.C.

Se le atribuye a Maimónides (c. 1200) otra tradición similar aunque existe cierta discusión acerca de su origen. En 1608 en Francia entre los practicantes farmacéuticos se instaura el Juramento de Galeno.

Recién en el 1900, se encuentra una de las primeras versiones de un juramento de compromiso ético aplicado en este caso a los ingenieros. En 1917, se hace una primera propuesta en Inglaterra de redactar un compromiso ético -del estilo del Juramento Hipocrático- para los científicos y florecen otras iniciativas similares para los ingenieros en distintas partes del mundo.

En la Tabla 5 y en su correspondiente anexo se detallan todos los textos de juramentos o compromisos éticos para científicos que han sido propuestos a lo largo de la historia. Este relevamiento resulta ser el más exhaustivo que existe, hasta este momento, publicado en la literatura especializada.

La Tabla 4 muestra que han existido unas 90 iniciativas, algunas de las cuales no propusieron un texto específico pero si la idea de instaurar un Juramento Ético para los Científicos. Estudios anteriores (por ej. Lemarchand 1990a y 1990b; AAAS Committee on Scien-tific Freedom and Responsibility 2000) llegaron a incluir en el análisis números inferiores a 20 textos.

Algunas de las propuestas, muchas de las cuales fueron efectivamente implementadas, sugerían que se utilizaran los juramentos durante las ceremonias de graduación. Margaret Mead (1966) consideraba a las ceremonias de graduación como la versión moderna de la vieja idea de los ritos de transición, en la que conviven el nacimiento y la muerte, cada uno al final de un estado y el comienzo de otro. Dentro de esta modalidad de ritos arcaicos, se han venido incorporando nuevas transiciones, un tanto artificiales, basadas, no ya en la maduración biológica del hombre y su envejecimiento, sino en el aprendizaje, en la competencia, y en la aceptación de nuevas responsabilidades. Tales ceremonias de transición, como en el pasado lo eran la coronación de un rey, la ordenación de un sacerdote, la instalación de una nueva autoridad en la universidad, forman parte del tipo de patrones de validación que la sociedad ya tiene asumidos.

Los hombres y mujeres que se gradúan proceden hacia el reconocimiento, expresado en forma de títulos académicos, como médicos, abogados, físicos, ingenieros, matemáticos, maestros. Una vez adquiridos, los nuevos graduados comienzan a ser dignos de confianza en la sociedad. En cada profesión existe un conjunto especial de derechos y responsabilidades elegantemente expresadas en fórmulas tradicionales. Los que reciben títulos son admitidos inmediatamente por la sociedad a gozar de todas las responsabilidades y privilegios que dichas profesiones les otorgan.

A diferencia de la tradición moderna, en que se toma el Juramento Hipocrático durante las ceremonias de graduación, en sus orígenes este era un juramento que los estudiantes/aprendices tomaban en el momento de iniciar sus estudios en medicina. En su formato ori-ginal, se pueden distinguir tres secciones del juramento bien diferenciadas:

En la primera parte el aprendiz reconoce sus obligaciones personales con respecto a sus maestros. Como tales se pueden considerar que estas obligaciones eran, en realidad, mutuas.

En la segunda parte, el aprendiz promete esmerarse en la práctica del arte de la medicina y mantener los más altos estándares profesionales que le sean posibles, comprometiéndose a transmitir y legar ese conocimiento luego a sus propios estudiantes/aprendices.

Finalmente, en la tercera parte se comprometen a utilizar su conocimiento únicamente para aliviar el sufrimiento, garantizar la confidencialidad con sus pacientes y esmerarse para que en todo momento se evite generar cualquier tipo de daño.

El propio Karl Popper (1968, 1971) consideraba también que los científicos deberían comprometerse ante la sociedad, a través de un Juramento Hipocrático para hacer buen uso de sus conocimientos. Propuso que dicho juramento debería organizarse en las siguientes tres etapas:

*Responsabilidad profesional:* La primera obligación de todo estudiante de ciencias es comprometerse a incrementar los conocimientos y saberes de la humanidad a través de una búsqueda implacable de la verdad. Obviamente, como humanos, nadie es perfecto y aun las mentes más brillantes de la historia se han equivocado en algún momento. En cierta forma se debe reconocer el viejo apotegma socrático que la certeza del conocimiento es finita, mientras que la ignorancia es infinita.

*Responsabilidades del estudiante:* Se asume una pertenencia a una tradición y una comunidad y por ello se debe infundir respeto a aquellos que han contribuido en el camino de la búsqueda de la verdad. El estudiante le debe lealtad a sus maestros quienes han compar-tido generosamente su conocimiento y saber, con él y al mismo tiempo debería empeñarse en no perder la actitud crítica con si mismo y su comunidad, evitando en todo momento sucumbir ante la arrogancia intelectual.

*Responsabilidad ante la humanidad:* El estudiante debe ser consciente que toda investigación que realice puede tener consecuencias que eventualmente afectarán la vida de otras personas y por ello debe hacerse responsable por las mismas. Deberá advertir, en la medida de sus posibilidades, cualquier mal uso que de ese conocimiento se pueda realizar. El imperativo categórico de todo investigador científico debe centrarse en la búsqueda de saberes que no sólo permitan comprender mejor las leyes que regulan el universo sino que también garanticen que ese conocimiento nunca lesione a la humanidad o la naturaleza.

Finalmente, Popper consideraba que los científicos naturales deberían considerar como parte de sus responsabilidades especiales, prever, dentro de sus posibilidades todas aquellas consecuencias no intencionales que se puedan derivar de su labor y ser conscientes de los caminos que deben ser evitados.

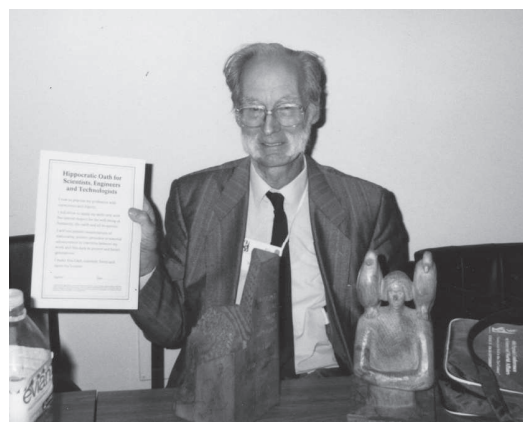

*Prof. Meredith W. Thring (1915-2006), ingeniero, inventor y humanista. En 1969 redactó y difundió un Juramento Hipocrático para Ingenieros y en 1987 fue responsable también de la composición del Juramento para Científicos, Ingenieros y Ejecutivos del Institute for Social Inventions (Londres), cuyo texto porta en su mano. Foto: Guillermo A. Lemarchand (c. 1990).*

**Tabla 4:** Lista de propuestas de Juramentos Hipocráticos, Códigos y Compromisos destinados a los científicos individuales en general. Fuente Elaboración propia.

| Fecha | Propuesta | Autor (es) | Notas en el Anexo |
|---|---|---|---|
| IV a.C. | Juramento Hipocrático | Escuela Pitagórica/Hipocrática | [1] |
| III a.C. | Juramento Indio de Iniciación a la Medicina | Charaka Samhita | [2] |
| C. 1200 | Oración de Maimónides† | Moisés Maimónides | [3] |
| 1608 | Juramento de Galeno para Farmacéuticos | 1608 versión original en Latín redactada por Jean de Renou, traducido al francés por Louis de Serres en 1624 | [4] |
| 1627 | Juramento Hipocrático de los Científicos de "La Nueva Atlántida" | Francis Bacon | [5] |
| 1807 | Juramento Hipocrático | Sociedad Médica del Estado de Nueva York | [6] |
| 1900 | Juramento de los Ingenieros de Canadá | Rudyard Kipling | [7] |
| 1917 | Propuesta de Juramento Hipocrático para Trabajadores Científicos | UK Association of Scientific Workers | [8] |
| 1929 | Juramento de Fe del Ingeniero | Michael Sullivan, Universidad de Michigan, clase 1929 | [9] |
| 1945 | Juramento Hipocrático para la Era Nuclear | Gene Weltfish (1945) | [10] |
| 1948 | Carta de los Trabajadores Científicos | Federación Mundial de Trabajadores Científicos (WFSW) | [11] |
| 1948 | Juramento Hipocrático Médico (versión modificada en la Declaración de Ginebra ) | Asamblea General de la Organización Mundial de la Salud | [12] |
| 1950 | Credo de los Ingenieros | Verein Deutscher Ingenieure | [13] |
| 1952 | Juramento para los Estadísticos | W.W.K. Freeman (1952) | [14] |
| 1954 | Credo de los Ingenieros | National Society of Professional Engineers (EEUU) | [15] |
| 1954 | Juramento de los Sicólogos | Edwin B Newman (1954) | [16] |
| 1954 | Juramento a la Bandera del Académico Americano | Edward J. Shoben (1954) | [17] |
| 1957 | Juramento Hipocrático para Científicos* | Fred W. Drecker (1957) | [18] |
| 1965 | Juramento de Graduación de la Universidad de Zagreb** | | [19] |
| 1966 | Juramento Hipocrático (actualizado) | Carl E. Taylor (1966) | [20] |
| 1967 | Juramento Hipocrático para Científicos | Harald Wergeland (1967) WG 6, 17 Conferencia Pugwash | [21] |
| 1968 | Juramento Hipocrático para Académicos | Eric Ashby (1968) | [22] |
| 1968 | Juramento Hipocrático para Científicos* | Karl R. Popper (1971) | [23] |
| 1969 | Juramento Hipocrático para Científicos* | J. Smittenberg (1969) | [24] |
| 1969 | Juramento Ético para Fitopatólogos | M. Jeuken (1969) | [25] |
| 1969 | Juramento para Ingenieros de Bratislava | Meredith W. Thring (1969) | [26] |
| 1970 | Juramento Ético para las Ciencias Naturales, Sociales y Humanidades | J.J. Groen (1970) | [27] |
| 1970 | Juramento de las Universidades de California en Berkeley y en Stanford | Charles Schwartz (1970) | [28] |

| Fecha | Propuesta | Autor (es) | Notas en el Anexo |
|---|---|---|---|
| 1970 | Juramento para Científicos Naturales | J. Dullaart (1970) | [29] |
| 1971 | Enmienda de la Sociedad de Física Americana (APS) | Robert March (1971) | [30] |
| 1971 | Cláusula Contractual de Responsabilidad Social de los científicos | International Society for Social Responsibility (Noruega) | [31] |
| 1972 | Juramento de Pugwash | Harald Wergeland y Philip Smith, WG 8, 22 Conferencia Pugwash | [32] |
| 1973 | Juramento Hipocrático para Investigadores Científicos* | Peter Sonntag (1973) | [33] |
| 1973 | Juramento del Ingeniero | Charles Susskind (1973) | [34] |
| 1974 | Juramento de la Universidad de Groningen | Philip Smith y texto de J. de Pugwash | [35] |
| 1975 | Juramento Hipocrático para Científicos* | W. Luck (1975) | [36] |
| 1978 | Juramento Hipocrático para los Físicos* | M. Beech (1978) | [37] |
| 1981 | Juramento Hipocrático para Académicos | Richard Davies (1981) | [38] |
| 1982 | Juramento Hipocrático para Científicos* | Daniel E. Harris (1982) | [39] |
| 1983 | Juramento Hipocrático adaptado a la Era Nuclear | AG OMS / IPPNW | [40] |
| 1983 | Juramento de Sarah | HTPFP | [41] |
| 1984 | Juramento en el Departamento de Física de la U. de Berkeley | Estudiantes y graduados | [42] |
| 1984 | Juramento en contra de las armas nucleares | M. Kellison (1984) | [43] |
| 1984 | Código de Ética de Uppsala | Bengt Gustafsson et al. (1984) | [44] |
| 1984 | Código de Ética de Wittemberg | Hans Peter Gensichen (1984) | [45] |
| 1985 | Compromiso para negarse a trabajar en la Iniciativa de Defensa Estratégica (IDE) | John Kogut et al. (1985) | [46] |
| 1986 | Juramento para Ingenieros y Científicos de Atenas* | Comisión Nacional de apoyo a la UNESCO de Grecia | [47] |
| 1986 | Propuesta de la Asociación de Estudiantes de Física Argentina (Reunión de Tucumán)* | Comisión de Astrofísica del CECEN-FCEN-UBA (1986) | [48] |
| 1987 | Endoso en las publicaciones científicas de Ciencia para la Paz | Peter Willis (1987) | [49] |
| 1987 | Compromiso del Comité para la Genética Responsable | Jonathan King (1987) | [50] |
| 1987 | Juramento de la Universidad Estatal de Humboldt | Matt Nicodemus y J. Berman (1987) | [51] |
| 1987 | Propuesta de la rama Argentina del IPPNW* | Emanuel Levin (1987) | [52] |
| 1987 | Juramento Hipocrático para Científicos, Tecnólogos y Ejecutivos | Institute for Social Inventions (1987) | [53] |
| 1987 | Juramento Hipocrático para Científicos (1) | David Krieger (1987) | [54] |
| 1987 | Juramento Hipocrático para Ingenieros (2) | David Krieger (1987) | [55] |
| 1987 | Juramento de Graduación para Científicos *** | Anatol Rapoport (1987) | [56] |
| 1988 | Juramento de Buenos Aires | Guillermo A. Lemarchand et al. (1988) | [57] |
| 1988 | Juramento Hipocrático para Científicos*** | Chandler Davies (1988) | [58] |
| 1988 | Juramento para el Ciudadano de la Tierra | David Krieger (1988) | [59] |
| 1989 | Juramento Hipocrático para Científicos* | André Baccard (1989) | [60] |

| Fecha | Propuesta | Autor (es) | Notas en el Anexo |
|-------|-----------|------------|-------------------|
| 1989 | Juramento Hipocrático para Científicos | Arnold Toynbee (1989) | [61] |
| 1990 | Propuesta de la 40ª Conferencia de Pugwash *** | WG 8, 40 Conferencia Pugwash | [62] |
| 1990 | Manifiesto sobre los derechos y responsabilidades de los trabajadores científicos WFSW y adopción del Juramento del Institute for Social Inventions (1987) | World Federation of Scientific Workers (WFSW) Science Policy committee | [63] |
| 1990 | Propuesta de Juramento Hipocrático para Científicos * | Conferencia de las Naciones Unidas en Sendal, Japón (1990) | [64] |
| 1990 | Juramento de Arquímedes (Primera versión) | Instituto Federal Politécnico de Grenoble | [65] |
| 1991 | Compromiso de los científicos en no tomar parte de investigaciones con financiamiento militar | Scientists Against Nuclear Arms (SANA), Londres | [66] |
| 1993 | Juramento Hipocrático para Científicos UCSB | John Ernest (1993) | [67] |
| 1995 | Juramento Hipocrático para Científicos | Jóvenes y Estudiantes de Pugwash | [68] |
| 1995 | Compromiso de Ingenieros y Científicos | INES International Network of Engineers and Scientists for Global Responsibility | [69] |
| 1998 | Juramento de Científicos para no trabajar en cuestiones bélicas* | Stephen Jay Gould (1998) | [70] |
| 1999 | Juramento Hipocrático para Científicos | Michel Serres (Correo de la UNESCO, 1999) | [71] |
| 1999 | Juramento Hipocrático para científicos en la Conferencia Mundial de la Ciencia de Budapest | Joseph Rotblat (1999) | [72] |
| 1999 | Juramento Hipocrático para Científicos propuesto en la Conferencia Mundial de la Ciencia por delegación Griega | Nicolas K. Artemiadis‡ | [73] |
| 1999 | Compromiso de Paz del Movimiento de Científicos de Japón | http://www.peacepledge.jp/ | [74] |
| 1999 | Juramento de los Científicos | Arnold Wolfendale (1999) | [75] |
| 2000 | Juramento de Arquímedes (Segunda versión) | Instituto Federal Politécnico de Grenoble | [76] |
| 2000 | Juramento de la AFA | Karen Hallberg et al. | [77] |
| 2000 | Juramento de Metz | Gilles E. Seralini et al. | [78] |
| 2001 | Compromiso de científicos e ingenieros para renunciar a trabajar en armamento de destrucción en masa | The Los Alamos Study Group, The Natural Resources Defense Council, TriValley CAREs y The Western States Legal Foundation | [79] |
| 2003 | Juramento para Biocientíficos | Daniel Fu-Chang Tsai y Ding-Shinn Chen | [80] |
| 2003 | Juramento Hipocrático para Científicos-COMEST | Henk ten Have (2003) | [81] |
| 2004 | Juramento Hipocrático para Desarrolladores de Software | Philip A. Laplante (2004) | [82] |
| 2005 | Código de Ética de la Sociedad de Energía Atómica de Japón | AESJ Ethics Committee | [83] |
| 2005 | Código de ética para evitar el bioterrorismo | M.A. Somerville y R. M. Atlas (2005) | [84] |
| 2006 | Juramento Hipocrático para Ciencias de la Vida | James Revill y Malcolm R. Dando (2006) | [85] |
| 2007 | Código de Ética y Juramento Hipocrático para Científicos | David King | [86] |
| 2007 | Los 10 Mandamientos de la Educación Superior | David Watson (2007) | [87] |
| 2007 | Juramento de Graduación de la Universidad de Toronto | K. D. Davies et a. (2008) | [88] |

| Fecha | Propuesta | Autor (es) | Notas en el Anexo |
|-------|-----------|------------|-------------------|
| 2009 | Juramento de "Aventura Espacial" IYA 2009 | Ciro Arévalo (COPOUS) | [89] |
| 2009 | Juramento Hipocrático para Ciencias de la Tierra | Eric C. Ellis y Peter K. Haff (2009) | [90] |

**Notas de la Tabla 4:**

* No se propuso ningún texto específico. †Se considera que probablemente la Oración de Moisés Maimónides, no haya sido compuesta por él mismo, sino por el médico judío alemán del siglo XVIII, Markus Herz, discípulo de Kant. ** Hay una incertidumbre en la fecha exacta que comenzó a usarse el Juramento de la Universidad de Zagreb. El mismo fue utilizado como modelo en la propuesta de Schwartz (1970) o sea que debería ser anterior a esa fecha. *** Se propusieron textos de ejemplo como el "Juramento de Buenos Aires", "Código de Ética de Uppsala" o el "Juramento Hipocrático para Científicos, Tecnólogos y Ejecutivos". ‡Si bien este texto se presentó en la CMC de Budapest en 1999, se hace referencia a que dicho Juramento ya era aplicado en una universidad griega que no se especificó (ver http://www.unesco.org/science/wcs/forum_3/greece.htm ). Es muy posible que el juramento haya sido establecido entre 1986 y 1999, de acuerdo a la propuesta [47]. Por esta razón hay una cierta incertidumbre en la fecha exacta en que dicho Juramento fue propuesto.

Como se detalló en la sección anterior, la detonación de las bombas atómicas en Hiroshima y Nagasaki, y la carrera armamentista nuclear que nació a partir de ese momento, desencadenó una alarma dentro de la propia comunidad científica acerca de uso y mal uso del conocimiento científico-tecnológico. Las primeras propuestas de Juramentos Hipocráticos para Científicos en donde estos últimos se comprometen a utilizar sus saberes en beneficio de la humanidad y a favor de la paz, comienzan en 1945. Es interesante constatar, que de acuerdo a los datos de la Tabla 4, la frecuencia con que aparecen las distintas propuestas de juramentos o compromisos éticos se ajustan muy bien tanto al crecimiento en el número de ojivas nucleares totales, como al gasto militar mundial.

La gráfica 6 muestra que el número acumulado de propuestas de Juramentos Éticos de Científicos en función del tiempo puede ser descripto a través de un crecimiento del tipo logístico. Esta curva, frecuente en la ecología de poblaciones, es similar también al tipo de curvas que se encuentran cuando se estudian los fenómenos de difusión de tecnologías en un nicho de mercado dado. De alguna manera, puede considerar a este tipo de compromisos éticos de los científicos como una especie de tecnología desincorporada, destinada a infundir consciencia entre los practicantes de la ciencia y tecnología, para que orienten sus saberes únicamente en benefició de la humanidad y a favor de la paz.

La derivada matemática de la curva logística representa la tasa de propuestas en función del tiempo (en la gráfica 6 se representa en líneas punteadas). Es interesante notar que el pico de propuestas de compromisos éticos (año 1987,4) coincide con el pico en el número de ojivas nucleares desplegadas y con el pico de gastos militares mundiales. Fue en ese período cuando se gestó el Juramento de Buenos Aires. Solo en 1987 se propusieron, en forma independiente, 8 iniciativas de juramentos éticos para científicos.

Aun en universidades tradicionalmente vinculadas con la I+D militar, como por ejemplo la Universidad de Stanford, en 1988, durante la 97 Ceremonia de Graduación, su entonces presidente, Donald Kennedy[12], pronunció un discurso en donde -hablando del Juramento Hipocrático para Científicos- dijo lo siguiente: *"debería ser aceptable, tanto para los conservadores como para los liberales, pues es algo que todos necesitamos promover entre los estudiantes, esto es enfocarnos en las consecuencias de lo que hacemos".*

---

[12]   *Kennedy Addresses Timeless Question: Life After Stanford, The Stanford University Campus Report, pp. 9-23, June 15, 1988.*

En 1988, Joseph Rotblat en su presentación para la Conferencia Anual del Grupo Pugwash, reconoce por primera vez que tal vez la idea de los Juramentos Hipocráticos para Científicos puede ser una iniciativa a contemplar y convoca, en la conferencia anual realizada en Egham en 1990, a un taller de trabajo específico de este tema (Lemarchand 1990b).

También en 1990, dentro de una conferencia organizada por las Naciones Unidas sobre ciencia, tecnología y seguridad mundial[13] en la cual participaron unos 50 científicos, expertos técnicos, académicos, diplomáticos y dirigentes políticos de unos 23 países se debatió la idea de aplicación de un código de conducta para científicos. En este caso, los participantes consideraron que era difícil lograr un equilibrio adecuado entre la libertad científica y la responsabilidad social, aunque abrigaron la esperanza de que las opciones

---

[13] *Conferencia de las Naciones Unidas sobre nuevas tendencias en material de ciencia y tecnología: consecuencias para la paz y la seguridad internacionales, Sendai, Japón, 16-19 de abril de 1990. Boletín de Desarme de las Naciones Unidas, Junio de 1990, p. 15.*

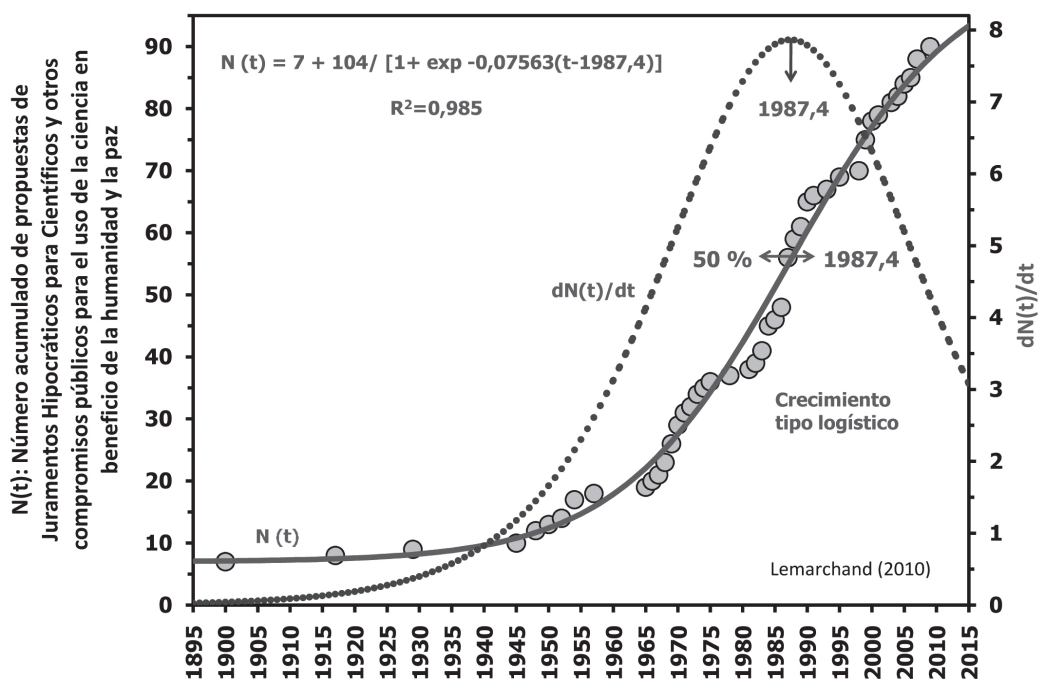

**Grafica 6:** Distribución N(t) del número acumulado de propuestas de juramentos y otros compromisos para científicos de acuerdo al listado de la Tabla 4. El mismo muestra un comportamiento tipo logístico con un coeficiente de Pearson $R^2 = 0,985$. Este es el mismo tipo de comportamiento que muestra cualquier tecnología que se difunde en un nicho cerrado. Asimismo, se representa la derivada temporal dN(t)/dt que indica la tasa de propuestas anuales. En este caso el pico de propuestas se encuentra en el año 1987,4. Fuente: Elaboración propia.

éticas podrían ser implementadas dentro del marco de una perspectiva mundial y con un punto de vista de vasto alcance.

La idea del uso generalizado de un juramento ético para científicos, llegó a introducirse en la agenda internacional a partir de la iniciativa de Joseph Rotblat propuesta durante la Conferencia Mundial de Ciencia de Budapest en 1999. Durante un tiempo, tanto el Comité de Libertad y Responsabilidad Científica de la Asociación Americana para el Avance de la Ciencia (AAAS Committee on Scientific Freedom and Responsibility 2000) como el COMEST (ten Have, en este volumen) analizaron la posibilidad de redactar un juramento universal para todos los científicos e ingenieros.

Estas iniciativas no tomaron en cuenta que en diversas universidades del planeta, varias fórmulas de juramento ya eran utilizadas durante sus respectivas ceremonias de graduación. Estas discusiones se centraron en la "forma" y contenido del texto, más que en la "esencia" de la idea subyacente, que es lo trascendente. El punto verdaderamente importante a lograr es que los estudiantes o jóvenes profesionales, mediten en algún momento de su vida acerca de la responsabilidad -en el sentido amplio- de la tarea que desempeñan y asuman un compromiso con la sociedad de hacer su mejor esfuerzo por trabajar únicamente en beneficio de la humanidad y a favor de la paz.

## 5. La génesis del Juramento de Buenos Aires

La idea de proponer un Juramento Hipocrático para Científicos surgió a iniciativa de algunos de los miembros del Comité Organizador del Simposio Internacional sobre los Científicos, la Paz y el Desarrollo a principios de 1986. Ese mismo año, sus promotores presentaron un proyecto de juramento ético de graduación, durante una reunión nacional de estudiantes universitarios de física celebrada

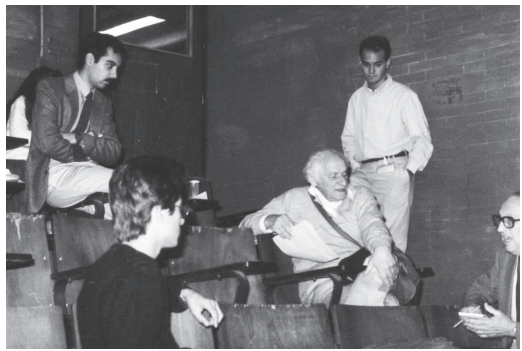

Imagen tomada durante las discusiones del taller de trabajo que tenía la responsabilidad de redactar el borrador del "Juramento de Buenos Aires". En la imagen de izquierda a derecha aparecen; Ernst Hamburger (Universidad de San Pablo), Guillermo A. Lemarchand (Comité Organizador), Gerardo Pozetti (estudiante de FCEN), Mischa Cotlar (Universidad Central de Caracas), Leonardo Graciotti (estudiante FCEN) y Emanuel Levín (IPPNW). Foto: Gabriela Bagalá (c. 1988).

en la ciudad de San Miguel de Tucumán (ver Anexo, nota [48]).

En las distintas convocatorias realizadas durante la preparación de la reunión internacional se explicitó que durante el Simposio se discutiría una propuesta de un Juramento Hipocrático para Científicos. Diversos representantes de distintas nacionalidades y áreas de la ciencia manifestaron su interés de participar del taller de trabajo en donde se elaboraría el texto del Juramento propuesto. Dentro de la asamblea del Simposio, la iniciativa del Juramento Hipocrático para Científicos fue presentada por Lemarchand (1988a).

Finalmente, durante la semana del 11 al 15 de abril de 1988, se reunió un Comité Redactor que estuvo integrado por Mischa Cotlar (Universidad Central de Venezuela), Ernst Hamburger (Universidad de San Pablo), Daniel Harris (Universidad de Harvard), Jean Marie Legay (Presidente de la Federación Internacional de Trabajadores Científicos, Francia), Emanuel Levín (IPPNW, Argentina), Jeremy Stone (Presidente de la Federación de Científicos Americanos, EEUU), y Guillermo

A. Lemarchand (Comité Organizador del Simposio, UBA). Durante las distintas reuniones participaron también una gran cantidad de estudiantes que estaban muy interesados en el tema. En una segunda etapa, Patricia Morales (Facultad de Filosofía y Letras de la UBA) colaboró en la edición de la versión final en castellano, que resultó ser un poco más precisa y mejor elaborada que su correspondiente versión en inglés.

Se debe señalar que cuando comenzaron las discusiones acerca de los lineamientos del Juramento, ninguno de los miembros del Comité Redactor conocía, por entonces, textos de otras propuestas análogas (por ej. Tabla 4). Se debe destacar que al menos tres de sus miembros (Harris, Levin y Lemarchand, ver Anexo

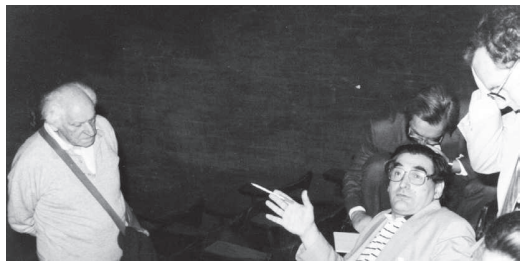

*Momento de debate sobre el contenido del texto del "Juramento de Buenos Aires", de izquierda a derecha, Mischa Cotlar (Universidad Central de Venezuela), Ernst Hamburger (Universidad de San Pablo), Daniel Harris (Universidad de Harvard) y Jeremy Stone (Presidente de la Federación de Científicos Americanos) y Paula da Cunha (estudiante de FCEN). Foto: Guillermo A. Lemarchand (c. 1988).*

notas [39], [52] y [48] respectivamente) ya habían realizado en el pasado manifestaciones públicas acerca de la necesidad de instaurar un juramento de estas características.

Dentro del Comité Redactor surgieron rápidamente dos posiciones, una que proponía redactar en forma explícita un compromiso para negarse a trabajar en cualquier actividad que estuviera financiada por el sector militar y otra que temía que la primera alternativa pudiera desencadenar discriminaciones y fracturas dentro de la comunidad científica. Incluso, dadas las distintas connotaciones semánticas de inglés y el español, se acordó que las versiones en dichos idiomas podrían ser ligeramente distintas. Por ejemplo en la versión en inglés se usa el sustantivo "sociedad", mientras que en español se prefirió utilizar "humanidad".

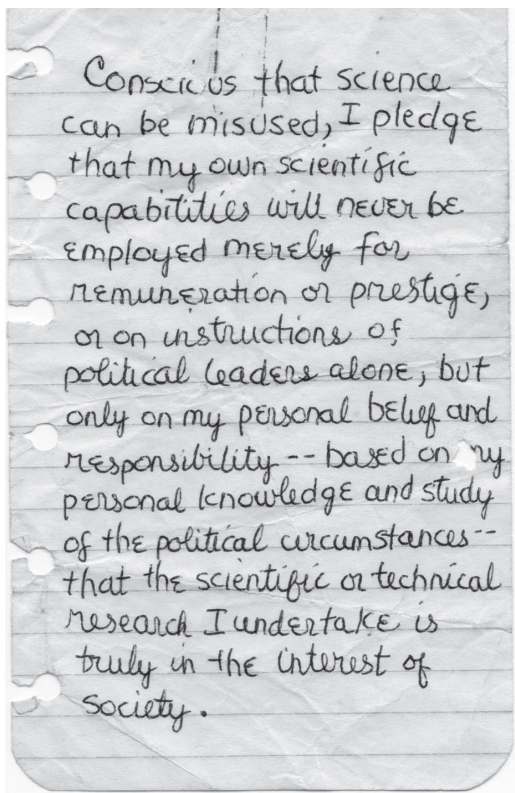

*Facsímil del manuscrito de Jeremy Stone (Presidente de la Federación de Científicos Americanos) con una de las primeras versiones (en inglés) del Juramento de Buenos Aires. Fuente: G.A. Lemarchand (c. 1988).*

El objetivo de la versión en inglés era disponer de un documento que circularía dentro de la comunidad científica internacional, para que sus miembros lo firmaran. Mientras que la versión en español debía estar preparada en un formato adecuado para que también pudiera ser utilizada en las ceremonias de graduación en las universidades.

JURAMENTO DE BUENOS AIRES

"Teniendo conciencia de que la ciencia y en particular sus resultados pueden ocasionar perjuicios a la sociedad y al ser humano cuando se encuentran ausentes los controles éticos: ¿Juráis que la investigación científica y tecnológica que desarrollareis será para beneficio de la humanidad y en favor de la paz, que os comprometéis firmemente a que vuestra capacidad como científico nunca servirá a fines que lesionen la dignidad humana, guiándoos por vuestras convicciones y creencias personales, asentadas en un auténtico conocimientos de las situaciones que os rodean y de las posibles consecuencias de los resultados que puedan derivarse de vuestra labor, no anteponiendo la remuneración o el prestigio, ni subordinándoos a los intereses de empleadores o dirigentes políticos? Si así no lo hicieris, vuestra conciencia os lo demande.

Buenos Aires, 15 de abril de 1988

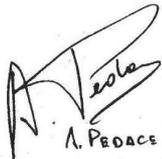
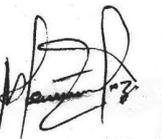

*Facsímil de la primera página de las firmas del "Juramento de Buenos Aires" recogidas el 15 de abril de 1988, durante la ceremonia de clausura del Simposio Internacional sobre los Científicos, la Paz y el Desarme. Se puede reconocer varias firmas de distinguidos científicos como Daniel Bes (CNEA-AFA), Raúl Boix Amat (ICSC World Laboratory), Mischa Cotlar (Universidad Central de Caracas), Daniel Harris (Universidad de Harvard), Amílcar Herrera (Distinguido geólogo y especialista latinoamericano en política científica y padre del Modelo Mundial de Bariloche), Carlos Mallmann (Director del Centro de Estudios Avanzados de la UBA), José Monserrat Filho (Ministerio de Ciencia y Tecnología de Brasil), Enrique Oteiza (Primer Director del Centro de Educación Superior para América Latina y el Caribe de la UNESCO), A. Roque Pedace (Secretario de Extensión Universitaria de FCEN, UBA), Fernando de Souza Barros (Sociedad Brasileña de Física), Jeremy Stone (Pte. de la Federación de Científicos Americanos), entre otros. Las membresías corresponden a las del año 1988. Fuente: G.A. Lemarchand (c. 1988).*

Finalmente, se consensuó una fórmula que hiciera hincapié en la conciencia individual, en donde los candidatos se comprometieran a dedicar su trabajo solo en beneficio de la humanidad y a favor de la paz, no anteponiendo ni la remuneración ni el prestigio, ni subordinarse ante la voluntad de empleadores o dirigentes políticos, asumiendo la responsabilidad total de sus propias acciones.

El 15 de abril de 1988 se presentó ante la Asamblea del Simposio Internacional sobre los Científicos, la Paz y el Desarme el texto final del "Juramento de Buenos Aires" que fue rápidamente aclamado. Espontáneamente el distinguido grupo de científicos participantes solicitó rubricar el acto con sus respectivas firmas.

Seguidamente se trabajó en desarrollar dos estrategias simultáneas. Por un lado dar a conocer el texto en inglés al mayor número de miembros de la comunidad científica internacional y por otro transformar el Juramento de Buenos Aires en una de las fórmulas optativas que los estudiantes pudieran acceder durante las ceremonias de graduación en la Facultad de Ciencias Exactas y Naturales de la Universidad de Buenos Aires.

Para la primera estrategia, se contó con el apoyo económico de la Federación de Científicos Americanos (FAS) que ayudó a difundir entre sus miembros el texto del nuevo juramento y a publicitarlo en los distintos medios académicos. Rápidamente, colegas de diversas partes del mundo comenzaron a hacer llegar traducciones del texto en inglés a sus respectivos idiomas nativos. Se recibieron traducciones al ruso, árabe, chino, sueco, alemán, etc. La repercusión fue casi instantánea y en pocos meses una docena de Premios Nobel en ciencias y centenares de destacadísimos científicos internacionales habían enviado ya el juramento con sus correspondientes firmas (ver por ej. las copias que aparecen en las imágenes contiguas).

A los tres meses se publicó una nota en la prestigiosa revista *Physics Today* que estaba encabezaba con las siguientes líneas "Científicos de Argentina y Gran Bretaña, seis años después de la Guerra por las Islas Malvinas, proponen independientemente sendos Juramentos Hipocráticos para Científicos" (Sweet 1988). Merced a la publicidad que el juramento iba recibiendo comenzaron a llegar noticias de otras propuestas de juramentos similares que habían aparecido en forma independiente en diversas partes del planeta, en el lapso de unos pocos meses (por ej. Tabla 4).

En agosto de 1988, se estaba desarrollando en la ciudad de Baltimore (EEUU), la XX Asamblea General de la Unión Astronómica Internacional (IAU). La misma contó con la presencia de unos 3.000 astrónomos de distintas partes del mundo. Allí, el Comité Nacional de Astronomía de la República Argentina, a través de su representante nacional, el Dr. Roberto H. Méndez (1988), presentó un proyecto de resolución de la Asamblea para que se invite a los miembros de la comunidad astronómica internacional y a las universidades del mundo a considerar implementar el uso voluntario de un compromiso ético en las ceremonias de graduación, similar al Juramento de Buenos Aires. Una iniciativa que parecía ser aparentemente inocente y poco comprometida. En un primer momento, se entendió que era adecuado hacer dicha propuesta en ese ámbito debido a que las ciencias astronómicas y astrofísicas parecían tener temas de investigación muy alejados de la I+D militar.

Simultáneamente y en forma independiente, la delegación sueca había presentado también una propuesta donde deploraba el incremento de las actividades militares en el espacio ultraterrestre y urgía a mantener el uso del espacio para fines científicos de carácter pacífico, en alusión tácita a la Iniciativa de Defensa Estratégica.

79

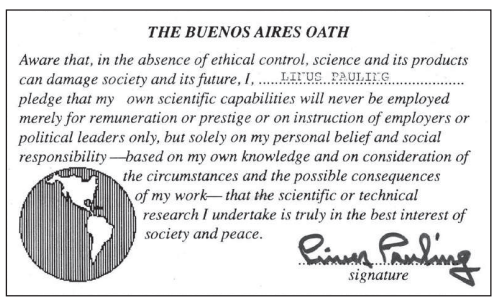

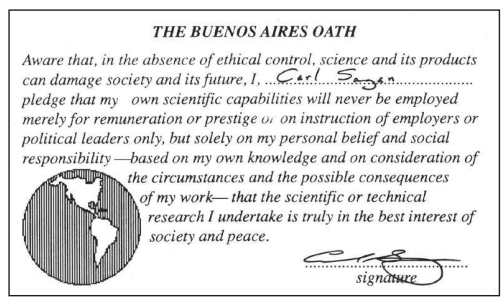

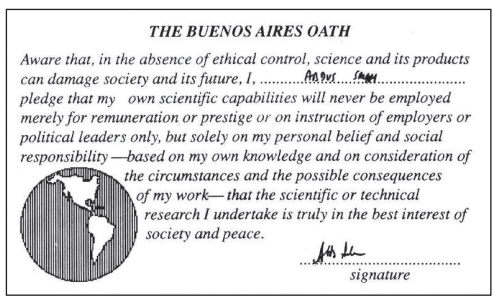

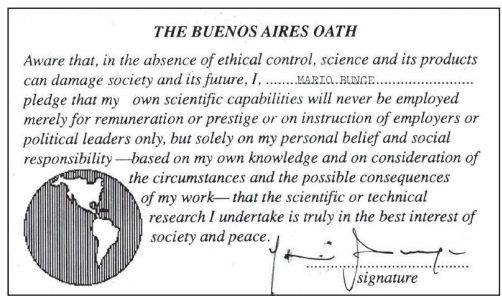

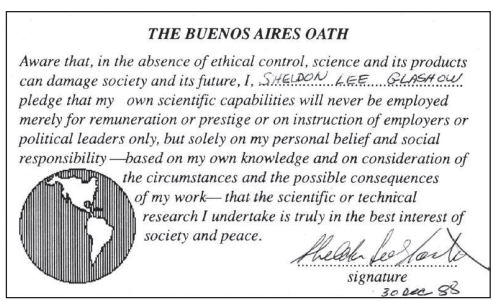

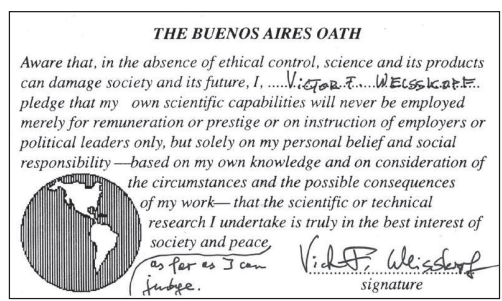

El Juramento de Buenos Aires (en su versión en inglés) firmada por tres científicos galardonados con el Premio Nobel: Linus Pauling (Premio Nobel de Química en 1954 y Premio Nobel de la Paz en 1962), Abdus Salam (Premio Nobel de Física en 1979) y Sheldon Lee Glashow (Premio Nobel de Física en 1979). Fuente: G.A. Lemarchand (c. 1988)

El Juramento de Buenos Aires (en su versión en inglés) firmada por otros tres científicos de fama internacional: Carl Sagan, Mario Bunge y Víctor F. Weisskopf (veterano del Proyecto Manhattan). Fuente: G.A. Lemarchand (c. 1988).

El proyecto del juramento había tenido una cálida recepción por parte de las bases de la comunidad astronómica que estaban reunidas en el lugar. Positivos comentarios sobre la propuesta de generalizar el Juramento de Buenos Aires a la comunidad astronómica internacional, aparecieron durante la Asamblea de la IAU publicados en *IAU Today* y *Sky & Telescope* (Méndez 1988; Tresch Fienberg 1988). Pese a ello, el Secretario General de la IAU mostró una gran resistencia para los dos proyectos (argentino y sueco), argumentando que "la IAU no suele tomar la iniciativa en cuestiones como éstas, que puede interpretarse como no estrictamente de su incumbencia, especialmente si se percibe que el tratamiento del tema podría provocar votaciones divididas en el Comité Ejecutivo" e instó a que los dos países retiraran sus respectivos proyectos y sugirió que eventualmente remitieran los mismos al ICSU (Méndez 1988). La respuesta recibida resultó ser un buen indicador de los tiempos que corrían.

En el año 1990, las Conferencias Pugwash sobre Ciencia y Asuntos Mundiales, convocaron a la reunión de un grupo de trabajo dedicado exclusivamente a tratar el tema de los Juramentos Hipocráticos para Científicos y en donde uno de los ejemplos considerados y sugeridos para su generalización fue el Juramento de Buenos Aires (Lemarchand 1990b).

Ese mismo año, la Federación Internacional de Trabajadores Científicos (WFSW), puso a consideración de su Asamblea General la incorporación de un Juramento Hipocrático destinado a sus miembros. Las fórmulas consideradas fueron por un lado el Juramento para Científicos, Ingenieros y Ejecutivos del *Institute for Social Inventions* de Londres y por otro la de la versión en inglés del Juramento de Buenos Aires. Finalmente, sus miembros optaron por el primer texto, cuya redacción en ingles (idioma original) presenta algunas ventajas con respecto al segundo.

En el año 1995, el Juramento de Buenos Aires fue también incluido como fuente para la redacción de un "Compromiso para Científicos e Ingenieros" propuesto por la Red Internacional de Ingenieros y Científicos por la Responsabilidad Global (INES).

Por otra parte, la estrategia local resultó ser mucho más fructífera. En el mes de mayo de 1988, a través del expediente 2.418/88, se hizo una presentación formal ante el Consejo Directivo (CD) de la Facultad de Ciencias Exactas y Naturales de la Universidad de Buenos Aires, solicitándole tuviera a bien considerar al texto del "Juramento de Buenos Aires" como una de las posibles fórmulas a utilizarse durante las ceremonias de graduación en dicha facultad. El 18 de diciembre de 1988, el CD mediante la Resolución 1.651 aprueba la propuesta y la remite al Consejo Superior (CS) de la UBA.

Finalmente, el 22 de marzo de 1989 el CS de la UBA a través de la Resolución 3.768 aprueba oficialmente el texto del Juramento de Buenos Aires como una de las fórmulas optativas para ser utilizadas en las ceremonias de graduación donde se reciben tanto los títulos de grado como los de posgrado.

Desde entonces, en cada ceremonia de graduación entre el 70 y el 80 % de los nuevos graduados deciden asumir públicamente las responsabilidades que el Juramento Hipocrático para Científicos de Buenos Aires establece.

## 6. Ciencia para la paz y el desarrollo

A lo largo de este trabajo se mostró cómo la ciencia y la tecnología fueron utilizadas para multiplicar cientos de millones de veces la capacidad destructiva del armamento disponible. También se detalló cómo, aun después de la caída del muro de Berlín, la humanidad se encuentra aún embarcada en procesos de militarización que consumen anualmente 10 veces el costo total que implica implementar totalmente los Objetivos de Desarrollo del Milenio propuestos por las Naciones Unidas. Los gastos anuales en tareas de investigación y desarrollo con objetivos militares, implican el presupuesto de la UNESCO para 412 años de funcionamiento o del Sector Ciencias Naturales de dicha organización durante 4.100 años.

A medida que una importante fracción de la comunidad científica internacional, hacía uso de esos fondos disponibles para tareas de I+D con objetivos militares, otro grupo alarmado por la escandalosa situación, operando casi de la misma manera que los anticuerpos ante una infección en un organismo enfermo, propusieron una larga lista de normas éticas (compromisos, juramentos, manifiestos, códigos, etc.) para garantizar que la ciencia no sea utilizada con fines que lesionen la dignidad humana y perjudiquen la continuidad de la vida en el planeta.

Cuando se adquiere una perspectiva cósmica de lo que representa nuestro planeta en la sinfonía universal, un pequeño punto azul contemplado desde los confines del sistema solar. Un solo píxel donde habitan, habitaron o habitarán todos los que conocemos, todos aquellos que alguna vez sentimos nombrar, cada ser humano que caminó o caminará por la faz de este mundo, cada ser que amamos, cada forma de vida que conocemos, de las bacterias a los elefantes, de los bebés a los ancianos, del ser más altruista al más egoísta, todos concentrados en un pequeño punto azul que se pierde en el cosmos infinito.

Desde esta perspectiva se muestra la fragilidad y vulnerabilidad terrestre, no para el planeta en sí, sino para la especie *Homo Sapiens*, organismos que se ven a sí mismos como los dominantes y rectores de la vida en el mundo. Únicos responsables de poner en peligro la habitabilidad de su hogar y aun incapaces de garantizar que su especie pueda seguir evolucionando dentro de los eones del tiempo.

## Referencias:

## Anexo. Textos de los Juramentos y Compromisos para científicos propuestos en distintas épocas

En este anexo se reproducen, en sus lenguas originales, los textos de los Juramentos, Compromisos y Códigos para el uso del conocimiento científico en beneficio de la Humanidad, destinados a científicos que se encuentran enumerados en la tabla 4. En general, siempre que ha sido posible, se ha transcripto los textos en su lengua original.

**[1]    Juramento Hipocrático (c. IV a.C.)**

*"Juro por Apolo médico, por Asclepio y por Higía, por Panacea y por todos los dioses y diosas, tomándolos por testigos, que cumpliré, en la medida de mis posibilidades y mi criterio, el juramento y compromiso siguientes:*

*Considerar a mi maestro en medicina como si fuera mi padre; compartir con él mis bienes y, si llega el caso, ayudarle en sus necesidades; tener a sus hijos por hermanos míos y enseñarles este Arte, si quieren aprenderlo, sin gratificación ni compromiso; hacer a mis hijos partícipes de los preceptos, enseñanzas y demás doctrinas, así como a los de mi maestro, y a los discípulos comprometidos y que han prestado juramento según la ley médica, pero a nadie más.*

*Dirigiré la dieta con los ojos puestos en la recuperación de los pacientes, en la medida de mis fuerzas y de mi juicio y les evitaré toda maldad y daño.*

*No administraré a nadie un fármaco mortal, aunque me lo pida, ni tomaré la iniciativa de una sugerencia de este tipo.*

*Asimismo, no recetaré a una mujer un pesario abortivo; por el contrario, viviré y practicaré mi arte de forma santa y pura.*

*No operaré con cuchillo ni siquiera a los pacientes enfermos de cálculos, sino que los dejaré en manos de quienes se ocupan de estas prácticas.*

*Al visitar una casa, entraré en ella para bien de los enfermos, manteniéndome al margen de daños voluntarios y de actos perversos, en especial de todo intento de seducir a mujeres o muchachos, ya sean libres o esclavos.*

*Callaré todo cuanto vea u oiga, dentro o fuera de mi actuación profesional, que se refiera a la intimidad humana y no deba divulgarse, convencido de que tales cosas deben mantenerse en secreto.*

*Si cumplo este juramento sin faltar a él, que se me conceda gozar de la vida y de mi profesión rodeado de la consideración de todos los hombres hasta el final de los tiempos, pero si lo violo y juro en falso, que me ocurra todo lo contrario."*

**Juramento Hipocrático · Versión moderna**

*"Juro (…) que yo, con todas mis fuerzas y con pleno conocimiento, cumpliré enteramente mi juramento (…), que dejaré participar en las doctrinas e instrucciones de toda la disciplina (…) a aquellos que con escrituras y juramentos se declaren discípulos míos, y a ninguno más fuera de estos.*

*Por lo que respecta a la curación de los enfermos, ordenaré la dieta según mi mejor juicio y mantendré alejado de ellos todo inconveniente. No me dejaré inducir por las súplicas de nadie, sea quien fuere, a administrar un veneno o a dar mi consejo en semejante contingencia.*

*Consideraré sagrados mi vida y mi arte (…) y cuando entre en una casa, entraré solamente para el bien de los enfermos y me abstendré de toda acción injusta (…).*

*Todo lo que vea y oiga durante la cura o fuera de ella en la vida común, lo callaré y conservaré siempre como secreto, si no me es permitido decirlo.*

*Si mantengo perfecta e intacta fe en este juramento, que me sea concedida una vida afortunada y la futura felicidad en el ejercicio del arte, de modo que mi fama sea alabada en todos los tiempos; pero si yo faltare al juramento o hubiere jurado en falso, que ocurra lo contrario."*

[2] **Juramento Médico del *Charaka Samhita* (c III a.C.)**

La siguiente traducción al inglés fue tomada de Menon y Haberman (1970)

**The Oath of Initiation**

1.  *The teacher then should instruct the disciple in the presence of the sacred fire, Brahmanas [Brahmins] and physicians.*

2.  *[saying] 'Thou shalt lead the life of a celebate, grow thy hair and beard, speak only the truth, eat no meat, eat only pure articles of food, be free from envy and carry no arms.*

3.  *There shall be nothing that thou should not do at my behest except hating the king, causing another's death, or committing an act of great unrighteousness or acts leading to calamity.*

4.  *Thou shalt dedicate thyself to me and regard me as thy chief. Thou shalt be subject to me and conduct thyself for ever for my welfare and pleasure. Thou shalt serve and dwell with me like a son or a slave or a supplicant. Thou shalt behave and act without arrogance, with care and attention and with undistracted mind, humility, constant reflection and ungrudging obedience. Acting either at my behest or otherwise, thou shalt conduct thyself for the achievement of thy teacher's purposes alone, to the best of thy abilities.*

5.  *If thou desirest success, wealth and fame as a physician and heaven after death, thou shalt pray for the welfare of all creatures beginning with the cows and Brahmanas.*

6.  *Day and night, however thou mayest be engaged, thou shalt endeavour for the relief of patients with all thy heart and soul. Thou shalt not desert or injure thy patient for the sake of thy life or thy living. Thou shalt not commit adultery even in thought. Even so, thou shalt not covet others' possessions. Thou shalt be modest in thy attire and appearance. Thou shouldst not be a drunkard or a sinful man nor shouldst thou associate with the abettors of crimes. Thou shouldst speak words that are gentle, pure and righteous, pleasing, worthy, true, wholesome, and moderate. Thy behavior must be in consideration of time and place and heedful of past experience. Thou shalt act always with a view to the acquisition of knowledge and fullness of equipment.*

7.  *No persons, who are hated by the king or who are haters of the king or who are hated by the public or who are haters of the public, shall receive treatment. Similarly, those who are extremely abnormal, wicked, and of miserable character and conduct, those who have not vindicated their honour, those who are on the point of death, and similarly women who are unattended by their husbands or guardians shall not receive treatment.*

8. *No offering of presents by a woman without the behest of her husband or guardian shall be accepted by thee. While entering the patient's house, thou shalt be accompanied by a man who is known to the patient and who has his permission to enter; and thou shalt be well-clad, bent of head, self-possessed, and conduct thyself only after repeated consideration. Thou shalt thus properly make thy entry. Having entered, thy speech, mind, intellect and senses shall be entirely devoted to no other thought than that of being helpful to the patient and of things concerning only him. The peculiar customs of the patient's household shall not be made public. Even knowing that the patient's span of life has come to its close, it shall not be mentioned by thee there, where if so done it would cause shock to the patient or to others. Though possessed of knowledge one should not boast very much of one's knowledge. Most people are offended by the boastfulness of even those who are otherwise good and authoritative.*

9. *There is no limit at all to the Science of Life, Medicine. So thou shouldst apply thyself to it with diligence. This is how thou shouldst act. Also thou shouldst learn the skill of practice from another without carping. The entire world is the teacher to the intelligent and the foe to the unintelligent. Hence, knowing this well, thou shouldst listen and act according to the words of instruction of even an unfriendly person, when his words are worthy and of a kind as to bring to you fame, long life, strength and prosperity.*

10. *Thereafter the teacher should say this-'Thou shouldst conduct thyself properly with the gods, sacred fire, Brahmanas, the guru, the aged, the scholars and the preceptors. If thou hast conducted thyself well with them, the precious stones, the grains and the gods become well disposed towards thee. If thou shouldst conduct thyself otherwise, they become unfavourable to thee'. To the teacher that has spoken thus, the disciple should say, 'Amen.'*

## [3] Oración de Moisés Maimónides (c. 1200)

*Dios todopoderoso, Tú creaste al ser humano con infinita sabiduría (…), Tú bendijiste Tu tierra, Tus ríos y Tus montañas con sustancias curativas. Ellas permiten a tus criaturas aliviar sus sufrimientos y sanar sus enfermedades. Tú has dotado al hombre de sabiduría para mitigar los padecimientos de su hermano, reconocer sus trastornos, extraer sustancias curativas, descubrir sus poderes y prepararlas y aplicarlas para combatir cualquier dolencia. En Tu eterna providencia, me has elegido para velar por la vida y la salud de Tus criaturas. Estoy ahora por iniciar los deberes de mi profesión. Socórreme, Dios Todopoderoso, en esta gran tarea que puede beneficiar a la humanidad, en la cual sin Tu ayuda no podré lograr ni el más mínimo éxito.*

*Inspírame amor por mi arte y por Tus criaturas. No permitas que el afán de lucro y la ambición de alcanzar fama y admiración perturben las labores de mi profesión, ya que son enemigos de la verdad y del amor a la humanidad y puede desviarme de la gran tarea de velar por el bienestar de Tus criaturas. Conserva las fuerzas de mi cuerpo y alma a fin que estén siempre gustosamente dispuestos a ayudar y apoyar a ricos y a pobres, buenos y malos, amigos y enemigos. Haz que en los enfermos sólo vea seres humanos. Ilumina mi inteligencia para que pueda reconocer lo que es evidente y para que pueda entender lo que está ausente u oculto (…).*

*Si los que son más sabios que yo quieren perfeccionarme y educarme, permite que mi alma siga su orientación con reconocimiento (…)*

*Induce dulzura y tranquilidad a mi alma (…).*

*Permíteme estar satisfecho con todo, salvo con mi dominio de la gran ciencia de mi profesión. Nunca permitas que llegue a pensar que he alcanzado un grado de saber suficiente, pero concédeme la fuerza, la posibilidad y la ambición de ampliar siempre mis conocimientos. Pues el arte es grande, pero la inteligencia del hombre no tiene límites.*

*¡Dios Todopoderoso! Me has elegido en Tu misericordia para velar por la vida y la muerte de Tus criaturas. Ahora voy a consagrarme a mi profesión. Socórreme en esta gran tarea a fin de que*

*pueda ser provechosa para la humanidad, pues sin Tu ayuda no podré lograr ni el más mínimo éxito.*

**[4]    Juramento de galeno para Farmacéuticos (c. 1627)**

*Le Serment Des Apothicaires chrétiens et craignant Dieu.*

*Je jure et promets devant Dieu, Auteur et Créateur de toutes choses, unique en essence et distingué en trois Personnes éternellement bienheureuses, que j'observerai de point en point tous ces articles suivants.*

*Et premièrement je jure et promets de vivre et mourir en la foi chrétienne.*

*Item d'aimer et d'honorer mes parents le mieux qu'il me sera possible.*

*Item d'honorer, respecter et faire service, en tant qu'en moi sera, non seulement aux Docteurs, Médecins qui m'auront instruit en la connaissance des préceptes de la Pharmacie, mais aussi à mes Précepteurs et Maîtres-Pharmaciens sous lesquels j'aurai appris mon métier.*

*Item de ne médire d'aucun de mes Anciens Docteurs, Maîtres-Pharmaciens ou autres, quels qu'ils soient.*

*Item de rapporter tout ce qui me sera possible pour l'honneur, la gloire, l'ornement et la majesté de la Médecine.*

*Item de n'enseigner point aux idiots et ingrats les secrets et raretés d'icelle.*

*Item de ne faire rien témérairement sans avis de Médecin, ou sous espérance de lucre tant seulement.*

*Item de ne donner aucun médicament purgatif aux malades affligés de quelque maladie aiguë, que premièrement je n'aie pris conseil de quelque docte Médecin.*

*Item de ne toucher aucunement aux parties honteuses et défendues des femmes, que ce ne soit par grande nécessité, c'est-à-dire lorsqu'il sera question d'appliquer dessus quelque remède.*

*Item de ne découvrir à personne les secrets qu'on m'aura fidèlement commis.*

*Item de ne donner jamais à boire aucune sorte de poison à personne et ne conseiller jamais à aucun d'en donner, non pas même à ses plus grands ennemis.*

*Item de ne donner jamais à boire aucune potion abortive.*

*Item de n'essayer jamais de faire sortir le fruit hors du ventre de sa mère, en quelque façon que ce soit, que ce ne soit par avis du Médecin.*

*Item d'exécuter de point en point les ordonnances des Médecins sans y ajouter ou diminuer, en tant qu'elles seront faites selon l'Art.*

*Item de ne me servir jamais d'aucun succédané ou substitut sans le conseil de quelqu'autre plus sage que moi.*

*Item de désavouer et fuir comme la peste la façon de pratiquer scandaleuse et totalement pernicieuse, de laquelle se servent aujourd'hui les charlatans empiriques et souffleurs d'alchimie, à la grande honte des Magistrats qui les tolèrent.*

*Item de donner aide et secours indifféremment à tous ceux qui m'emploieront.*

*Et finalement de ne tenir aucune mauvaise et vieille drogue dans ma boutique.*

*Le Seigneur me bénisse toujours, tant que j'observerai ces choses.*

**Versión Moderna (c. siglo XX)**

*Je jure, en présence des maîtres de la faculté, des conseillers de l'ordre des pharmaciens et de mes condisciples :*

*D'honorer ceux qui m'ont instruit dans les préceptes de mon art et de leur témoigner ma reconnaissance en restant fidèle à leur enseignement ;*

*D'exercer, dans l'intérêt de la santé publique, ma profession avec conscience et de respecter non seulement la législation en vigueur, mais aussi les règles de l'honneur, de la probité et du désintéressement ;*

*De ne jamais oublier ma responsabilité et mes devoirs envers le malade et sa dignité humaine.*

*En aucun cas, je ne consentirai à utiliser mes connaissances et mon état pour corrompre les m'urs et favoriser des actes criminels.*

*Que les hommes m'accordent leur estime si je suis fidèle à mes promesses. Que je sois couvert d'opprobre et méprisé de mes confrères si j'y manque.*

**[5]   Cita del Juramento Hipocrático de los científicos de "La Nueva Atlántida" (c.1627)**

*...We have consultations, which of the inventions and experiences which we have discovered shall be published, and which not; and take all an oath of secrecy for the concealing of those which we think fit to keep secret; though some of those we do reveal sometime to the State, and some not.*

**[6]   Código de la Sociedad Médica del Estado de Nueva York (c.1807)**

*I do solemnly declare, that I will honestly, virtuously and chastely conduct myself in the practice of physic and surgery, with the privileges of exercising which profession I am now to be invested; and that I will, with fidelity and honor, do everything in my power for the benefit of the sick committed to my charge.*

**[7]   Juramento de los Ingenieros de Canadá (c.1900)**

*The Ritual of the Calling of an Engineer: Obligation. I, ____________, in the presence of these my betters and my equals in my calling, bind myself upon my honor and Cold Iron, that, to the best of my knowledge and power, I will henceforward suffer or pass, or be privy to the passing of, Bad Workmanship or Faulty Material in aught that concerns my work before men as an Engineer, or in my dealings with my own soul before my Maker.*

*MY TIME I will not refuse; MY THOUGHT I will not grudge; MY CARE I will not deny towards the honor, use, stability and perfection of any works to which I may be called to set my hand. MY FAIR WAGES for that work I will openly take.*

*MY REPUTATION in my calling I will honorably guard; but I will in no way go about to wrest judgment or gratification from anyone with whom I may deal.*

*And further, I will early and warily strive my uttermost against professional jealousy or the belittling of my working bothers in any fields of their labor.*

*For my assured failures and derelictions, I ask pardon beforehand of my betters and my equals in my Calling here assembled; praying that in the hour of my temptations, weakness and weariness, the memory of this, my Obligation and of the company before whom it was entered into, may return to me to aid, comfort and restrain.*

*Camp... To all these things you have subscribed upon Honor and Cold Iron God helping me, by these things I purpose to abide.*

**[8]**   En 1917 la  Asociación de Trabajadores Científicos del Reino Unido (entidad fundadora de la Federación Mundial de Trabajadores Científicos) intentó promover la difusión de un Juramento Hipocrático para científicos, aunque luego desistió por encontrar su implementación no práctica. Citado por S. Davinson, *Scientific World*, vol. 34 (3): 7, 1990.

**[9]**  Michael Sullivan de la Clase 1929 de la Universidad de Michigan propuso el siguiente texto de compromiso de fe del ingeniero que luego fue adoptado por el *Engineering Council for Professional Development* de los EEUU y luego por ABET *(Accrediting Board for Engineering and Technology).*

**Faith of the Engineer (1929)**

*I am an Engineer. In my profession I take deep pride but without vainglory; to it I owe solemn obligations that I am eager to fulfill.*

*As an engineer I will participate in none but honest enterprise to him that has engaged my services, as employer or client, I will agree the utmost of performance and fidelity.*

*When needed, my skills and knowledge shall be given without reservation for the public good. From special capacity springs the obligation to use it well in the service of humanity; and I accept the challenge that this implies.*

*Jealous of the high repute of my calling, I will strive to protect true interests and the good name of any engineer that I know to be deserving; but I will not shrink should duty dictate, from disclosing the truth regarding anyone that, by unscrupulous act, has shown himself unworthy of the profession.*

*Since the age of the stone, human progress has been conditioned by the genius of my professional forbearers. By them have been rendered usable to mankind nature's vast resources of material and energy. By them have been vitalized and turned to practical account the principles of science and the revelations of technology. Except for this heritage of accumulated experience, my efforts would be feeble. I dedicate myself to the dissemination of engineering knowledge, and especially, to the instruction of younger members of my profession in all its arts and traditions.*

*To my fellows I pledge in the same full measure I ask of them in integrity and fair dealing tolerance and respect and devotion to the standards and the dignity of our professionals with the consciousness always that our special expertness carries with the obligation to serve humanity with complete sincerity.*

**[10]  Juramento Hipocrático para la Era Nuclear (1945)**

*I pledge that I will use my knowledge for the good of humanity and against the destructive forces of the world and the ruthless intent of man; and that I will work together with my fellow scientists of whatever nation, creed, or color, for these, our common ends.*

**[11]  Carta de los Trabajadores Científicos (1948)**

*The profession of science, due to the special importance of the consequences of its good or bad use, carries with it special responsibilities over and above those of the ordinary duties of citizenship. In particular, the scientific worker, because he has or can easily acquire knowledge inaccessible to the public, must do his utmost to ensure that that knowledge is employed for good.*

**[12]  Declaración de Ginebra (1948)**

*At the time of being admitted as a Member of the medical profession, I solemnly pledge myself to consecrate my life to the service of humanity; I will give to my teachers the respect and gratitude which is their due; I will practice my profession with conscience and dignity; The health and life of my patient will be my first consideration; I will respect the secrets which are confided in me; I will maintain by all means in my power, the honor and the noble traditions of the medical profession; My colleagues will be my brothers; I will not permit considerations of religion, nationality, race, party politics or social standing to intervene between my duty and my*

*patient; I will maintain the utmost respect for human life, from the time of its conception, even under threat, I will not use my medical knowledge contrary to the laws of humanity; I make these promises solemnly, freely and upon my honor.*

**[13]** Credo de los Ingenieros de Verein Deutsher Ingenieure (1950) aparece citado en E. S. Ferguson, "Unassuaged Alarms," *Science*, vol. 182: 815-816, 1973.

**[14] Juramento para Estadísticos (1952)**

*Like Euclid and all the other great thinkers who have used symbols to reveal the truths of nature, I will be a seeker of truth.*

*Realizing that numbers are only a shorthand convention for describing past events and fore-casting trends, I will search for those fats expressed in numbers which show relationships and events most truly.*

*Though surrounded by clamor of the marketplace or of the political arena, I will not be a fraud, who selects figures to prove chicanery and misnamed conclusion.*

**[15] Credo de los ingenieros adoptado por la *National Society of Professional Engineers* (EEUU), Junio de 1954.**

*As a Professional Engineer, I dedicate my professional knowledge and skill to the advancement and betterment of human welfare*

*I pledge:*

*To give the utmost of performance;*

*To participate in none but honest enterprise;*

*To live and work according to the laws of man and the highest standards of professional conduct;*

*To place service before profit, the honor and standing of the profession before personal advantage, and the public welfare above all other considerations.*

*In humility and with the need for Divine Guidance, I make this pledge.*

**[16] Juramento de los Sicólogos (1954)**

1. *As a scientist my highest value is truth. I am obliged constantly to search for the truth, and to say what I believe to be the truth whenever I have the occasion.*

2. *As an intellectual I believe that the greatest human weal may be attained by using knowledge and understanding.*

3. *As a teacher I have to place foremost the wise development of my students, not making their growth subservient to my ends.*

4. *As a member of a profession I am bound by certain rules of conduct on which the majority agree. I accept the fact that to violate these rules may injure the group to which I belong, and this I should not knowingly do.*

5. *As a citizen I have certain work to do and responsibilities for the common good, and when called upon I shall not evade my duty to state, country, and mankind.*

**[17] Juramento a la Bandera del Académico Americano (1954).**

Edward J. Shoben (1954), de la Escuela de Maestros de la Universidad de Columbia, en una carta a la revista *The American Psychologist*, respondiendo a la propuesta de Juramento de los Sicólogos propone el siguiente juramento para terminar con las propuestas de juramentos, que es un contraejemplo del resto de las propuestas listadas.

*As an American scholar, I pledge allegiance:*

*To the United States, the country that has given me opportunities for life, livelihood, and the pursuit of truth;*

*To the U. S. Constitution, embodying as it does the principles of democratic government, and in accordance with our history, to the legal and orderly means of amending it as changing times and circumstances may require for the maximum satisfaction and safety of the nation;*

*To the principle of the free worship of God by all citizens in their several ways so long as the chosen devotional forms of some do not interfere with the same inalienable religious rights of others;*

*To the belief in the ultimate primacy of reason and therefore to the principle of unrestricted public competition among ideas on the ground that "error of opinion may be tolerated where reason is left free to combat it";*

*To the untrammeled search for knowledge and understanding in classrooms, libraries, and laboratories everywhere in the United States;*

*To the conviction that scholarly fitness is best determined only by other scholars rather than by those who merely disagree with a scholar's uttered or published opinions; and*

*To the role of the scholar as a responsible citizen who is ready to serve the United States in any capacity in which he may be useful to it, whose possible criticism of American institutions is based upon his love for the fundamental American ways of life, and whose actions are always based on his most sincere interpretation of what constitutes the demands of loyalty to his nation and its posterity.*

*As an American scholar, I make this pledge solemnly and without reservation upon my sacred honor.*

**[18]** Fred W. Decker (1957) propone, en la prestigiosa revista *Science,* la necesidad de que los científicos cultiven los aspectos éticos dentro de sus actividades y destaca la necesidad de que se implemente un Juramento similar al Hipocrático en donde se defina claramente las obligaciones del científico para con la sociedad.

**[19]** Juramento de Graduación de la Universidad de Zagreb. Supek y Malecki (1982: 183) mencionan la existencia de un juramento de graduación de la Universidad de Zagreb (Croacia) en donde los candidatos a doctor en ciencias se comprometen a *"adherir siempre a la verdad científica y trabajar para el beneficio de toda la humanidad"*.

**[20] Juramento Hipocrático -Propuesta de actualización en la revista *Science* (1966)**

*I will share the science and art by precept, by demonstration, and by every mode of teaching with other physicians regardless of their national origin. I will try to help secure for the physicians in each country the esteem of their own people, and in collaborative work see that they get full credit.*

*I will strive to eliminate sources of disease everywhere in the world and not merely set up barriers to the spread of disease to my own people.*

*I will work for understanding of the diverse causes of disease, including the social, economic, and environmental. I will promote the well-being of mankind in all its aspects, not merely the bodily, with sympathy and consideration for- a people's culture and beliefs.*

*I will strive to prevent painful and untimely death, and also to help parents to achieve a family size conforming to their desires and to their ability to care for their children. In my concern with whole communities I will never forget the needs of its individual members.*

**[21]  Juramento Hipocrático para Científicos (1967)**

*I will not use my scientific training for any purpose which to my knowledge is intended to harm human beings.*

**[22]**  El destacado naturalista y educador británico Eric Ashby (1968) propuso la implementación de un compromiso o Juramento Hipocrático para todos aquellos que desempeñen una vida académica de educación e investigación.

**[23]**  Karl Popper durante una sesión especial de "Ciencia y Ética" realizada durante el Congreso Internacional de Filosofía desarrollado en Viena en 1968, propuso la necesidad de que exista un Juramento Hipocrático para científicos. Una versión editada y actualizada de esa propuesta fue publicada respectivamente en inglés y alemán en Popper (1971, 1975).

**[24]**  J. Smittenberg (1969) reconoce que tanto la ciencia básica como la aplicada desencadena una serie de dilemas éticos para los cuales sería necesario implementar un Juramento Hipocrático análogo para ser empleado por los científicos naturales.

**[25]**  Un grupo de jóvenes estudiantes de la Universidad Agrícola de Wageningen (Países Bajos) un Juramento Ético para Fitopatólogos. Citado por M. Jeuken (1969).

**[26]  Juramento para Ingenieros propuesto en la reunión de Bratislava de la WFSW (1969)**

*I vow to strive to apply my professional skills only on projects which, after conscientious examination, I believe to contribute to the goal of coexistence of all human beings in peace, human dignity and self-fulfillment. I believe that the goal requires the provision of an adequate supply of the necessities of life (good food, air, water, clothing and housing, access to natural and man-made beauty), education and opportunities to enable each person to work out for himself, his life objectives and to develop creativeness and skill in the use of the hands as well as the head. I vow to struggle through my work the minimize danger, noise, strain or invasion of privacy of the individual, pollution of Earth, air or water, destruction of natural beauty and resources and of wildlife.*

**[27]**  El destacado biólogo neerlandés J. J. Groen (1970) propuso un Juramento para las Ciencias Naturales, Sociales y Humanidades, aunque esta versión no hace mención explícita a las consecuencias negativas del mal uso del conocimiento.

**[28]  Juramento en las Universidades de Berkeley y Stanford (1970)**

*The purpose of science should be the general enhancement of life and not the causing of harm to man. I affirm that I will uphold this principle, in teaching and in practice of my science, to the best of my ability and judgment.*

**[29] Juramento para Científicos Naturales (1970)**

*Being admitted to the practice of the natural sciences I pledge to put my knowledge completely at the service of mankind. I shall prosecute my profession conscientiously and with dignity. I shall never collaborate in research aimed at the unjustified extermination of living organisms or the disturbance of the biological equilibrium which is harmful to mankind, neither shall I support such research in any way.*

*Guide of my scientific work will be the promotion of the common welfare of mankind and in this context I shall not kill organisms nor shall I allow the killing of organisms for inferior, short-sighted, opportunistic reason.*

*I accept responsibility for unforeseen, harmful results directly originating from my work; I shall undo these results as far as lies in my power.*

*This I vow voluntarily and on my word of honour.*

**[30] Enmienda de la Sociedad Americana de Física (1971)**

*Article II: The object of the Society shall be the advancement and diffusion of the knowledge of physics in order to increase man's understanding of nature and to contribute to the enhancement of the quality of life for all people. The Society shall assist its members in the pursuit of these humane goals and it shall shun those activities which are judged to contribute harmfully to the welfare of mankind.*

**[31] Cláusula Contractual de Responsabilidad Social y Medio Ambiente (1971)**

Durante la conferencia anual de la *International Society for Social Responsibility*, realizada en Trondheim, Noruega, en agosto de 1971, se propuso la redacción de un proyecto de resolución para ser presentado durante la Conferencia de las Naciones Unidas del Medio Ambiente (Estocolmo, 1972). El punto 2 establecía lo siguiente:

*That all scientists and engineers be offered a pledge to sign which expresses the fundamental responsibility of the professional man for all future consequences of his work, direct or indirect.*

*...When they sign a contract a contract of employment it should contain a clause with the following sense:*

*"To honor my pledge to abide by the code of ethics of the scientific community, mu acceptance of this contract is based in the understanding that I be given freedom to disclose the results of my works where this becomes necessary to comply with the spirit of my pledge."*

**[32] Juramento de Pugwash – Conferencia Anual celebrada en Oxford (1972)**

*...I will not use my scientific training for any purpose which I believe is intended to harm human beings. I shall strive for peace, justice, freedom and the betterment of the human conditions...*

**[33]** P. Sonntag (1973) hace un interesante planteo desde las ciencias exactas y técnicas sugiriendo la aplicación de un análogo al Juramento Hipocrático para Científicos y Técnicos para reducir los riesgos sociales de la actividad de investigación y desarrollo.

**[34] Juramento del Ingeniero (1973)**

*I solemnly pledge myself to consecrate my life to the service of humanity. I will give to my teachers the respect and gratitude which is their due; I will be loyal to the profession of engi-*

*neering and just and generous to its members; I will lead my life and practice my profession in uprightness and honor; whatever project I shall undertake, it shall be for the good of mankind to the utmost power; I will keep far aloof from wrong, from corruption, and from tempting others to vicious practice; I will exercise my profession solely for the benefit of humanity and perform no act for a criminal purpose, even if solicited, far less suggest it; I will speak out against evil and unjust practice where so ever I encounter it; I will not permit considerations of religion, nationality, race, party politics, or social standing to intervene between my duty and my work; even under threat, I will not use my professional knowledge contrary to the laws of humanity; I will endeavor to avoid waste and the consumption of nonrenewable resources. I make these promises solemnly, freely, and upon my honor.*

**[35]** En 1974, Philip Smith introduce el texto del Juramento de Pugwash **[32]** en las ceremonias de graduación en la Universidad de Groningen.

**[36]** W. Luck (1975) retoma las discusiones de los años 1962 y 1963 sobre la aplicación del Juramento Hipocrático para Científicos de las que él había formado parte cuando fundó la Sociedad de Responsabilidad en Ciencia. En este artículo llama la atención del renacimiento de la idea del Juramento Hipocrático para Científicos y discute los alcances de las propuestas de Charles Schwartz y Anatol Rapoport en las reuniones de la Sociedad Americana de Física en 1971.

**[37]** M. Beech (1978) en un ensayo publicado en una revista de educación en física, propone la necesidad de que los físicos y los científicos en general asuman públicamente su responsabilidad, a través de un Juramento tipo Hipocrático, con la raza humana de la misma manera que los doctores la asumen con sus pacientes. No propone ninguna fórmula específica.

**[38] Juramento Hipocrático para Académicos (1981)**

1. *I undertake, throughout my academic career, regardless of any position I hold, to base all judgements of others on an objective analysis of the available facts. When called upon to do so, I shall cite the exact information upon which my judgements are based.*

2. *My judgements of others shall, wherever possible, be made publicly. When anonymity is essential, I shall provide arguments worthy of public scrutiny.*

3. *I shall endeavour always to distinguish between my judgements on academic grounds and any personal interests. I shall always declare the latter, even if they require my withdrawal from a particular decision.*

4. *I shall at all times avoid in word and deed any form, of discrimination against others in race, religion, ethnic background, sex, marital status, age, political affiliation, nationality, and physical condition. I shall encourage my students and colleagues to do likewise in my presence.*

5. *All requests for judgement of the work of others will be treated by me as a matter for the utmost priority. If, for any reason beyond my control, I am unable to give them urgent attention I shall request to be relieved of my responsibility.*

6. *I recognise as an academic a basic obligation to use my intellect and training by teaching or writing for the illumination of the community, following the truth fearlessly wherever it may lead, regardless of vested interests. I accept the intellectual requirements of my students as a prior claim on my time, taking precedence over all other activities. I undertake the continual re-evaluation of my teaching techniques to ensure that students obtain the best education available.*

[39] Daniel E. Harris (1982) introduce el concepto de la necesidad de un Juramento Hipocrático para científicos que sirva de compromiso moral con la sociedad.

[40] **Juramento Hipocrático de la Era Nuclear (1983)**

*As a physician of the 20th century, I recognize that nuclear weapons have presented my profession with a challenge of unprecedented proportions, and that a nuclear war would be the final epidemic for the humankind. I will do all in my power to work for the prevention of a nuclear war...*

[41] **Juramento de Sarah (1983), propuesto por High Technology Professionals for Peace (HTPFP) de Cambridge, MA, EEUU.**

*The work I do shall be for the benefit of the world according to my ability and my judgment, and not for the harm to peoples of the world or for any wrong. I will make no deadly weapon, though it be asked of me, nor will I council such, and especially I will not aid in the development of nuclear weapons. Whatsoever project I undertake, I will work for the benefit of people, refraining from all wrong-doing or corruption, and especially from the developing technology which would be used for the detriment of people.*

[42] En mayo de 1984 los profesores, graduados y estudiantes del Departamento de Física de la Universidad de California en Berkeley organizaron un simposio sobre las conexiones entre la "física y los militares" (ver Schwartz, 1984). Luego, los estudiantes comenzaron a hacer campaña para formalizar una ceremonia oficial de compromiso de graduación voluntario en donde se negarían a usar sus conocimientos con fines militares y/o aceptar trabajos vinculados a estos temas. Luego de realizar una consulta que mostró una abrumadora mayoría de estudiantes a favor de tomar este tipo de compromiso, el Director del Departamento de Física, anuló la consulta y rechazó la idea de implementar ese Compromiso de Graduación, argumentando que los padres de varios estudiantes se habían mostrado irritados con este asunto.

[43] **Juramento en contra de las armas nucleares (1984)**

*I pledge that will never be involve in any way with the development or production of nuclear arms*

[44] **Código de Ética de Uppsala (1984)**

*Scientific research is an indispensable activity of great significance to mankind - for our description and understanding of the world, our material conditions, social life and welfare. Research can contribute to solving the great problems facing humanity, such as the threat of nuclear war, damage to the environment, and the uneven distribution of the Earth's resources. In addition, scientific research is justified and valuable as a pure quest for knowledge, and it should be pursued in a free exchange of methods and findings. Yet research can also, both directly and indirectly, aggravate the problems of mankind.*

*This code of ethics for scientists has been formulated as a response to a concern about the applications and consequences of scientific research. In particular it appears that the potential hazards deriving from modern technological warfare are so overwhelming that it is doubtful whether it is ethically defensible for scientists to lend any support to weapons development.*

*The code is intended for the individual scientist; it is primarily he or she who shall assess the consequences of his/her own research. Such an assessment is always difficult to make, and may not infrequently be impossible. Scientists do not as a rule have control over either research*

*results or their application or even in many cases over the planning of their work. Nevertheless this must not prevent the individual scientist from making a sincere attempt to continually judge the possible consequences of his/her research, to make these judgments known, and to refrain from such research as he/she deems to be unethical.*

*In this connection the following should particularly be considered:*

1. *Research shall be so directed that its applications and other consequences do not cause significant ecological damage.*

2. *Research shall be so directed that its consequences do not render it more difficult for present and future generations to lead a secure existence. Scientific efforts shall therefore not aim at applications or skills for use in war or oppression. Nor shall research be so directed that its consequences conflict with basic human rights as expressed in international agreements on civic, political, economic, social and cultural rights.*

3. *The scientist has a special responsibility to assess carefully the consequences of his/her research, and to make them public.*

4. *Scientists who form the judgment that the research which they are conducting or participating in is in conflict with this code, shall discontinue such research, and publicly state the reasons for their judgment. Such judgments shall take into consideration both the probability and the gravity of the negative consequences involved.*

*It is of urgent importance that the scientific community support colleagues who find themselves forced to discontinue their research for the reasons given in this code.*

**[45]** En 1984, el Grupo de Investigación Wittenberg, por entonces en la República Democrática Alemana (DDR), un grupo liderado por Hans Peter Gensichen, publicó, en forma independiente, un Código de Ética muy similar al de la Universidad de Uppsala. (H.P. Gensichen, Wissenschaftethik heute, KF19-84, Für innerkirchlichen Gebrauch). Una versión en sueco de este código se la puede también encontrar en Rydén (1990).

**[46]** **Compromiso sobre la Iniciativa de Defensa Estratégica (1985)**

*The pledge of non-participation*

*We, the undersigned science and engineering faculty, believe that the Strategic Defense Initiative (SDI) program (commonly known as Star Wars) is ill-conceived and dangerous.*

*Anti-ballistic missile defense of sufficient reliability to defend the population of the United States against a Soviet attack is not technically feasible. A system of more limited capability will only serve to escalate the arms race by encouraging the development of both additional offensive overkill and an all-out competition in anti ballistic missile weapons. The program will jeopardize existing arms control agreements and make arms control negotiation even more difficult than it is at present. The program is a step toward the type of weapons and strategy likely to trigger a nuclear holocaust.*

*For these reasons, we believe that the SDI program represents, not an advance toward genuine safety, but rather a major step backwards. The likelihood that SDI funding will restrict academic freedom and blur the distinction between classified and unclassified research is greater than for other sources of funding. The structure of SDI research programs makes it likely that groups doing only unclassified research will be part of a Research Consortium and will therefore work closely with other universities and industries doing classified research. SDI officials openly concede that any successful unclassified project may become classified. Moreover, the potentially sensitive nature of the research may invoke legal restrictions required by the Export Administration Act. Participation in SDI by individual researchers would lend their institution's name to*

*a program of dubious scientific validity, and give legitimacy to this program at a time when the involvement of prestigious research institutions is being sought to increase congressional support. Researchers who oppose the SDI program yet choose to participate in it should therefore recognize that their participation would contribute to the political acceptance of SDI.*

*Accordingly, as working scientists and engineers, we pledge neither to solicit nor accept SDI funds, and encourage others to join us in this refusal. We hope together to persuade the public and Congress not to support this deeply misguided and dangerous program.*

[47] Entre el 20 y el 24 de enero de 1986, en la ciudad de Atenas, organizada por la UNESCO y la Comisión Nacional de Apoyo a la UNESCO de Grecia, se realizó una consulta internacional sobre la manera de mejorar la educación superior en materia de paz, y el impacto social de la ciencia y la tecnología. Allí, Marie Françoise Farge, consideró que la mayor parte de la tecnología desarrollada entonces era una consecuencia de los procesos de investigación y desarrollo militares, no pudiendo hacerse una clara distinción entre la investigación civil y militar. Aseguró que la calidad e impacto de la investigación dependía de la eficiencia de la difusión de sus resultados y en este aspecto el secreto de la investigación militar conspiraba en contra. Por esta razón propuso establecer en las universidades y los establecimientos de educación superior un juramento para los jóvenes científicos e ingenieros que se gradúan recordándoles las consecuencias éticas y humanas de las investigaciones científicas y de realizaciones tecnológicas.

[48] En el mes de septiembre de 1986, E.S. Santini y G.A. Lemarchand, estudiantes de física pertenecientes a la Comisión de Astrofísica del Centro de Estudiantes de Ciencias Exactas y Naturales (CE-CEN) de la Facultad de Ciencias Exactas y Naturales de la Universidad de Buenos Aires, presentaron un proyecto de Juramento Hipocrático para Científicos dentro de la Reunión Anual de la Asociación de Estudiantes de Física de la República Argentina (AEFA) realizada en la ciudad de Tucumán (Argentina). La propuesta estaba endosada por A.R. Pedace, Secretario de Extensión Universitaria de la FCEN-UBA y estaba acompañada por un artículo escrito por el matemático Mischa Cotlar. Este proyecto que fue aprobado por la asamblea correspondiente, fue el que se presentó en 1988 durante el *Simposio Internacional sobre los Científicos, la Paz y el Desarme,* y que se transformó luego en el llamado "Juramento de Buenos Aires". A los pocos meses, el grupo de estudiantes de la Comisión de Astrofísica distribuyó, entre más de 2000 científicos pertenecientes al Consejo Internacional de la Ciencia (ICSU) un documento con 6 puntos que describían los contenidos a ser incluidos en un Juramento Hipocrático para Científicos.

[49] **Endoso de Ciencia para la Paz (1987)**

*It is the author's wish that no agency should ever derive military benefit from the publication of this paper. Authors who cite this work in support of their own are requested to qualify similarly the availability of their results.*

[50] **Compromiso del Comité para la Genética Responsable (1987)**

*We the undersigned, biologists and chemists, oppose the use of our research for military purposes. Rapid advances in biotechnology have catalyzed a growing interest by the military in many countries in chemical and biological weapons and in the possible development of a new and novel chemical and biological warfare agents. We are concerned that this may lead to another arms race. We believe that biomedical research should support rather than threaten life. Therefore, WE PLEDGE, not to engage knowingly in research and teaching that will further the development of chemical and biological warfare agents.*

99

**[51] Juramento de la Universidad de Humboldt (1987)**

*I pledge to thoroughly investigate and take into account the social and environmental conse-
quences of any job or opportunity I consider.*

**[52]** En 1987, los miembros del Capítulo Argentino del Grupo Internacional de Médicos para la Preven-
ción de la Guerra Nuclear (IPPNW), liderados por Emanuel Levin propusieron en su reunión anual
la necesidad de desarrollar una fórmula de Juramento Hipocrático para Científicos a los fines de ser
utilizada en las ceremonias de graduación.

**[53] Juramento Hipocrático para Científicos, Tecnólogos y Ejecutivos (1987)**

*I vow to practice my profession with conscience and dignity; I will strive to apply my skills only
with the utmost respect for the well-being of humanity, the earth and all its species; I will not
permit considerations of nationality, politics, prejudice or material advancement to intervene
between my work and this duty to present and future generations; I make this Oath solemnly,
freely and upon my honor.*

**[54] Juramento Hipocrático para Científicos de la *Nuclear Age Peace Foundation* (1987)**

*As a scientist, I am a seeker of truth and explorer of our universe; recognizing and affirming the
responsibilities which accompany the privilege of my training, I pledge:*

- *To use my intellect and employ my skills for the benefit of life, placing the humanity above
  all nations;*
- *To limit my work to socially and environmentally constructive ends;*
- *Never to use my intellect or employ my skills for the development of weapons of mass
  destruction.*
- *I commit myself, without reservation, to the obligation of this oath.*

**[55] Juramento Hipocrático para Ingenieros de la Nuclear Age Peace Foundation (1987)**

*As an engineer, I am a builder of bridges to the future; recognizing and affirming the responsi-
bilities which accompany the privilege of my training, I pledge:*

- *To use my intellect and employ my skills for the benefit of life, placing the humanity above
  all nations;*
- *To limit my work to socially and environmentally constructive ends;*
- *Never to use my intellect or employ my skills for the development of weapons of mass
  destruction.*
- *I commit myself, without reservation, to the obligation of this oath.*

**[56]** Uno de los fundadores, el sociólogo Johan Galtung (1988), en su discurso durante una ceremonia de
graduación en la Universidad de Hawaii el 20 de diciembre de 1987, habló de los problemas de la
carrera armamentista, del número de investigadores científicos que trabajan en proyectos militares
en el mundo, del desbalance que existe con los investigadores científicos que trabajan en temas
de paz, sobre la necesidad de implantar estudios de posgrado en temas de paz y la posibilidad que
tienen las ceremonias de graduación para asumir un compromiso ético para usar sus conocimientos
solo a favor de la paz. En ese caso sugirió el juramento propuesto por la Nuclear Age Peace Founda-
tion [54].

**[57]** **Juramento de Buenos Aires, versión en inglés (1988).** La versión en castellano está en el texto principal, la versión en inglés (ligeramente distinta) reza así:

*Aware that, in the absence of ethical control, science and its products can damage society and its future, I pledge that my own scientific capabilities will never be employed merely for remuneration or prestige or on instruction of employers or political leaders only but solely on my personal belief and social responsibility – based on my own knowledge and on consideration of the circumstances and the possible consequences of my work – that the scientific or technical research I undertake is truly in the best interest of society and peace.*

**[58]** Chandler Davis, destacado matemático canadiense, propagó la idea de los Juramentos Hipocráticos para Científicos y en particular matemáticos en una serie de escritos y conferencias internacionales (ver Davis, 1988, 1990).

**[59]** **Juramento para Ciudadano de la Tierra de la Nuclear Age Peace Foundation (1988)**

*Aware of the vastness of the universe and the uniqueness of life, I accept and affirm my responsibility as an Earth citizen to nurture and care for our planet as a peaceful, harmonious home where life may flourish. Believing that each of us can make a difference, I pledge to persevere in Waging Peace. With my spirit, intellect and energy I shall strive to: Reverse the nuclear arms race, and this omnicidal threat to the continuation of life; Redirect scientific and economic resources from the destructive pursuit of weapon technologies to the beneficial tasks of ending hunger, disease and poverty; Break down barriers between people and nations, and preserving the natural beauty and profound elegance of our land, mountains, oceans and sky; And, teach others, by my words and deeds, to accept all members of the human family, and to love the Earth and live with dignity and justice upon it.*

**[60]** André Baccard (1989) convocó a las sociedades científicas y de ingeniería a implementar un Juramento Hipocrático que sea aplicable a químicos orgánicos, biólogos, físicos e ingenieros.

**[61]** Arnold J. Toynbee (1989) propone la necesidad de un Juramento Hipocrático para profesionales:

*"En la era de la civilización tecnológica, la educación en la forma correcta de vivir tiene que ser completada con una formación profesional en las distintas ramas especiales del conocimiento y en el desarrollo de diversas habilidades. Pero antes de entrar en su profesión, todos los que han recibido una formación profesional deberían tomar un Juramento Hipocrático similar al que se prescribe para aquellos que entran a la profesión médica. Cualquiera sea la profesión, todas las personas deberían comprometerse a utilizar sus conocimientos especiales y sus habilidades para servir a sus semejantes y no para explotarlos. Deberían priorizar sus obligaciones de servicio por sobre sus necesidades incidentales de ganarse la vida para sí y su familia. Maximizar el servicio y no el maximizar el beneficio, debería ser el objetivo al cual deberían dedicar su vida."*

**[62]** En 1990, durante la 40ª Conferencia de Pugwash, celebrada en Egham, Reino Unido, se organizó un Grupo de Trabajo (WG 8) para tratar el tema de los Juramentos Hipocráticos para Científicos y del impacto que tiene el trabajo de los científicos en el desarrollo de armamento nuclear, químico y biológico. Entre sus participantes se encontraban Meredith Thring (autor del Juramento de Bratislava), Lars Rydén (coautor del Código de Ética de Uppsala), Guillermo A. Lemarchand (co-autor del Juramento de Buenos Aires), Sergei Kapitza, Martin Rees, Maurice Wilkins, John Avery, Hugh

DeWitt, David Parnas, etc. El ganador del Premio Nobel, Maurice Wilkins, introdujo a las implicaciones éticas del trabajo en el Proyecto de Genoma Humano (una década antes de su desarrollo). En el documento final, preparado por el cosmólogo Martin Rees, se sugirieron las fórmulas de los Juramentos de Buenos Aires y del Institute for Social Inventions de Londres.

**[63]** En 1990, la Federación Internacional de Trabajadores Científicos (WFSW) publicó un extenso Manifiesto sobre los Derechos y Responsabilidades de los Trabajadores Científicos. El texto completo se reproduce en *Scientific World*, vol. 34 (3): 2-5, 1990.

**[64]** En 1990, dentro de una conferencia organizada por las Naciones Unidas sobre ciencia, tecnología y seguridad mundial en la cual participaron unos 50 científicos, expertos técnicos, académicos, diplomáticos y dirigentes políticos de unos 23 países se debatió la idea de aplicación de un código de conducta para científicos. En este caso, los participantes consideraron que era difícil lograr un equilibrio adecuado entre la libertad científica y la responsabilidad social, aunque abrigaron la esperanza de que las opciones éticas se basarían en una perspectiva mundial y en un punto de vista de vasto alcance. Citado en Conferencia de las Naciones Unidas sobre nuevas tendencias en material de ciencia y tecnología: consecuencias para la paz y la seguridad internacionales, Sendal, Japón, 16-19 de abril de 1990. *Boletín de Desarme de las Naciones Unidas*, Junio de 1990, p. 15.

**[65]  Juramento de Arquímedes - Primera Versión (1990)**

*Considérant la vie d'Archimède de Syracuse qui illustra dès l'Antiquité le potentiel ambivalent de la technique,*

*Considérant la responsabilité croissante des ingénieurs et des scientifiques à l'égard des hommes et de la nature,*

*Considérant l'importance des problèmes éthiques que soulèvent la technique et ses applications,*

*Aujourd'hui, je prends les engagements suivants et m'efforcerai de tendre vers l'idéal qu'ils représentent :*

- *Je pratiquerai ma profession pour le bien des personnes, dans le respect des Droits de l'Homme1 et de l'environnement.*
- *Je reconnaîtrai, m étant informé au mieux, la responsabilité de mes actes et ne m'en déchargerai en aucun cas sur autrui.*
- *Je m'appliquerai à parfaire mes compétences professionnelles.*
- *Dans le choix et la réalisation de mes projets, je resterai attentif à leur contexte et à leurs conséquences, notamment des points de vue technique, économique, social, écologique... Je porterai une attention particulière aux projets pouvant avoir des fins militaires.*
- *Je contribuerai, dans la mesure de mes moyens, à promouvoir des rapports équitables entre les hommes et à soutenir le développement des pays économiquement faibles.*
- *Je transmettrai, avec rigueur et honnêteté, à des interlocuteurs choisis avec discernement, toute information importante, si elle représente un acquis pour la société ou si sa rétention constitue un danger pour autrui. Dans ce dernier cas, je veillerai à ce que l'information débouche sur des dispositions concrètes.*
- *Je ne me laisserai pas dominer par la défense de mes intérêts ou ceux de ma profession.*
- *Je m'efforcerai, dans la mesure de mes moyens, d'amener mon entreprise à prendre en compte les préoccupations du présent Serment.*
- *Je pratiquerai ma profession en toute honnêteté intellectuelle, avec conscience et dignité.*

*Je le promets solennellement, librement et sur mon honneur.*

**[66]** En Londres, en 1991, el Grupo de Científicos en Contra del Armamento Nuclear (SANA: Scientists Against Nuclear Arms) propone formalizar un compromiso para no tomar parte de las investigaciones financiadas por laboratorios militares.

**[67] Juramento Hipocrático para Científicos UCSB (1993)**

*Recognizing the responsibilities which accompany the privilege of my training, I pledge to give constant personal consideration to the morality of the means and the ends of any activities on which I engage my skills and intellect. I accept personal accountability for the social, environmental and human consequences of my scientific and technical work.*

**[68] Juramento del grupo de Jóvenes y Estudiantes de Pugwash, EEUU (1995)**

*I promise to work for a better world, where science and technology are used in socially responsible ways. I will not use my education for any purpose intended to harm human beings or the environment. Throughout my career, I will consider the ethical implications of my work before I take action. While the demands placed upon me may be great, I sign this declaration because I recognize that individual responsibility is the first step on the path to peace.*

**[69] Compromiso de Ingenieros y Científicos de INES (1995)**

1. *I acknowledge as a scientist or engineer that I have a special responsibility for the future of humankind. I share a duty to sustain life as a whole. I therefore pledge to reflect upon my scientific work and its possible consequences in advance and to judge it according to ethical standards. I will do this even though it is not possible to foresee all possible consequences and even if I have no direct influence on them.*

2. *I pledge to use my knowledge and abilities for the protection and enrichment of life. I will respect human rights, and the dignity and importance of all forms of life in their interconnectedness. I am aware that curiosity and pressure to succeed may lead me into conflict with that objective. If there are indications that my work could pose severe threats to human life or to the environment, I will abstain until appropriate assessment and precautionary actions have been taken. If necessary and appropriate, I will inform the public.*

3. *I pledge not to take part in the development and production of weapons of mass destruction and of weapons that are banned by international conventions. Aware that even conventional arms can contribute to mass destruction, I will support political efforts to bring arms production, arms trade, and the transfer of military technology under strict international control.*

4. *I pledge to be truthful and to subject the assumptions, methods, findings and goals of my work, including possible impacts on humanity and on the environment, to open and critical discussion. To the best of my ability I shall contribute to public understanding of science. I shall support public participation in a critical discussion of the funding priorities and uses of science and technology. I will carefully consider the arguments from such discussions which question my work or its impact.*

5. *I pledge to support the open publication and discussion of scientific research. Since the results of science ultimately belong to humankind, I will conscientiously consider my participation in secret research projects that serve military or economic interests. I will not participate in secret research projects if I conclude that society will be injured thereby. Should I decide to participate in any secret research, I will continuously reflect upon its implications for society and the environment.*

103

6. *I pledge to enhance the awareness of ethical principles and the resulting obligations among scientists and engineers. I will join fellow scientists and others willing to take responsibility. I will support those who might experience professional disadvantages in attempting to live up to the principles of this pledge. I will support the establishment and the work of institutions that enable scientists to exercise their responsibilities more effectively according to this pledge.*

7. *I pledge to support research projects, whether in basic or applied science, that contribute to the solution of vital problems of humankind, including poverty, violations of human rights, armed conflicts and environmental degradation.*

8. *I acknowledge my duty to present and future generations, and pledge that the fulfillment of this duty will not be influenced by material advantages or political, national or economic loyalties.*

[70] Stephen Jay Gould (1998) propone la necesidad de implementar el concepto básico de "no dañar por sobre todas las cosas" implícito en el Juramento Hipocrático, a modo de compromiso para evitar que los científicos dediquen sus habilidades a cuestiones bélicas.

### [71] Juramento Hipocrático para científicos - Michel Serres (1999)

*"I swear that, in whatever falls within my responsibility, I will never use my knowledge, my inventions and the applications that I might find for them, to promote destruction or death, to increase poverty and ignorance, to enslave people or to promote inequality, but instead to dedicate them to achieving equality between people, to help them live, to enhance their lives and make them more free."*

### [72] Propuesta del Juramento Hipocrático para Científicos en la Conferencia Mundial de la Ciencia en Budapest (1999)

*I promise to work for a better world, where science and technology are used in socially responsible ways. I will not use my education for any purpose intended to harm human beings or the environment. Throughout my career, I will consider the ethical implications of my work before I take action. While the demands placed upon me might be great, I sign this declaration because I recognize that individual responsibility is the first step on the path to peace.*

### [73] Juramento de Graduación para Científicos de la Universidad Griega (1986 -1999)

*I pledge: To foster science to the best of my ability and to strive and to glorify it all the time, and not to use it for gain or for the pursuit of vain glory, but so that the light of divine truth be further diffused to illuminate the many. Also to perform readily all that shall lead to piety, orderly behavior and dignified manners and never to disparage the teachings of others out of foolish vanity nor to attempt to refute their tenets with fallacies or to profess the opposite of what I know myself, and not to trade upon science and put the dignity of the disciples of the Muses to shame by disorderly conduct. As I accomplish this pledge of mine, so help me God.*

### [74] Peace Pledge Movement for Scientists in Japan (1999)

*I, undersigned below, pledge with honor and dignity: To the best of my knowledge, I will not participate in research, development, manufacture, acquisition and utilization of nuclear weapons as well as of other weapons of mass destruction.*

**[75]  Juramento de los Científicos (1999)**

*I will not, knowingly, carry out research which is to the detriment of humanity. If, in the event, research to which I have contributed is used, in my view, to the detriment of the human race then I shall work actively to combat its development.*

**[76]  Juramento de Arquímedes - Versión actualizada (2000)**

1. *I will practice my profession abiding by the ethics of human rights and I will be aware of my responsibility for mankind's natural heritage.*
2. *In all acts of my professional life I will assume my responsibility towards my institution, towards society and towards future generations.*
3. *I will pay special attention to promoting fair relations between all men and supporting the development of economically underprivileged countries.*
4. *I commit myself to explaining my choices to decision-makers and citizens, making these choices as transparent as possible.*
5. *I will give priority to the forms of management permitting broad cooperation between all the actors with a view to making everyone's work and innovations meaningful.*
6. *I pledge myself to respecting ethical codes as well as examining and using means of infor-mation and communication critically.*
7. *I will take special care to honing my professional skills in all aspects of technological, economic, human and social sciences involved in my work.*

**[77]  Juramento de la Asociación de Física Argentina (2000)**

*Juro trabajar por un mundo mejor, en el cual la ciencia y la tecnología sean empleadas en formas socialmente responsables. No usaré mi educación para ningún fin encaminado a dañar a seres humanos o al entorno y antes de actuar, consideraré las implicaciones éticas de mi tra-bajo. Realizo este juramento porque acepto que la responsabilidad individual es el primer paso en el sendero hacia la paz.*

**[78]  Juramento de Metz (2000)**

*El 16 de noviembre 2000, se celebró en Metz (Francia) la 4ª edición de la Conferencia sobre el Estado General de la Ética. En la misma se discutió la pertinencia de una carta que esboza los principios de responsabilidad de los científicos. Entre los editores del juramento, se encontraba el profesor de biología molecular Gilles-Eric Seralini,, el botánico Jean-Marie Pelt, el sociólogo Edgar Morin y Corinne Lepage, abogado y ex ministro de Medio Ambiente.*

*"Je jure d'être fidèle à l'éthique du respect des personnes et des vies humaines et de contribuer au développement de la connaissance et à la plus large diffusion du savoir. Je respecterai toutes les espèces dans leur biodiversité : ce respect inspirera mes actes et mes projets, notamment au cours de mes expérimentations sur les animaux ou les tissus humains. Je m'efforcerai de soulager les souffrances de tous les êtres vivants. Admis(e) à avoir accès à l'intimité tissulaire ou génétique des personnes, je tairai leur identité et m'astreindrai au secret médical. Même sous la contrainte, je ne ferai pas usage de mes connaissances contre les lois de l'humanité. Je préserverai l'indépendance nécessaire à l'accomplissement de ma mission. Je m'informerai et réfléchirai au sens de mes expérimentations et à leurs conséquences.*

*Je veillerai à ce que mes travaux et recherches ne soient pas utilisés à des fins de destruction ou de manipulation. Je respecterai les savoirs des ethnies et des sociétés traditionnelles. Je n'aurai garde d'oublier mes responsabilités à l'égard des générations présentes et futures. Je n'accepterai pas que des considérations de nationalité, de culture, de politique ou d'avantages*

*matériels me détournent de mes devoirs. J'interviendrai pour défendre, s'il m'en est donné l'occasion, l'ensemble de ces règles. Que les hommes et mes confrères m'accordent leur estime si je suis fidèle à mes promesses. Que je sois déshonoré(e) et méprisé(e) si j'y manque."*

**[79] Compromiso de Científicos e Ingenieros para renunciar a trabajar en armamento de destrucción en masa (2002)**

*I pledge never to participate in: the design, development, testing, production, maintenance, targeting, or use of nuclear, biological, or chemical weapons or their means of delivery; or in research or engineering that I have reason to believe will be used by others to do so.*

**[80] Juramento para Biocientíficos (2003)**

*At the moment of my becoming a member of the bioscience community, I do solemnly declare that I will respect the value and dignity of life, and conduct myself to honor this profession. I acknowledge that I have a special responsibility for promoting the welfare of humankind, and will so behave as to pursue and exercise my bioscience knowledge in an ethical and a socially responsible way. Never will I use my training to do harm to others or the environment; neither will I do anything to diminish social justice. Whatever action I take and career I choose, I will consider their moral implications. Since I realize that only ethically responsible bioscientists can hope to contribute to peace and security to people, thus promote genuine human flourishing. I make this declaration whole heartedly and upon my honor.*

**[81]** En el 2003, el Director de la División Ética de la Ciencia y la Tecnología del Sector Ciencias Sociales y Humanas de la UNESCO durante la Tercera Sesión de COMEST en Rio de Janeiro entre el 1-4 de diciembre de 2003, propone el establecimiento de un Juramento Hipocrático para Científicos basado en los distintos intentos anteriores. Esta presentación ha sido reproducida en este volumen (ver H. ten Have, "Hacia un Juramento Ético Universal para Científicos" en este volumen).

**[82] Juramento Hipocrático para Desarrolladores de Software (2004)**

*I solemnly pledge, first, to do no harm to the software entrusted to me; to not knowingly adopt any harmful practice, nor to adopt any practice or tool that I do not fully understand.*

*With fervor, I promise to abstain from whatever is deleterious and mischievous. I will do all in my power to expand my skills and understanding, and will maintain and elevate the standard of my profession. With loyalty will I endeavor to aid the stakeholders, to hold in confidence all information that comes to my knowledge in the practice of my calling, and to devote myself to the welfare of the project committed to my care.*

**[83] Código de Ética de la Sociedad de Energía Atómica de Japón (2005)**

*Preamble: We the members of the Atomic Energy Society of Japan (AESJ) amply recognize that nuclear technology brings tremendous benefits to humans but also raises the possibility of catastrophe. Based on that premise of recognition, with pride and a sense of mission of being directly engaged in the peaceful use of atomic energy, we energetically pursue human welfare and sustainable development while conserving global and local environments through the use of atomic energy.*

*Whenever we conduct atomic energy research, development, utilization, and education, under the principle of information disclosure, we the members of the AESJ make constant efforts to enhance our knowledge and skills, to keep pride and responsibility in our work, to keep a spirit*

*of self restraint, to maintain a harmonious relationship with society, to comply with laws and regulations, and to secure nuclear safety.*

*In order to implement these ideals, we the members of the AESJ have established herein fundamental canons of attitude and conduct.*

*Fundamental Canons:*

- *We shall restrict the use of atomic energy to peaceful purposes while endeavoring to solve the problems confronting humans.*

- *We shall hold the safety of the public paramount in the performance of our professional duties and through our conduct strive to obtain the public trust.*

- *We shall strive to improve our own professional competence and simultaneously to improve the professional competence of persons involved.*

- *We shall make every effort to be fully aware of our own professional capabilities. If a job requires an extraordinary proficiency beyond our capability, we shall pursue a course that will not cause serious damage to society.*

- *We shall strive to assure that all information we utilize is accurate and fulfill the obligation to disclose all information to the public in order to obtain the public trust.*

- *We shall respect truth and make our own judgments with fairness, justice, and impartiality.*

- *To the extent that contract clauses do not conflict with the provisions of all the laws as well as the norms in society, we shall seriously consider and faithfully fulfill the contracts related to our work.*

- *We shall conduct our work related to atomic energy with pride, and make sincere efforts to increase the esteem of that work.*

[84] **Código de Ética para evitar el Bioterrorismo (2005).**

**Code of ethics for the life sciences**

*All persons and institutions engaged in any aspect of the life sciences must:*

1. *Work to ensure that their discoveries and knowledge do no harm. (i) by refusing to engage in any research that is intended to facilitate or that has a high probability of being used to facilitate bioterrorism or biowarfare; and (ii) by never knowingly or recklessly contributing to development, production, or acquisition of microbial or other biological agents or toxins, whatever their origin or method of production, of types or in quantities that cannot be justified on the basis that they are necessary for prophylactic, protective, therapeutic, or other peaceful purposes.*

2. *Work for ethical and beneficent advancement, development, and use of scientific knowledge.*

3. *Call to the attention of the public, or appropriate authorities, activities (including unethical research) that there are reasonable grounds to believe are likely to contribute to bioterrorism or biowarfare.*

4. *Seek to allow access to biological agents that could be used as biological weapons only to individuals for whom there are reasonable grounds to believe that they will not misuse them.*

5. *Seek to restrict dissemination of dual-use information and knowledge to those who need to know in cases where there are reasonable grounds to believe that the information or knowledge could be readily misused through bioterrorism or biowarfare.*

6. *Subject research activities to ethics and safety reviews and monitoring to ensure that (i) legitimate benefits are being sought and that they outweigh the risks and harms; and (ii)* *involvement of human or animal subjects is ethical and essential for carrying out highly important research.*

7. *Abide by laws and regulations that apply to the conduct of science unless to do so would be unethical and recognize a responsibility to work through societal institutions to change laws and regulations that conflict with ethics.*

8. *Recognize, without penalty, all persons' rights of conscientious objection to participation in research that they consider ethically or morally objectionable.*

9. *Faithfully transmit this code and the ethical principles upon which it is based to all who are or may become engaged in the conduct of science.*

**[85] Juramento Hipocrático para Ciencias de la Vida (2006)**

*Knowledge of the life sciences is a privilege and with such privilege comes responsibility as the life sciences can be used for both benign and malign purposes. In entering into the community of life scientists, I pledge: to be honest, fair and as open as possible in my work: to act with due skill and diligence in all scientific work to ensure that the agents and equipment used in dangerous work are kept safe, not knowingly to engage in the development and production of biological and toxin weapons prohibited by international law, to give consideration to the potentially negative ramifications of my work, particularly before commencement and prior to publication, and to contribute to the development of safeguards and oversight mechanisms.*

**[86] Código de Ética y Juramento para Científicos (2007)**

*The seven principles are:*

1. *Act with skill and care, keep skills up to date;*

2. *Prevent corrupt practice and declare conflicts of interest;*

3. *Respect and acknowledge the work of other scientists;*

4. *Ensure that research is justified and lawful;*

5. *Minimize impacts on people, animals and the environment;*

6. *Discuss issues science raises for society;*

7. *Do not mislead; present evidence honestly.*

**[87] Los 10 Mandamientos para la Educación Superior (2007)**

1. *Strive to tell the truth.*
   *'Academic freedom', in the sense of following difficult ideas wherever they may lead, is possibly the fundamental 'academic' value.*

2. *Take care in establishing the truth.*
   *Adherence to the scientific method is critical here (as in the use of evidence and 'falsifiability' principle), but so too is the concept of social scientific 'warrant' and the search for 'authenticity' in the humanities and arts (leading, in particular, to concerns about rhetoric and persuasion independently of the grounds for conviction).*

3. *Be fair.*
   *This is about equality of opportunity, non-discrimination and perhaps even affirmative action. As has been pointed out, along with 'freedom' in the academic value-system goes 'respect for persons'.*

4. *Always be ready to explain.*
   *Academic freedom is a 'first amendment' and not a 'fifth amendment' right; it is about freedom of speech and not about protection from self-incrimination. It does not absolve*

*any member from the obligation of explaining his actions and, as far as possible, their consequences. Accountability is inescapable and should not be unreasonably resisted.*

5. *Do no harm.*
   *This is where the assessment of consequences cashes out (and presents our nearest equivalent to the Hippocratic Oath, to strive 'not to harm but to help'). It is about non-exploitation of either human subjects or the environment. It underpins other notions like 'progressive engagement'. It helps with really wicked issues like the use of animals in medical experiments.*

6. *Keep your promises.*
   *As previously suggested, 'business' excuses for retreating from or unreasonably seeking to renegotiate agreements are much less acceptable in an academic context.*

7. *Respect your colleagues, your students and especially your opponents.*
   *Working in an academic community means listening as well as speaking, seeking always to understand the other point of view and ensuring that rational discourse is not derailed by prejudice, by egotism, or by bullying of any kind.*

8. *Sustain the community.*
   *All of the values expressed so far are deeply communal. Obligations that arise are not just to the subject or to the professional community, or even to the institution in which you might be working at any one time, but to the family of institutions that make up the university sector, both nationally and internationally.*

9. *Guard your treasure.*
   *University and college communities, and those responsible for leading and managing them, are, in the traditional sense, 'stewards' of real and virtual assets and of the capacity to continue to operate responsibly and effectively.*

10. *Never be satisfied.*
    *Academic communities understood the principles of 'continuous improvement' long before it was adopted by 'management' literature. They also understand its merciless and asymptotic nature. The academic project will never be complete or perfect.*

**[88] Juramento de Graduación de la Universidad de Toronto (Canadá) (2007)**

*"I, [NAME], have entered the serious pursuit of new knowledge as a member of the community of graduate students at the University of Toronto.*

*"I declare the following:*

*"Pride: I solemnly declare my pride in belonging to the international community of research scholars.*

*"Integrity: I promise never to allow financial gain, competitiveness, or ambition cloud my judgment in the conduct of ethical research and scholarship.*

*"Pursuit: I will pursue knowledge and create knowledge for the greater good, but never to the detriment of colleagues, supervisors, research subjects or the international community of scholars of which I am now a member.*

*"By pronouncing this Graduate Student Oath, I affirm my commitment to professional conduct and to abide by the principles of ethical conduct and research policies as set out by the University of Toronto."*

**[89] Juramento de "Aventura Espacial" --Barranquilla, Colombia (2009)**

*Toma de juramento de mi Aventura Espacial;*
*Teniendo en cuenta que la conquista del espacio es de toda la humanidad, que como persona puedo contribuir a estos logros, inspirándome en las grandes perspectivas que se ofrecen a mi*

*país y para el mundo la utilización pacífica del espacio ultraterrestre, y teniendo en cuenta los Tratados del Espacio Ultraterrestre de las Naciones Unidas, juro y me comprometo a: Preservar el medio ambiente de la Tierra y del espacio ultraterrestre con fines pacíficos para mantener la Paz en la Tierra, contribuir con mis ideas y esfuerzos a la cooperación internacional en la exploración del espacio ultraterrestre, hacer lo posible para que las aplicaciones espaciales ayuden al desarrollo de mi gente, mi región, mi país y el mundo entero.*

**[90] Juramento Hipocrático para Ciencias de la Tierra (2009)**

*I vow to always:*

*Advise against any intervention into the functioning of Earth systems that I believe might harm humanity, the biosphere, atmosphere or other Earth systems upon which our well being depends.*

*Make clear to the public that scientific understanding of Earth systems is limited and that this makes all alterations of Earth systems inherently risky.*

*Describe, to the best of my knowledge and that of my discipline, the specific risks incurred by any intentional alteration of an Earth system, including the risks to humans, other organisms, and the systems that support life on Earth.*

*Ensure that whatever advice I give, I give for the benefit of humanity, remaining free of intentional distortion or personal bias.*